\documentclass[twocolumn]{aastex631}
\usepackage{rotating}

\newcommand\tstrut{\rule{0pt}{5.5ex}}
\newcommand\tstrutsmall{\rule{0pt}{5.5ex}}
\newcommand\bstrut{\rule[-6.0ex]{0pt}{0pt}}
\newcommand\bstrutsmall{\rule[-2.5ex]{0pt}{0pt}}
\newcommand{\msun}{$M_{\odot}$}
\newcommand{\logg}{$\log{g}$}
\newcommand{\teff}{${T}_{\mathrm{eff}}$}

\shorttitle{WD+WD Spectroscopic vs. Photometric Ages}
\shortauthors{Heintz et al.}

\graphicspath{{./}}
\begin{document}

\title{A Test of Spectroscopic Age Estimates of White Dwarfs using Wide WD+WD Binaries}

\correspondingauthor{Tyler M. Heintz}
\email{tmheintz@bu.edu}

\author[0000-0003-3868-1123]{Tyler M. Heintz}
\affiliation{Department of Astronomy \& Institute for Astrophyiscal Research, Boston University, 725 Commonwealth Ave, Boston, MA, 02215, USA}

\author[0000-0001-5941-2286]{J. J. Hermes}
\affiliation{Department of Astronomy \& Institute for Astrophyiscal Research, Boston University, 725 Commonwealth Ave, Boston, MA, 02215, USA}

\author[0000-0001-9873-0121]{P.-E. Tremblay}
\affiliation{Department of Physics, University of Warwick, Coventry, CV4 7AL, UK}

\author[0009-0002-6065-3292]{Lou Baya Ould Rouis}
\affiliation{Department of Astronomy \& Institute for Astrophyiscal Research, Boston University, 725 Commonwealth Ave, Boston, MA, 02215, USA}

\author[0000-0003-1862-2951]{Joshua S. Reding}
\affiliation{Department of Physics and Astronomy, University of North Carolina at Chapel Hill, Chapel Hill, NC 27599, USA}

\author[0000-0003-1970-4684]{B. C. Kaiser}
\affiliation{Department of Physics and Astronomy, University of North Carolina at Chapel Hill, Chapel Hill, NC 27599, USA}

\author[0000-0002-4284-8638]{Jennifer L. van Saders}
\affiliation{Institute for Astronomy, University of Hawai’i, 2680 Woodlawn Drive, Honolulu, HI 96822, USA}

\begin{abstract}
White dwarf stars have been used for decades as precise and accurate age indicators. This work presents a test of the reliability of white dwarf total ages when spectroscopic observations are available. We conduct follow-up spectroscopy of 148 individual white dwarfs in widely separated double-white-dwarf (WD+WD) binaries. We supplement the sample with 264 previously published white dwarf spectra, as well as 1292 high-confidence white dwarf spectral types inferred from their Gaia XP spectra. We find that spectroscopic fits to optical spectra do not provide noticeable improvement to the age agreement among white dwarfs in wide WD+WD binaries. The median age agreement is $\approx$$1.5\sigma$ for both photometrically and spectroscopically determined total ages, for pairs of white dwarfs with each having a total age uncertaintiy $<$ 20\%. For DA white dwarfs, we further find that photometrically determined atmospheric parameters from spectral energy distribution fitting give better total age agreement ($1.0\sigma$, 0.2 Gyr, or 14\% of the binary's average total age) compared to spectroscopically determined parameters from Balmer-line fits (agreement of $1.5\sigma$, 0.3 Gyr, or 28\% of binary's average total age). We find further evidence of a significant merger fraction among wide WD+WD binaries: across multiple spectroscopically identified samples, roughly 20\% are inconsistent with a monotonically increasing initial-final mass relation. We recommend the acquisition of an identification spectrum to ensure the correct atmospheric models are used in photometric fits in order to determine the most accurate total age of a white dwarf star.
\end{abstract}


\section{Introduction}\label{sec:intro}
Stellar age is a notoriously difficult parameter to measure, especially for stars that experience little observable evolution in a color-magnitude diagram (CMD) over their lifetimes. The ages of field stars are vital for understanding the evolution of stellar and galactic systems, but there is no single method suitable for the wide range of stellar types observed. Thus, stellar ages are determined through a variety of indirect methods, including their position on the CMD (isochrone fitting, e.g., \citealt{2018ApJ...866...99B}, \citealt{2022Natur.603..599X}), rotation periods (gyrochronology, e.g., \citealt{2007ApJ...669.1167B}, \citealt{2019ApJ...871...39M}, \citealt{2020ApJ...888...43C}), activity (e.g., \citealt{2008ApJ...687.1264M}), stellar pulsations (asteroseismology, e.g., \citealt{2014ApJS..210....1C}), and kinematic properties (e.g., \citealt{2009A&A...501..941H, 2019ApJ...886..100C}, \citealt{2021AJ....161..189L}). The method most widely used to determine precise stellar ages is isochrone fitting to stellar clusters (e.g., \citealt{2018AJ....156..165C}), which provides the most direct way of measuring stellar ages. Unfortunately this method cannot be widely applied to stars in the field, especially cool main-sequence stars, but can be used as a calibrator for the previously mentioned techniques.

All previously mentioned methods can only be applied to specific classes of stars and are not universally useful. Isochrone fitting is most useful for populations of stars born at the same time or stars that experience relatively rapid evolution on the CMD (e.g., subgiant stars, \citealt{2022Natur.603..599X}). Gyrochronology works well for young and intermediate-mass solar-type stars, as their rotations are slowed through interactions with their stellar winds \citep{1972ApJ...171..565S}, but becomes less effective in old Sun-like stars \citep{2016Natur.529..181V} that undergo weakened magnetic braking, and in low-mass stars at a wide range of ages that undergo core-envelope decoupling \citep{2019ApJ...877..157L, 2020ApJ...904..140C}. It fails entirely in stars without convective envelopes \citep{1967ApJ...150..551K}.

White dwarf stars are ubiquitous throughout the Galaxy and are the final endpoint of 97\% of all stars \citep{2001PASP..113..409F}. Their relatively simple and characteristic cooling allows for accurate and precise age determinations through robust cooling models (e.g., \citealt{2020ApJ...901...93B}). Their total ages (the combined time as a white dwarf and the time in the main-sequence and giant phases) can be inferred through the use of an initial-final mass relation (IFMR, \citealt{1983A&A...121...77W,2008MNRAS.387.1693C,2008ApJ...676..594K,2009ApJ...692.1013S,2009ApJ...693..355W,2018ApJ...866...21C,2024MNRAS.527.3602C}), which relates the zero-age main-sequence (ZAMS) mass to the final white dwarf mass in tandem with stellar evolutionary grids (e.g., MESA, \citealt{2016ApJS..222....8D}). Thus, white dwarfs both serve as additional calibrators to the methods mentioned previously (e.g. \citealt{2023MNRAS.526.4787R}) and provide an understanding of the limitations of white dwarf total ages. These are crucial for their future use as stellar age indicators.

The total age of a white dwarf is strongly dependent on how much mass it lost during its giant phase, and thus the IFMR chosen, especially for the lowest-mass white dwarfs (M~$<$ 0.63\,\msun). The IFMR is poorly constrained at these low masses due to the lack of white dwarfs with lower-than-average masses in observable stellar clusters. This problem has been investigated using both white dwarfs in wide binaries \citep{2012ApJ...746..144Z, 2021ApJ...923..181B, 2024MNRAS.527.9061H} and the distribution of white dwarfs in the CMD \citep{2018ApJ...860L..17E,2024MNRAS.527.3602C}. However, the resulting IFMRs from these studies provide a wide range of predicted initial masses, especially for lower-mass white dwarfs. The IFMR is also likely affected by the initial rotation and metallicity of the white dwarf progenitors \citep{2019ApJ...871L..18C}.

With the advent of the Gaia mission \citep{2016A&A...595A...1G, 2018A&A...616A...1G, 2021A&A...649A...1G}, samples of widely separated binaries have increased by an order of magnitude \citep{2015ApJ...815...63A, 2018MNRAS.480.4884E, Tian}. In Gaia Early Data Release 3 (EDR3), \citet{2021MNRAS.506.2269E} found more than a million wide ($>$100\,au separation) binaries. These wide binaries provide unique opportunities to test the levels of precision and accuracy obtainable from white dwarf total ages. 

In \cite{2021MNRAS.506.2269E}, there are more than $18{,}000$ high-confidence wide white dwarf plus main-sequence binaries (WD+MS). Previous work used white dwarfs in wide binaries to constrain the ages of their binary companions (e.g., \citealt{2019ApJ...870....9F}, \citealt{2021ApJS..253...58Q}). These works provide large catalogs (92 and 4050, respectively) of stellar ages derived from coeval white dwarf companions using their available broad-band photometry. To fully realize the potential of the large sample of wide WD+MS binaries from \cite{2021MNRAS.506.2269E}, a thorough understanding of the accuracy and precision achievable with white dwarf total ages is needed.

In a previous study \citep{2022ApJ...934..148H}, we tested the accuracy and precision of white dwarf total ages using the sample of more than 1500 widely separated double-white-dwarf binaries (WD+WD) from \cite{2021MNRAS.506.2269E}, as well as two dozen new triple systems with at least two white dwarfs. In that work, we used broad-band photometry to determine the total ages of the white dwarfs. We found that that roughly $20-35$\% of wide WD+WD binaries showed more massive white dwarfs to have shorter cooling ages than their wide companions, implying that the less massive white dwarf came from a more massive main-sequence star, which is inconsistent with the current model of a monotonically increasing IFMR. With only photometric data available for the majority of the sample, it was difficult to determine the exact reason for these discrepancies. We aim to test here whether spectroscopic observations can help illuminate the intrinsic causes of the previously found discrepancies.

In recent years, there has been evidence of disagreements between atmospheric parameters derived from photometry and spectroscopy \citep{2019MNRAS.482.5222T, 2019ApJ...876...67B, 2021MNRAS.508.3877G, 2023MNRAS.518.3055O, 2023MNRAS.526.5800S}. The disagreement for an individual white dwarf can be on the order of a few percent in temperature and on the order of 0.1 dex in surface gravity (\logg). Although these disagreements appear marginal, they can translate into large differences in derived total ages for individual white dwarfs, especially for the lowest-mass white dwarfs where the progenitor lifetime is a steep function with respect to mass.

In this work, we present follow-up spectroscopy for 148 individual white dwarfs in wide WD+WD binaries along with previously published spectroscopic observations of 264 individual white dwarfs, and investigate the effects on the accuracy and precision of their derived total ages compared to photometrically determined ages. Section~\ref{sec:obs} presents the observations and reduction steps, Section~\ref{sec:wd_ages} discusses the methods used to determine the white dwarf atmospheric parameters and ages, Section~\ref{sec:comparisons} presents comparisons of the derived atmospheric parameters and ages from photometry and spectroscopy, and Section~\ref{sec:conclusions} summarizes and discusses the implications of this work.

\section{Observations and Reduction}\label{sec:obs}
\subsection{LDT+DeVeny Spectroscopy}\label{sec:LDT_reduce}
We obtained follow-up, low-resolution optical spectra over 12 nights of 84 individual white dwarfs in wide WD+WD binaries with the DeVeny Spectrograph \citep{2014SPIE.9147E..2NB} on the 4.1-m Lowell Discovery Telescope (LDT) in Happy Jack, Arizona, United States. We used a 300~line~mm$^{-1}$ grating with a 3-arcsec slit, which resulted in wavelength coverage from $3500-7000$\,\AA\ and a seeing-limited resolution, typically $13-15$\,\AA.

For the LDT observations, we were restricted to sufficiently separated systems ($\gtrsim 3$~arcseconds) and sufficiently bright targets ($G\lesssim 19.0$~mag) to obtain enough signal to perform spectroscopic fits. In addition, observations were not always collected at the parallactic angle; instead when both stars were in the field of view, we rotated the slit to the appropriate position angle to capture spectra of both white dwarfs simultaneously, substantially cutting our acquisition time. LDT+DeVeny does not have an atmospheric dispersion corrector (ADC),  so we planned our observations to occur when the slit position angle was as close to the parallactic angle as possible. A log of our observations is shown in Table~\ref{tab:new_obs}.

The LDT observations were reduced and the spectra were optimally extracted using the spectroscopic data reduction pipeline PypeIt \citep{2020JOSS....5.2308P}. A flexure correction is applied using available sky lines in the spectra. The 1-D spectra extracted from each exposure were co-added to be fit for atmospheric parameters (see Section~\ref{sec:specfits}). We observed one flux standard per night from the ESO X-Shooter spectrophotometric standard stars to flux calibrate the spectra. 

\subsection{SOAR+Goodman Spectroscopy}
For our southern targets, we obtained follow-up optical spectra over 17 nights of 64 individual white dwarfs in wide WD+WD binaries using the Goodman High-Throughput Spectrograph \citep{2004SPIE.5492..331C} on the 4.1-m Southern Astrophysical Research (SOAR) Telescope located at Cerro Pach\'{o}n, Chile. We used the 930~line~mm$^{-1}$ grating with grating and camera angles of 13.0 and 24.0 degrees, respectively, which yield a wavelength coverage of $3600-5200$\,\AA. We used slit widths of $1.0$ arcsecond and $3.2$ arcseconds depending on seeing or weather conditions. The 1.0-arcsecond slit setup resulted in a resolution of 3\,\AA, while the resolution for the 3.2-arcsecond slit was seeing-limited, resulting in resolutions ranging from $3.5-8$\,\AA. Observations are detailed in Table~\ref{tab:new_obs}.

As with our LDT+Deveny observations, we often rotated our slit during SOAR+Goodman observations in order to simultaneously capture both white dwarfs when possible. In these cases we used the ADC on SOAR \citep{2018SPIE10700E..3ZB}, minimizing color-dependent slit losses that could negatively affect flux calibration. We selected targets for this follow-up randomly from the observable wide WD+WD binary sample from \cite{2018MNRAS.480.4884E}.

The SOAR+Goodman observations were also reduced and optimally extracted using the same spectroscopic data reduction pipeline PypeIt \citep{2020JOSS....5.2308P} mentioned in Section~\ref{sec:LDT_reduce}. For the SOAR observations, no flexure correction is applied since there are not sufficient sky lines in the wavelength range observed. Thus, the wavelength solutions are unreliable and likely are systematically shifted. This should not affect our fitting described in Section~\ref{sec:specfits} since the radial velocity of the white dwarf is a free parameter that can account for these systematic shifts. We adapted the existing \texttt{soar\_goodman\_red} and \texttt{soar\_goodman\_blue} PypeIt classes for our setup, which are suited for the 400~line~mm$^{-1}$ grating. The only change to the classes was to not use the template wavelength solution for the 400~line~mm$^{-1}$ grating, and instead use the ``holy-grail'' algorithm in PypeIt to generate the wavelength solution for our 930~line~mm$^{-1}$ setup.

\subsection{Spectral Types}\label{sec:spectypes}
With the 148 extracted LDT and SOAR spectra, we determine the spectral type of each white dwarf by visual inspection. We find that 104 (67\%) are hydrogen-dominated (DA) white dwarfs, one (0.7\%) is a helium-dominated (DB) white dwarf, 34 (23\%) exhibit flat continua (DC), three (2\%) are metal-polluted (DAZ/DZ) white dwarfs, six (4\%) are magnetic hydrogen-dominated (DAH) white dwarfs, and one (0.7\%) is a carbon-dominated (DQ) white dwarf. As a comparison, in the volume-limited 40 pc sample \citep{2024MNRAS.527.8687O}, 50\% are DA white dwarfs, 0.8\% are DB white dwarfs, 27\% are DC white dwarfs, 11\% show metal-pollution, 6\% are DAH white dwarfs, and 4\% are DQ white dwarfs. Our sample is more similar to a magnitude-limited sample; in \cite{2013ApJS..204....5K}, 65\% are DA white dwarfs, 4.7\% are DB white dwarfs, 2.8\% are DC white dwarfs, 2\% are metal-polluted, 3.2\% are DAH white dwarfs, and 1.1\% are DQ white dwarfs. The sample in \cite{2013ApJS..204....5K} from the Sloan Digital Sky Survey (SDSS) is biased towards bluer objects, which explains the discrepancy in percentage of DC white dwarfs between SDSS and our follow-up. 

\begin{figure}
    \centering
    \includegraphics[width=0.95\columnwidth]{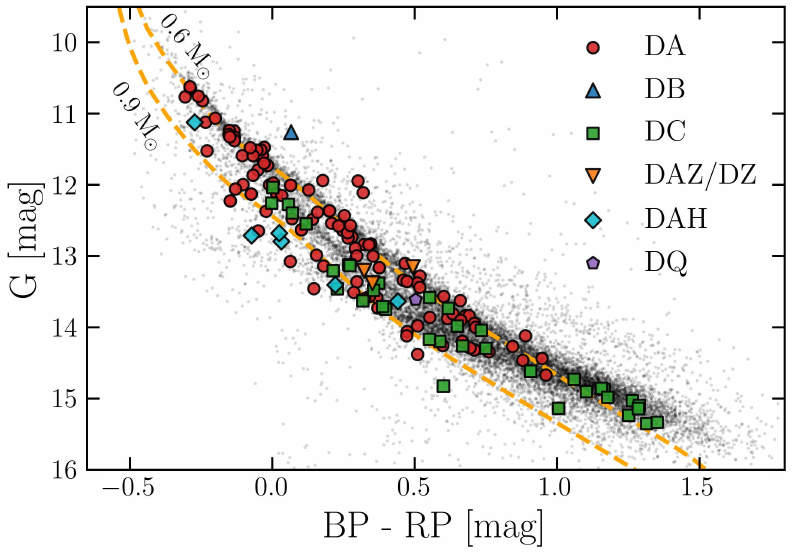}
    \caption{A Gaia CMD with spectral types of the newly observed white dwarfs taken from LDT and SOAR spectra. The grey points in the background are a sample of white dwarfs within 100 pc from \cite{2021A&A...649A...1G} with $<$$5$\% parallax uncertainties. The yellow dashed curves are white dwarf cooling sequences for a 0.6 \msun\ and 0.9 \msun\ DA white dwarf from \cite{2020ApJ...901...93B}}
    \label{fig:cmd}
\end{figure}

A CMD of the 148 newly observed white dwarfs with discernible spectral types is shown in Figure~\ref{fig:cmd}. As expected, the coolest objects at the lower-right-hand region of the CMD are DC white dwarfs. Many of the DAH white dwarfs fall in the massive regions of the CMD, as expected \citep[e.g.,][]{2020IAUS..357...60K}. The full list of newly observed white dwarf spectral types can be found in Table~\ref{tab:new_obs}.

\subsection{Spectroscopic Observations from the Literature}

To compliment our new spectroscopic observations from LDT and SOAR, we also take spectral types and atmospheric parameters from the literature. We find 261 white dwarfs with previously published spectral types, and 161 which have accompanying atmospheric parameters from fitting their spectra\footnote{We have not included spectral types from the recent result exploring the IFMR by \citet{2024MNRAS.527.9061H}, but note that 40 of their objects are represented in our new LDT and SOAR data; see Section~\ref{sec:specfits}.}. The previously published spectral types and atmospheric parameters used in this work can be found in Table~\ref{tab:lit_obs} along with the relevant references. 

The previously published atmospheric parameters compiled in this work come from a variety of sources that use different sets of spectral models; mainly from \cite{2010MmSAI..81..921K} and \cite{2011ApJ...730..128T}. There could be systematic effects introduced into our results based on two different sets of models and fitting procedures. However, we find no discernible systematic when comparing the two subsets of white dwarf atmospheric parameters determined through the two different models to photometric parameters as discussed in Section~\ref{sec:spec_vs_phot}, and do not believe the atmospheric parameters from disparate sources systematically affect our results.

We also cross-match all white dwarfs in wide WD+WD binaries with newly determined spectral types from \citet{2024A&A...682A...5V}, whose spectral typing is determined through a neural-network-based pipeline to select white dwarf candidates and classify their Gaia XP spectra. The pipeline is trained on previously labeled white dwarf spectra from SDSS DR16 \citep{2020ApJS..249....3A} and achieves an accuracy greater than 90\%. We find 1525 white dwarfs that have spectral types determined in this way, with 1292 high-confidence determinations (P~$>$~0.65). We only use the high-confidence determinations in this work.

We test the spectral types determined from the Gaia XP spectra by comparing the spectral types from \cite{2024A&A...682A...5V} to the spectral types gathered in this work from the literature and our follow-up spectra. This consists of 322 individual white dwarfs with high-confidence determinations from \cite{2024A&A...682A...5V}. We find that 8.7\% of the spectral types disagree between the Gaia XP determinations and the higher-resolution determinations. This percentage ignores complex spectral type determinations which were not considered in \cite{2024A&A...682A...5V} (e.g. DAH, DBA, DAZ, etc.). The majority of the disagreements ($\approx 70\%$) come from the higher-resolution spectrum being labelled as a DC while the Gaia XP spectra claim a different spectral type (e.g. DA, DB, or DQ). In our analysis, we ignore the spectral type from the Gaia XP spectra when another spectral type has been determined from higher-resolution spectra.

\section{White Dwarf Age Determinations}\label{sec:wd_ages}
We measure the total ages of the white dwarfs in our sample from both photometric and spectroscopic determinations of their atmospheric parameters. 

We take photometric ages from \citet{2022ApJ...934..148H}, which were determined from fits to their spectral energy distributions (SEDs), and assumed a DA spectral type for all white dwarfs in the sample. In brief, a Markov chain Monte Carlo (MCMC) approach is used to fit synthetic fluxes calculated from model DA spectra from \cite{2010MmSAI..81..921K} to the observed fluxes in all-sky surveys including Gaia, SDSS, the Panoramic Survey Telescope and Rapid Response System (Pan-STARRS), the SkyMapper Southern Sky Survey (SkyMapper), and the Two Micron All-Sky Survey (2MASS). The weighted-mean parallax of the binary is used to get more precise determinations of the components' atmospheric parameters and subsequent total ages, often improving on the mass uncertainties determined from Gaia photometric fitting by \citet{2021MNRAS.508.3877G}.

The spectroscopic ages in this work are derived from atmospheric parameters fit to DA spectra from LDT and SOAR, supplemented by literature determinations. Literature values come from a variety of sources (see Table~\ref{tab:lit_obs}), with some atmospheric parameters determined before 3-D corrections to convective-atmosphere fits were readily available (usually $<$\,$15{,}000$\,K for DAs). We apply 3-D corrections from \cite{2013A&A...559A.104T} to the effective temperatures and surface gravities for all DA white dwarfs that have not been previously corrected.

\subsection{Spectral Fitting Methods} \label{sec:specfits}
We determine spectroscopic atmospheric parameters for only the DA white dwarfs in our newly observed sample. To determine the ages of these DA white dwarfs, we first estimate their effective temperatures and surface gravities by fitting their Balmer absorption lines using DA models from \cite{2011ApJ...730..128T}, with 3-D corrections from \cite{2013A&A...559A.104T}. The fitting procedure we employ has been outlined in previous works (e.g., \citealt{2011ApJ...743..138G},  \citealt{2020MNRAS.497..130T}). The individual Balmer lines are normalized to a continuum determined by fitting a full model spectrum to the observations with a polynomial with 10 free parameters. Then, a $\chi^{2}$ minimization is performed between the observed and model line profiles, convolved with the instrumental resolution. In this work, we conduct this minimization on the lines H$\beta$ through H9. This process produces hot and cold solutions on either side of the Balmer maximum ($\approx$$13{,}000$\,K). We choose the solution with the closest effective temperature to the photometrically determined atmospheric parameters. We impose a minimum uncertainty of 1.2\% on temperature and 0.038 dex on the surface gravity to account for any systematics in the flux calibration and fitting procedure \citep{2005ApJS..156...47L}. The corresponding fitted temperatures and surface gravities can be found in Table~\ref{tab:new_obs}.

To validate our spectroscopic fits to the newly observed DAs, we compare derived temperatures and surface gravities from an overlapping sample of 40 white dwarfs observed in common with \citet{2024MNRAS.527.9061H}, who used spectroscopic observations of wide WD+WD binaries to constrain the IFMR. \cite{2024MNRAS.527.9061H} use a different approach to derive the temperatures and surface gravities as compared to our work: they simultaneously fit the photometry and spectroscopy of both white dwarfs in the binary. We do not find any significant differences between our derived parameters and those derived in \cite{2024MNRAS.527.9061H}; their effective temperatures are on average 2\% cooler, with a standard deviation of 15\% about that value, and their surface gravities are on average 0.004 dex smaller (below our imposed minimum uncertainty), with a standard deviation of 0.02 dex about that value. We do not include their atmospheric parameters since they simultaneously include both photometric and spectroscopic information in the fitting procedure.

In rare cases, our independent fitting procedures reveal possible constraints on system histories and architectures. Gaia DR3 4311973098757316608 (WDJ185400.72+104854.69) shows a significant discrepancy between photometric and spectroscopic atmospheric parameters ($\Delta \mathrm{T_{eff}} = 2100$~K and $\Delta$\logg $= 0.4$~dex). The low photometric surface gravity (7.64 dex) and large temperature discrepancy ($T_{\mathrm{phot}} = 9100$ K; $T_{\mathrm{spec}} = 11{,}200$ K) suggest Gaia DR3 4311973098757316608 could be a double-degenerate binary and the system is, in fact, a heirarchical triple. The low photometric surface gravity from the source being overluminous in the Gaia CMD suggests that the photometry is blended by a nearby star. This is also supported by the photometric temperature being less than the spectroscopic temperature where the photometric temperature has been influenced by a cooler, fainter companion. Additionally, Gaia DR3 6642323156995855232 (WDJ194406.24-534220.66) and Gaia DR3 4855870439806768512 (WDJ035012.13-383057.28) both show broad line cores which could be indicative of the presence of a modest magnetic field (e.g., \citealt{1998AA...338..612K}). We remove all three objects from future analysis as a result; the corresponding fits can be seen in Figure~\ref{fig:DA_fits}.

\subsection{White Dwarf Age Determinations}\label{sec:tot_age_determ}
To derive a total age of a white dwarf, a cooling age and progenitor lifetime must be determined. To derive cooling ages and masses from the spectroscopic temperatures and surface gravities of the DAs in our sample, we use the thick H-layer DA cooling sequences from \citet{2020ApJ...901...93B}.

To determine the progenitor ZAMS lifetimes of the white dwarfs in our sample, we use a theoretically motivated IFMR to connect the initial ZAMS mass of a star to its final white dwarf mass. We use the IFMR described in \cite{2022ApJ...934..148H}, which uses a theoretical IFMR from \cite{2016ApJ...823...46F} with an offset that is fit using white dwarfs in solar-metallicity clusters \citep{2008MNRAS.387.1693C, 2015ApJ...807...90C,2016ApJ...818...84C} and zero-rotation, solar-metallicity stellar models from Modules for Experiments in Stellar Astrophysics (MESA, \citealt{2011ApJS..192....3P}, \citealt{2013ApJS..208....4P}, \citealt{2015ApJS..220...15P}).

In all cases, the dominant contribution to our measured uncertainty on the total white dwarf age hinges on the precision of the white dwarf mass (e.g., \citealt{2004ApJS..155..551H,2012Natur.486...90K}).

\section{Age Comparisons and Results}\label{sec:comparisons}

\subsection{Spectroscopic vs. Photometric Parameters}\label{sec:spec_vs_phot}
\begin{figure}[t]
    \centering
    \includegraphics[width=0.95\columnwidth]{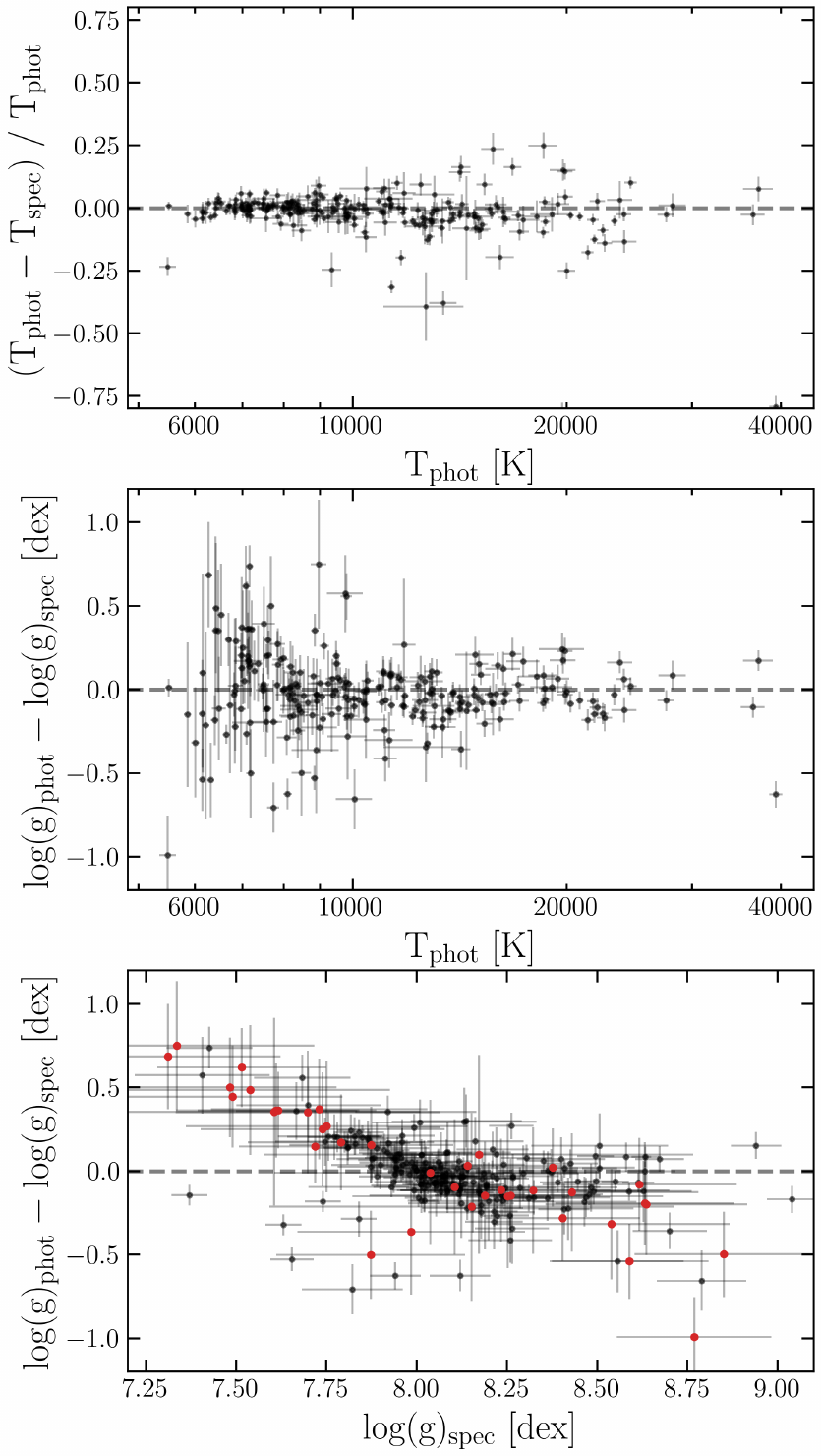}
    \caption{A comparison of derived effective temperatures and surface gravities from photometry and spectroscopy. The top panel shows that the temperature agreement is within 1\% across all temperatures. The middle panel shows that there is no trend with surface gravity, and that there is a slight systematic offset of 0.03 dex, where spectroscopy systematically results in slightly more massive solutions. The bottom panel compares photometric and spectroscopic surface gravities, and shows an interesting trend discussed more at length in Appendix~\ref{sec:appendix} that arises simply from a sample strongly peaked at \logg=8.0 with larger uncertainties in spectroscopic fits. The red points in the bottom panel highlight spectroscopic fits with \logg\ uncertainties $>0.2$~dex.}
    \label{fig:phot_vs_spec}
\end{figure}

To assess the total ages derived from spectroscopy, we first compare in Figure~\ref{fig:phot_vs_spec} the derived white dwarf atmospheric parameters to those determined from photometry by \cite{2022ApJ...934..148H}, which utilized the DA white dwarf model atmospheres of \cite{2010MmSAI..81..921K}. For our spectroscopic fits we use the DA model atmospheres of \cite{2011ApJ...730..128T} (see Section~\ref{sec:specfits}). 

Comparing the atmospheric parameters derived from spectroscopy and photometry, we find the effective temperatures derived from both methods are consistent; across the entire sample, the photometric temperatures are on average 1\% cooler than the spectroscopic temperatures, with a standard deviation of 10\%. This systematic shift is comparable to the minimum uncertainty imposed on the spectroscopic effective temperatures, and is encouraging given photometric fits to SEDs are highly sensitive to the effective temperature. In addition, we find the photometric surface gravities in our sample are systematically 0.03 dex smaller than those derived from spectroscopy, with the \logg\ differences between photometry and spectroscopy having a standard deviation about that value of 0.23 dex (see middle panel of Figure~\ref{fig:phot_vs_spec}). A difference of 0.23 dex results in mass differences of order 0.2\msun. This has a drastic effect on the derived progenitor masses, and subsequently the progenitor lifetimes.

To make sure the use of photometric parameters derived from a different set of models from spectroscopy does not change the trends seen in this section, we fit the photometry of a subset of our sample of 1200 individual white dwarfs with the DA models from \cite{2011ApJ...730..128T}. These fits use the same fitting technique described in \cite{2022ApJ...934..148H}. 

We find the atmospheric parameters from \citet{2022ApJ...934..148H} derived using photometric fits to \cite{2010MmSAI..81..921K} models are entirely consistent, within the reported uncertainties, to photometric fits using the \citet{2011ApJ...730..128T} models. We only see notable discrepancies below $<$5000\,K, where opacities likely differ between the models.

The photometric fits from \cite{2022ApJ...934..148H} use photometry from a variety of all-sky surveys, but not all white dwarfs fit have the same photometric bands available. \cite{2019ApJ...876...67B} found that when using the photometric method, SDSS u-band likely yields the most accurate \teff\ measurements. However, this is still debated in the literature, since SDSS u-band has an ad-hoc atmospheric transmission correction to align it on the AB system \citep{2006ApJS..167...40E}; Gaia space-based photometry may be more accurate \citep{2019MNRAS.482.5222T,2019MNRAS.482.4570G,2020MNRAS.499.1890M}.

For white dwarfs with and without SDSS u-band, we find the same median temperature difference between photometry and spectroscopy as above (1\%) and conclude that the absence of SDSS u-band in the photometric fits is not significantly biasing the temperatures. We do find that the inclusion of SDSS u-band improves the agreement of surface gravities between spectroscopy and photometry. For the sample of white dwarfs without SDSS u-band, the surface gravities from spectroscopy are 0.06 dex larger on average which could explain the apparent systematic shift of 0.03 dex seen for the full sample. We do not find that the subset of white dwarfs with SDSS u-band measurably changes the conclusions from the total age agreements from photometric fits discussed in Section~\ref{sec:acc_prec}.

We also find an apparent systematic trend wherein the lowest surface gravities determined from spectroscopy deviate more strongly from those determined from photometry (see bottom panel of Figure~\ref{fig:phot_vs_spec}). We find that the best explanation is not an actual physical difference in the fits, but rather that the smaller \logg\ values from spectroscopy are statistical outliers with larger uncertainties than the photometrically determined values, drawn from a population with surface gravities strongly peaked near \logg=8.0 (see more detailed discussion and simulations Appendix~\ref{sec:appendix}). We find the white dwarfs with higher \logg\ uncertainties from spectroscopy ($>0.2$~dex) are mostly cool white dwarfs with effective temperatures $<$8000~K where the Balmer lines are weak. For these lower effective temperatures, the photometric method is better able to constrain the true surface gravity and should be implemented instead.

These temperature and surface gravity differences are explored further based on a total age comparison for our hypothetically coeval population of wide binaries in the following sections.

\subsection{White Dwarf Age Accuracy and Precision} \label{sec:acc_prec}

To assess the accuracy and precision of white dwarf parameters determined from both spectroscopy and photometry, we compare the agreement of the total ages using the parameters from each method.

To start, we look at the sample of white dwarfs with photometric ages that have average total age uncertainties less than 20\%, chance alignment factors less than 0.1 \citep{2021MNRAS.506.2269E}, and binary separations greater than 2 arcseconds from \cite{2022ApJ...934..148H}. This results in a subsample of 192 binaries in which these conditions are met for both white dwarfs. In \cite{2022ApJ...934..148H} a DA atmosphere was assumed for all white dwarfs, and this sample will include DA and non-DA white dwarfs. The 384 white dwarfs in 192 binaries have a mean Gaia $G$ magnitude of 19.2, a mean mass of 0.81\,\msun\,, and span the full temperature range of white dwarfs from 4700~K to $30{,}000$~K. The relatively high mean mass is due to the fact that low-mass white dwarfs have larger uncertainties on their progenitor lifetimes and are preferentially removed by the restriction on total age uncertainties. The mean Gaia $G$ magnitude is slightly brighter than the mean magnitude of the full wide WD+WD sample from \cite{2022ApJ...934..148H} (19.5~mag). When comparing the photometric ages of the two components of the binary, we find good agreement. We find that 58\% of the sample has total ages that agree within 28.3\% (20\% added in quadrature to reflect, at most, 20\% total age uncertainties for each white dwarf) of the system's weighted average age. The median absolute total age difference ($\Delta \tau$) is 0.6~Gyr (or 27\% of the system's average total age). Another way to quantify the agreement between the total ages in each binary is to look at the uncertainties ($\sigma_\Delta\tau$) on the total age differences of each system and quote the median $\sigma_\Delta\tau$ of the sample. When looking at the uncertainties on the total ages, we find average agreement within $2.2\sigma$. It was found in \cite{2022ApJ...934..148H} that for more massive white dwarfs ($>0.67$ \msun), the reported uncertainties were too small to account for total age differences between the components in wide WD+WD binaries. If we use the recommended inflation factors for the uncertainties, we find average age agreement within $1.4\sigma$.

To compare to the accuracy of white dwarf ages from spectroscopy, we use the sample of 70 DA+DA systems with atmospheric parameters from spectroscopic fits. We perform the same cuts on average total age uncertainties, chance alignment factors, and binary separations, which produce a sample of 14 DA+DA systems with $<$20\% spectroscopic total age uncertainty on each component. These systems are comparatively much brighter, with a mean Gaia $G$ magnitude of 17.5 (and mean spectroscopic S/N~=~35). The mean mass of 0.85\,\msun\ is similar to that of the sample of precise photometric ages discussed above, and temperatures span a large range from 6500~K to $28{,}000$~K. Of these 14 systems, seven (50\%) have total age agreement within 28.3\% of the weighted average system age. The median $\Delta \tau$ is 0.3~Gyr (or 28\% of the system's average age). In terms of $\sigma_\Delta\tau$, the average agreement is within $2.1\sigma$. If we apply the same inflation factors that were determined for photometric ages, this agreement improves to $1.5\sigma$. Among this same sample of systems we compare photometrically determined total ages: eight out of 14 systems (57\%) have total age agreement within 28.3\% of the average age. The average agreement in measured uncertainty is $1.2\sigma$, and improves to $0.6\sigma$ on average when inflation factors are applied. The photometric total ages of stars in this sample do not necessarily have $<$20\% total age uncertainty, which is responsible for the increased level of $\sigma_\Delta\tau$ agreement. The sample size of 14 is small, which makes it easy for statistical outliers to affect our conclusions. An increase in the number of systems with high-resolution and high-S/N spectra will help provide stronger conclusions in the future. Since this sample is small, we tabulate these systems in Table~\ref{tab:spec_precise}.

One might expect the spectroscopic method to provide better age constraints for hotter DA white dwarfs, since cooler DAs show weaker lines and thus, may provide less reliable constraints on the atmospheric parameters. Due to the small sample size, we are unable to split the sample into two statistically significant temperature regimes of cool ($<$$10{,}000$~K) and hot/warm ($>$$10{,}000$~K) but when doing so, we find that the four systems with two cool white dwarfs show stronger total age agreement than the nine systems with two hot white dwarfs.

We conduct a similar investigation into any temperature dependence on our conclusions for the photometric sample. We find that the systems with two cool ($<$$10{,}000$~K) white dwarfs show similar total age agreement to systems with two hot/warm ($>$$10{,}000$~K) white dwarfs (in both cases, 71\% agreeing within 28.3\% of the system's average age). However, we find that systems with one cool white dwarf and one hot/warm white dwarf show abnormally high disagreements, with only 12\% agreeing within 28.3\% of the system's average age. The cause of this discrepancy is unclear.

\begin{deluxetable*}{cccccccccc}
\tabletypesize{\footnotesize}
\tablecaption{Wide WD+WD Binaries Where Both Components are DA and Both Have Precise Spectroscopic Ages}\label{tab:spec_precise}
\tablehead{
\colhead{Gaia DR3 source\_id1} & \colhead{Gaia DR3 source\_id2} & & \colhead{$T_1$ [K]} & \colhead{$M_1$ [$M_\odot$]} & \colhead{$T_2$ [K]} & \colhead{$M_2$ [$M_\odot$]} & \colhead{$\tau_1$ [Gyr]} & \colhead{$\tau_2$ [Gyr]}
}
\startdata
 \tstrut  & &\bf{Phot} & $18{,}660^{+300}_{-320}$ & $0.86^{+0.01}_{-0.01}$ & $13{,}340^{+170}_{-160}$ & $0.75^{+0.01}_{-0.01}$ & $0.45^{+0.01}_{-0.01}$ & $0.78^{+0.02}_{-0.02}$ \\ [-0.2cm]
1874954641491354624 & 1874954645786146304 &  &  &  &  &  &  &  & \\ [-0.2cm]
 & &\bf{Spec} & $19{,}680\pm240$ & $0.89^{+0.02}_{-0.02}$ & $14{,}400\pm440$ & $0.77^{+0.02}_{-0.02}$ & $0.39^{+0.02}_{-0.03}$ & $0.69^{+0.04}_{-0.03}$ \\
 & &\bf{Phot} & $16{,}320^{+350}_{-350}$ & $0.65^{+0.02}_{-0.02}$ & $13{,}700^{+270}_{-270}$ & $0.63^{+0.02}_{-0.02}$ & $1.05^{+0.36}_{-0.22}$ & $1.54^{+1.13}_{-0.36}$ \\ [-0.2cm]
1911420636118031744 & 1911420636119326976 &  &  &  &  &  &  &  & \\ [-0.2cm]
 & &\bf{Spec} & $16{,}960\pm200$ & $0.69^{+0.02}_{-0.02}$ & $14{,}430\pm470$ & $0.71^{+0.02}_{-0.03}$ & $0.70^{+0.10}_{-0.07}$ & $0.76^{+0.07}_{-0.05}$ \\
 & &\bf{Phot} & $21{,}860^{+280}_{-300}$ & $0.68^{+0.01}_{-0.01}$ & $14{,}900^{+180}_{-180}$ & $0.90^{+0.01}_{-0.01}$ & $0.63^{+0.08}_{-0.04}$ & $0.62^{+0.02}_{-0.02}$ \\ [-0.2cm]
2142145385907166208 & 2142145179748734976 &  &  &  &  &  &  &  & \\ [-0.2cm]
 & &\bf{Spec} & $24{,}590\pm300$ & $0.77^{+0.02}_{-0.02}$ & $16{,}070\pm290$ & $0.98^{+0.02}_{-0.02}$ & $0.38^{+0.03}_{-0.03}$ & $0.55^{+0.03}_{-0.03}$ \\
 & &\bf{Phot} & $12{,}550^{+560}_{-600}$ & $0.82^{+0.03}_{-0.03}$ & $9560^{+410}_{-430}$ & $0.60^{+0.06}_{-0.06}$ & $0.85^{+0.05}_{-0.04}$ & $6.45^{+37.47}_{-4.97}$ \\ [-0.2cm]
2359214952893455872 & 2359214952893456000 &  &  &  &  &  &  &  & \\ [-0.2cm]
 & &\bf{Spec} & $13{,}320\pm220$ & $0.90^{+0.03}_{-0.02}$ & $10{,}110\pm120$ & $0.67^{+0.02}_{-0.03}$ & $0.79^{+0.03}_{-0.03}$ & $1.36^{+0.31}_{-0.09}$ \\
 & &\bf{Phot} & $22{,}080^{+560}_{-570}$ & $0.64^{+0.02}_{-0.02}$ & $9820^{+150}_{-150}$ & $0.69^{+0.02}_{-0.02}$ & $1.25^{+1.46}_{-0.39}$ & $1.34^{+0.12}_{-0.08}$ \\ [-0.2cm]
2773334982315549824 & 2773334978019335040 &  &  &  &  &  &  &  & \\ [-0.2cm]
 & &\bf{Spec} & $21{,}470\pm440$ & $0.69^{+0.02}_{-0.02}$ & $10{,}160\pm160$ & $0.79^{+0.04}_{-0.04}$ & $0.59^{+0.10}_{-0.08}$ & $1.28^{+0.08}_{-0.06}$ \\
 & &\bf{Phot} & $14{,}150^{+500}_{-580}$ & $0.72^{+0.03}_{-0.03}$ & $12{,}460^{+440}_{-510}$ & $0.91^{+0.03}_{-0.03}$ & $0.75^{+0.10}_{-0.06}$ & $0.90^{+0.06}_{-0.04}$ \\ [-0.2cm]
282878984339974528 & 282878984339976320 &  &  &  &  &  &  &  & \\ [-0.2cm]
 & &\bf{Spec} & $12{,}140\pm240$ & $0.74^{+0.04}_{-0.04}$ & $11{,}300\pm330$ & $0.85^{+0.08}_{-0.08}$ & $0.92^{+0.06}_{-0.04}$ & $1.09^{+0.13}_{-0.07}$ \\
 & &\bf{Phot} & $24{,}100^{+980}_{-1000}$ & $0.85^{+0.03}_{-0.03}$ & $28{,}200^{+1300}_{-1400}$ & $1.04^{+0.03}_{-0.03}$ & $0.33^{+0.03}_{-0.04}$ & $0.17^{+0.02}_{-0.01}$ \\ [-0.2cm]
3072961070640767488 & 3072961074934467200 &  &  &  &  &  &  &  & \\ [-0.2cm]
 & &\bf{Spec} & $27{,}310\pm450$ & $0.93^{+0.04}_{-0.03}$ & $27{,}860\pm490$ & $0.98^{+0.04}_{-0.04}$ & $0.20^{+0.05}_{-0.02}$ & $0.18^{+0.01}_{-0.01}$ \\
 & &\bf{Phot} & $18{,}540^{+870}_{-1000}$ & $1.06^{+0.03}_{-0.03}$ & $15{,}140^{+480}_{-500}$ & $0.98^{+0.02}_{-0.02}$ & $0.45^{+0.02}_{-0.02}$ & $0.62^{+0.02}_{-0.02}$ \\ [-0.2cm]
3330855616042214912 & 3330855616042218112 &  &  &  &  &  &  &  & \\ [-0.2cm]
 & &\bf{Spec} & $13{,}930\pm440$ & $1.00^{+0.02}_{-0.02}$ & $16{,}400\pm360$ & $0.92^{+0.03}_{-0.03}$ & $0.81^{+0.08}_{-0.06}$ & $0.52^{+0.03}_{-0.03}$ \\
 & &\bf{Phot} & $8260^{+370}_{-410}$ & $0.79^{+0.06}_{-0.06}$ & $8390^{+480}_{-550}$ & $0.84^{+0.08}_{-0.08}$ & $2.02^{+0.15}_{-0.20}$ & $2.25^{+0.27}_{-0.25}$ \\ [-0.2cm]
5281331588075295232 & 5281331583776780416 &  &  &  &  &  &  &  & \\ [-0.2cm]
 & &\bf{Spec} & $8830\pm110$ & $0.89^{+0.04}_{-0.04}$ & $8560\pm100$ & $0.90^{+0.05}_{-0.05}$ & $2.28^{+0.26}_{-0.27}$ & $2.55^{+0.28}_{-0.35}$ \\
 & &\bf{Phot} & $8980^{+520}_{-570}$ & $0.71^{+0.09}_{-0.08}$ & $10{,}460^{+880}_{-1010}$ & $1.00^{+0.10}_{-0.08}$ & $1.53^{+0.98}_{-0.16}$ & $1.82^{+0.16}_{-0.09}$ \\ [-0.2cm]
5490676742283052800 & 5490676707920375936 &  &  &  &  &  &  &  & \\ [-0.2cm]
 & &\bf{Spec} & $9060\pm110$ & $0.85^{+0.05}_{-0.05}$ & $9640\pm120$ & $1.00^{+0.05}_{-0.05}$ & $1.87^{+0.28}_{-0.20}$ & $2.32^{+0.26}_{-0.29}$ \\
 & &\bf{Phot} & $7090^{+90}_{-90}$ & $0.59^{+0.01}_{-0.01}$ & $6150^{+70}_{-70}$ & $0.60^{+0.01}_{-0.01}$ & $11.43^{+1.74}_{-1.44}$ & $8.71^{+1.93}_{-1.31}$ \\ [-0.2cm]
5564028981196462336 & 5564029702750970112 &  &  &  &  &  &  &  & \\ [-0.2cm]
 & &\bf{Spec} & $7220\pm90$ & $0.75^{+0.03}_{-0.04}$ & $6400\pm110$ & $0.94^{+0.11}_{-0.12}$ & $2.73^{+0.33}_{-0.26}$ & $5.22^{+0.32}_{-0.64}$ \\
 & &\bf{Phot} & $9380^{+230}_{-240}$ & $0.66^{+0.03}_{-0.04}$ & $9000^{+260}_{-260}$ & $0.77^{+0.04}_{-0.04}$ & $1.64^{+1.19}_{-0.26}$ & $1.59^{+0.07}_{-0.06}$ \\ [-0.2cm]
5780478664144123264 & 5780478664144129152 &  &  &  &  &  &  &  & \\ [-0.2cm]
 & &\bf{Spec} & $9350\pm110$ & $0.67^{+0.02}_{-0.02}$ & $8520\pm100$ & $0.75^{+0.04}_{-0.04}$ & $1.52^{+0.23}_{-0.07}$ & $1.77^{+0.09}_{-0.06}$ \\
 & &\bf{Phot} & $24{,}080^{+570}_{-590}$ & $0.88^{+0.02}_{-0.02}$ & $22{,}620^{+550}_{-610}$ & $1.14^{+0.01}_{-0.01}$ & $0.31^{+0.02}_{-0.03}$ & $0.34^{+0.01}_{-0.01}$ \\ [-0.2cm]
630770819920096640 & 630770819920096768 &  &  &  &  &  &  &  & \\ [-0.2cm]
 & &\bf{Spec} & $24{,}670\pm390$ & $0.84^{+0.03}_{-0.03}$ & $25{,}780\pm590$ & $1.21^{+0.03}_{-0.03}$ & $0.33^{+0.02}_{-0.02}$ & $0.32^{+0.04}_{-0.03}$ \\
 & &\bf{Phot} & $10{,}870^{+130}_{-130}$ & $0.87^{+0.01}_{-0.01}$ & $8230^{+100}_{-100}$ & $0.71^{+0.01}_{-0.01}$ & $1.20^{+0.02}_{-0.02}$ & $1.82^{+0.06}_{-0.04}$ \\ [-0.2cm]
6898489884295412352 & 6898489884295407488 &  &  &  &  &  &  &  & \\ [-0.2cm]
 & &\bf{Spec} & $10{,}580\pm130$ & $0.86^{+0.02}_{-0.02}$ & $8140\pm100$ & $0.72^{+0.05}_{-0.05}$ & $1.26^{+0.05}_{-0.05}$ & $1.95^{+0.20}_{-0.08}$ \bstrut \\ 
\enddata
\end{deluxetable*}

To understand if an identification spectrum ensures a more accurate total age we use photometric total ages of a sample of 295 DA+DA systems, including ones identified in \cite{2024A&A...682A...5V} through their Gaia XP spectra. Out of the 295 DA+DA systems, 52 have average photometric total age uncertainties less than 20\%, chance alignment factors less than 0.1, and separations greater than 2 arcseconds. This sample of DA+DA binaries is slightly brighter than the photometric sample discussed previously, with a mean Gaia G magnitude of 18.5. Of these 52 systems, 35 (67\%) have total ages that agree within 28.3\%. The median $\Delta \tau$ is 0.2~Gyr (or 14\% of the system's average age). Incorporating their uncertainties, they agree within $1.9\sigma$ on average without inflated errors and $1.0\sigma$ with inflated errors. Thus, we find discernible improvement in the accuracy of white dwarf ages when the sample is cleaned of non-DA white dwarfs, demonstrating that collecting an identification spectrum is a crucial step for determining the best ages of an ensemble of white dwarfs. A summary of the level of total age agreement for each sub-sample can be found in Table~\ref{tab:acc_prec}.

\begin{deluxetable*}{cccccc}
\tablecaption{Age Agreement for Different Samples of Wide WD+WD Binaries}\label{tab:acc_prec}

\tablehead{\colhead{Sample} & \colhead{N} & \colhead{$\Delta \tau < 28.3\% \ \tau_{avg}$} & \colhead{Median $\Delta \tau$} & \colhead{Median $\sigma$} & \colhead{Median $\sigma_\mathrm{inflated}$}\\ & \colhead{} & \colhead{} & \colhead{(Gyr, \% of $\tau_{avg}$)} & \colhead{Agreement} & \colhead{Agreement}}
\startdata
 \tstrutsmall Phot Ages with $\sigma_\tau < 20\%$ & 192 & 58\% & 0.6 (23\%) & 2.2 & 1.4 \bstrutsmall \\
 Spec Ages with $\sigma_\tau < 20\%$ & 14 & 50\% & 0.3 (28\%) & 2.1 & 1.5 \bstrutsmall \\
 Phot Ages with ID Spectrum and $\sigma_\tau < 20\%$ & 52 & 67\% & 0.2 (14\%) & 1.9 & 1.0 \bstrutsmall 
 \enddata
\tablecomments{~$\tau_{avg}$ refers to the weighted average age of the binary and $\sigma_{\mathrm{inflated}}$ references the total age agreement when inflation factors from \cite{2022ApJ...934..148H} are applied to the total age uncertainties.}
\end{deluxetable*}

It is important to understand when spectroscopy is needed for accurate and precise white dwarf ages. Photometry is readily available for hundreds of thousands of white dwarfs, while spectroscopy can be observationally expensive. From our investigation across the multiple subsets discussed, we find no discernible improvement in age agreement between white dwarfs in wide binaries when spectroscopic fits to Balmer lines are used. Instead, we find that photometric determinations of white dwarf total ages perform better when an identification spectrum is provided and the proper atmospheric models are applied. For spectroscopically identified DA white dwarfs, photometrically determined atmospheric parameters give slightly better total age agreement than spectroscopic fits to the Balmer lines. The sample size of precise spectroscopic total ages is small (14), which makes these conclusions very preliminary. It is important to caveat that this is valid for our spectroscopic S/N; higher S/N and higher-resolution spectra may provide more accurate age constraints.

\subsection{Inconsistencies with the Monotonic IFMR}\label{sec:inconsist}
As compared to previous work, many systems in the wide WD+WD binary sample appear to be in contention with our understanding of the IFMR. In \cite{2022ApJ...934..148H}, we found that $21-36$\% of the systems had a more massive white dwarf with a shorter cooling age. This would imply that the more massive white dwarf had a longer progenitor lifetime and, thus, came from a less massive progenitor. This contradicts the generally accepted monotonic form of the IFMR (e.g., \citealt{2018ApJ...866...21C}, \citealt{2018ApJ...860L..17E}), but there has been evidence of a non-monotonic form of the IFMR over a narrow mass range \citep{2020NatAs...4.1102M}.

\begin{figure*}[t]
    \centering
    \includegraphics[width=0.95\textwidth]{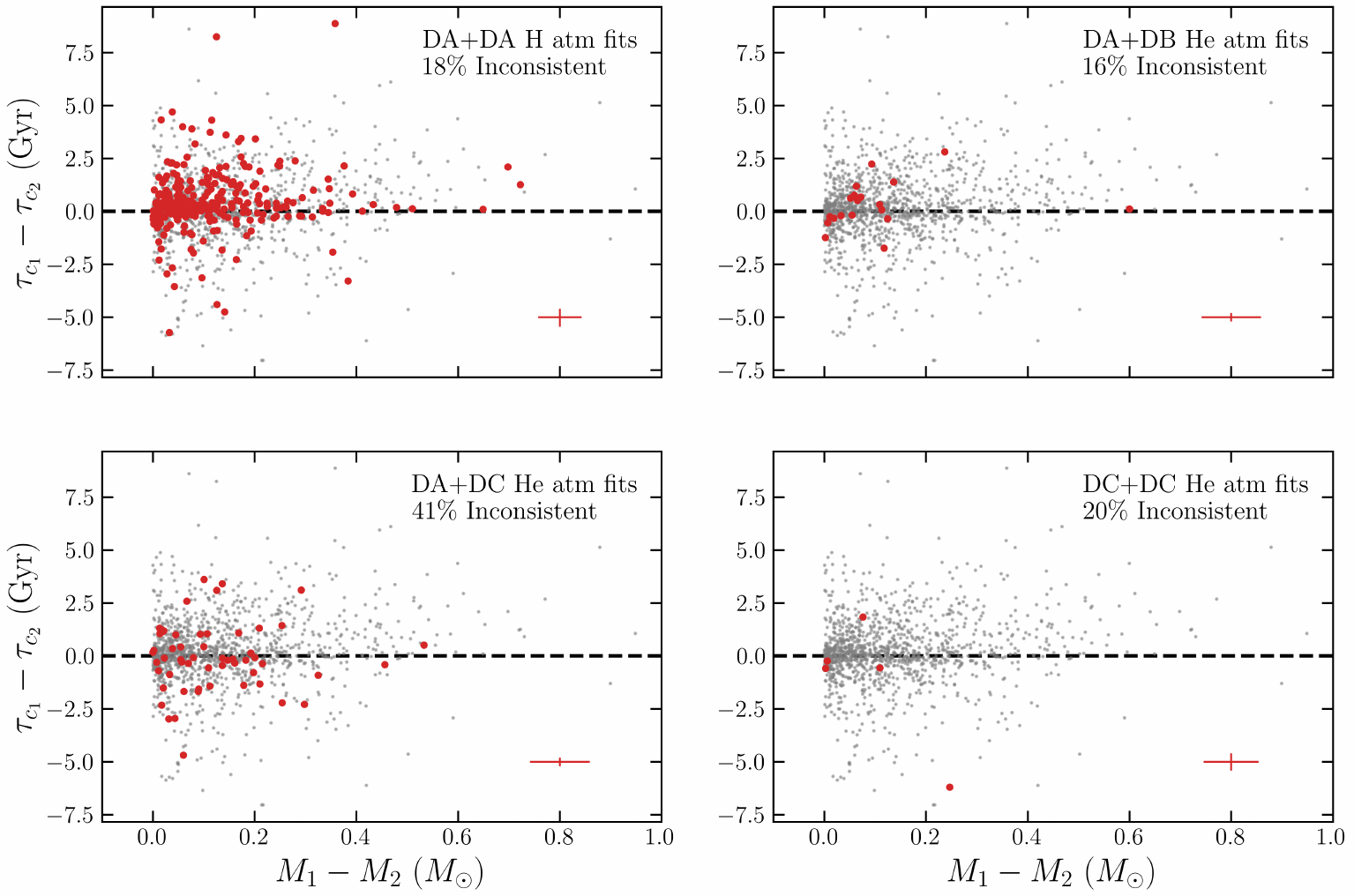}
    \caption{Comparison of cooling age difference versus mass difference using the correct model atmospheres for photometric fitting (red points). The grey points in each panel are from all WD+WD binaries in \cite{2022ApJ...934..148H}. The percentage of systems in contention with the monotonic IFMR are shown in each panel as well. The number of systems in contention with a monotonic IFMR are determined as the number of systems more than $1\sigma$ away from having $\tau_{\mathrm{c1}}>\tau_{\mathrm{c2}}$.} An average uncertainty is shown in the bottom right of each panel. Using the correct atmospheric parameters for DA+DB systems produces a roughly similar number of ill-behaved systems as the known DA+DA systems; a larger number of systems with at least one DC appear in contention with the monotonic IFMR.
    \label{fig:dm_dtc_He}
\end{figure*}

We investigate whether such a trend still exists with the addition of spectroscopic information. We take the cooling ages and masses determined in \cite{2022ApJ...934..148H}, which assumed a DA atmosphere for all white dwarfs, in conjunction with the spectral types gathered in this work. We find 34\% of the DA+DA systems are in contention with a monotonic IFMR, while 18\% are more than $1\sigma$ away from being consistent with the monotonic IFMR (see top left panel of Figure~\ref{fig:dm_dtc_He}). In addition, we find that DA+non-DA systems show 66\% of the systems are inconsistent with a monotonic IFMR while 51\% show inconsistencies that are greater than $1\sigma$. For non-DA+non-DA systems, we find 71\% of the systems are inconsistent with 43\% having inconsistencies greater than $1\sigma$.

 The larger number of inconsistent systems among those with a non-DA component could result from incorrect atmospheres being applied to their photometry. Fitting a DA model to a non-DA can introduce systematic errors on the order of 10-15\% on WD mass \citep{2012ApJS..199...29G}. To alleviate this problem, we conduct SED fits to all the spectroscopically confirmed DB and DC white dwarfs using the appropriate atmospheric models. We perform the same fitting methods and quality control checks outlined in \cite{2022ApJ...934..148H} {(see summary in Section~\ref{sec:spec_vs_phot})}.

For DB white dwarfs, we use the pure helium-atmosphere models from \cite{2021MNRAS.501.5274C} and the thin H-layer cooling models from \cite{2020ApJ...901...93B}. For the DC white dwarfs in the sample, we follow \citealt{2024A&A...682A...5V}: For DCs with $T_{\mathrm{eff}} < 5500$\,K, we use the DA fits done previously in \cite{2022ApJ...934..148H}. For DCs with $5500 < T_{\mathrm{eff}} < 11{,}000$\,K, we use mixed-atmosphere models with H/He ratios of $10^{-5}$ from \cite{2021MNRAS.501.5274C} and the corresponding thin H-layer cooling models from \cite{2020ApJ...901...93B}. For DCs with $T_{\mathrm{eff}} > 11{,}000$\,K, we use the same pure He models used in the DB fits.

Figure~\ref{fig:dm_dtc_He} shows the distribution of the binaries' mass difference and cooling age difference when using the correct models. We see significant improvement in the number of inconsistent systems for the DB white dwarfs, but 42\% are still inconsistent with a monotonically increasing IFMR, with 16\% being more than $1\sigma$ inconsistent. Most of the IFMRs to date have been constrained only with DA white dwarfs and an IFMR for DB white dwarfs remains elusive \citep{2021AJ....162..162B}.

The number of systems with a DB white dwarf that are inconsistent with the monotonic IFMR is in line with the fraction of known DA+DA systems with inconsistent cooling ages and masses. This leaves two possibilities: either roughly 20\% of wide WD+WD binaries suffer from merger contamination or blended photometry and were likely born as triple systems, or the IFMR is not monotonic. We do not find that white dwarfs in inconsistent systems are preferentially found at any mass range when compared to the full sample. 58\% of binaries in the DA+DA sample shown in Figure~\ref{fig:dm_dtc_He} that are inconsistent with a monotonic IFMR have at least one white dwarf with a mass between 0.6 \msun and 0.7\msun. This is not surprising given that the white dwarf mass distribution peaks in this range \citep{2019MNRAS.482.5222T}.

For systems with at least one DC white dwarf, we find that the inconsistency reaches 40\% even using non-DA atmospheres (see bottom panels of Figure~\ref{fig:dm_dtc_He}). We suspect this reflects the inaccuracy of photometrically determined parameters for DC white dwarfs due to the inexact understanding of their atmospheric compositions, although it may also point to a higher fraction of merger contaminants among DC white dwarfs. While it has been widely thought that hotter DCs ($>11{,}000$~K) are strongly magnetic and a signpost of a past merger (e.g., \citealt{2015MNRAS.447.1713B}), we find that cooler DCs in our sample still show significant disagreement (59\% of the 62 DA+DC and DC+DC systems with DC temperatures $<11{,}000$~K are inconsistent with a monotonically increasing IFMR).

To determine if the percentages of inconsistent systems we find are not due to random errors, we generate a mock sample of 5000 binaries with masses in the range of $1-8$ \msun\ determined through a Salpeter initial-mass function (IMF, \citealt{1955ApJ...121..161S}). We set the criteria that their total ages must be the same and use the IFMR discussed in Section~\ref{sec:tot_age_determ} to convert the ZAMS mass to the final white dwarf mass. The cooling age is determined by subtracting the progenitor lifetime for the given ZAMS mass (determined through the MESA evolutionary sequences discussed in Section~\ref{sec:tot_age_determ}). Then, the white dwarf cooling age and mass are converted into the white dwarf's temperature using the cooling models from \cite{2020ApJ...901...93B}. We introduce random errors into the white dwarf mass and temperatures which are drawn from the distribution of reported uncertainties on temperature and mass from the photometric fits conducted in \cite{2022ApJ...934..148H}. Finally, we count the number of systems in contention with the monotonic IFMR. We find that for the sample of 5000 binaries, 18\% are inconsistent with a monotonic IFMR simply due to random errors with only 1.3\% of those systems being more than $1\sigma$ away from being consistent. To better represent the sample size of the DA+DA binaries in our sample, we conduct the same test 20 times with 100 binaries. Through this test, we find $19^{+4}_{-5}$\% to be inconsistent with the monotonic IFMR with only $1.5\pm1.5$\% are more than $1\sigma$ away from the being consistent. 

For the DA+DA and DA+DB binaries, we find 20\% of the systems are more than $1\sigma$ inconsistent with a monotonic IFMR. The number of systems more than $1\sigma$ away from being consistent is minimally affected by random statistical errors, and thus, we conclude that the number of systems inconsistent with a monotonic IFMR is roughly 20\% for the DA+DA and DA+DB binaries.

\subsection{Other White Dwarf Spectral Types}
Using our follow-up observations and collected literature spectra we identified several components of wide WD+WD systems that are noteworthy, including systems with metal-polluted white dwarfs (DZ), white dwarfs with carbon features in their spectra (DQ), and highly magnetic white dwarfs (DAH) (see Section~\ref{sec:spectypes}).

\subsubsection{Metal-Polluted Systems}
Metal pollution can provide insights into the remnants of planetary systems around white dwarfs. In our follow-up work, we discover a new DZ+DA binary system. The primary (Gaia DR3 4471738941493617408, WDJ181332.20+060412.61) is a 0.66\,\msun\ DZ white dwarf at 7600 K, and the secondary (Gaia DR3 4471739010213094272, WDJ181330.47+060412.51) is a 6500 K, 0.62\,\msun\ DA white dwarf. The mean weighted age of the system is $2.3\pm0.2$~Gyr, and the total age difference between the two stars is $1.9^{+3.2}_{-0.9}$ Gyr. Systems like this provide a great opportunity to test DZ atmospheric models against their DA counterparts.

We identify a total of 22 wide WD+WD systems with at least one DZ. Of these 22 systems, 17 are identified through their Gaia XP spectra alone \citep{2024A&A...682A...5V}. For these DZ white dwarfs, we employ the same criteria used for DC white dwarfs to determine their atmospheric parameters from photometry (see Section \ref{sec:inconsist}). If the white dwarf has observable H (DAZ) or He (DBZ) lines, we use the appropriate pure H and He models, respectively. Of the 22 systems containing a metal-polluted white dwarf, 13 suggest the more massive white dwarf has a shorter cooling age. This level of inconsistency for the DZ white dwarfs parallels the DC white dwarf population. This adds more evidence that the problem with the DC white dwarfs is an improper application of mixed- and He-atmosphere models and not an intrinsically higher number of mergers among their population, although it is possible that a large fraction of the 17 identified by their Gaia XP spectra could be improperly labelled as DZ white dwarfs.

\subsubsection{Magnetic White Dwarfs}
We identify 16 wide WD+WD systems with at least one DAH white dwarf. \cite{2024A&A...682A...5V} do not make attempts to identify and classify magnetic white dwarfs through Gaia XP spectra. Thus, all magnetic white dwarfs in the sample are manually classified from their spectra. 

From our follow-up observations, we identify 3 new DA+DAH systems and one new DAH+DAH system. The first system contains a $10{,}500$\,K, 0.86\,\msun\ DAH primary (Gaia DR3 4940155500795290624, WDJ022440.28-461133.91) with an $11{,}700$\,K, 0.71\,\msun\ DA secondary (Gaia DR3 4940155500795290880, WDJ022440.65-461140.65). The second system contains an 8100\,K, 0.88\,\msun\ DA primary (Gaia DR3 5062948237833763840, WDJ024051.59-324837.30) with a 7500 K, 0.79\,\msun\ DAH secondary (Gaia DR3 5062948340912201088, WDJ024051.94-324814.09). The last DA+DAH system contains a 9700\,K, 1.02\,\msun\ DAH primary (Gaia DR3 6093462311215003776, WDJ135006.24-502534.22) and a 6500\,K, 0.66\,\msun\ DA secondary (Gaia DR3 6093462315507278208, WDJ135006.04-502540.07). The new DAH+DAH system contains a $14{,}000$\,K, 1.09\,\msun\ DAH primary (Gaia DR3 5551794951535528448, WDJ064715.56-470733.36) with an $11{,}500$\,K, 0.99\,\msun\ DAH secondary (Gaia DR3 5551794951535526144, WDJ064716.66-470744.24).

Of the 16 systems with at least one DAH, four have a more massive white dwarf with a shorter cooling age. The percentage of inconsistent systems is comparable to the sample of DA+DA systems. This indicates that the merger fraction among magnetic white dwarfs is not noticeably higher in our sample, which is not surprising given 11 of the 16 DAHs are cooler than $10{,}000$~K. These cooler magnetic white dwarfs are generally not massive and do not require dynamo generation from a merger, supporting similar conclusions from field white dwarfs within 40\,pc by \citet{2022ApJ...935L..12B}. When labelling hot DCs ($>11{,}000$~K) as magnetic white dwarfs and using H atmosphere fits, we find the magnetic white dwarfs do have higher inconsistencies with the monotonic IFMR (52\% inconsistent with 39\% more than $1\sigma$ inconsistent). The implication is that these hot DCs may have H or He lines, but very strong (often $>$100\,MG) magnetic fields have shifted these absorption lines all over the spectrum (e.g., \citealt{2021Natur.595...39C}). With the inclusion of hot DCs, we find that the most strongly magnetic white dwarfs are ill-behaved, in line with the idea that strong magnetism is a signpost for mergers (e.g., \citealt{2023MNRAS.518.2341K}).

\subsubsection{DQ White Dwarfs}
Hot ($>$$15{,}000$~K) and warm ($10{,}000$~$<$~$T_{\mathrm{eff}}$~$<$~$15{,}000$~K) DQ white dwarfs are hypothesized to form in binary mergers (e.g., \citealt{2015ASPC..493..547D,2016ApJ...817...27W,2020MNRAS.491L..40K}). If the DQs in our sample did, in fact, form from a binary interaction, their cooling ages would be inconsistent with an increasing IFMR.

We identify 11 wide WD+WD systems with at least one DQ. Of these 11 systems, 4 are identified from their Gaia XP spectra alone \citep{2024A&A...682A...5V}. For atmospheric parameters and cooling ages of the DQs in the sample, we use the He-atmosphere models that we use for the DB and DC white dwarfs described in Section~\ref{sec:inconsist}. Of the 11 systems with a DQ white dwarf component, nine have the more massive white dwarf with a shorter cooling age. Of the nine systems in which the more massive white dwarf has a shorter cooling age, we do not see any preference for higher temperatures with the sample spanning a temperature range of 6700-9500~K. These DQs are not part of the hot/warm population theorized to form from binary mergers and there is no clear evidence of DQs in this temperature range having a higher expected merger fraction \citep[e.g., ][]{2019ApJ...885...74C, 2019A&A...628A.102K}. Improved modeling of this sub-sample could help discern if this inconsistency arises purely from incorrect model atmospheres, or if DQ white dwarfs in this temperature range are preferentially formed in mergers.

\section{Discussion and Conclusions}\label{sec:conclusions}
In this work, we collected spectral types for 419 individual white dwarfs in wide (separated by $>$100\,au) WD+WD binaries, as well as spectral types of 1292 individual white dwarfs from \cite{2024A&A...682A...5V}, identified from their Gaia XP spectra. We additionally derived atmospheric parameters from spectra for 256 of the 419 white dwarfs, with 94 new atmospheric parameters determined from our spectroscopic follow-up.

The main conclusions of this work can be summarized as follows:
\begin{itemize}

\item We find that fits to optical spectra of white dwarfs do not provide any noticeable improvement in agreement between the total ages of white dwarfs in wide WD+WD binaries. 50\% of wide WD+WD systems with precise total ages from spectroscopy ($<$20\% uncertainties) agree within 28.28\% (20\% added in quadrature) of the weighted mean age of the system. This nearly matches, albeit with slightly worse performance, what is found in photometrically determined ages (57\%). The mean magnitude of the sample with $<$20\% total age uncertainties from spectroscopy is $G=17.5$\,mag, which is much brighter than the sample with $<$20\% total age uncertainties from photometry (mean of $G=19.2$\,mag), highlighting the capability of photometry combined with Gaia parallaxes to provide precise total ages for fainter objects.

\item We find that collecting an identification spectrum is a crucial step in determining the most accurate white dwarf total ages. For a sample of photometric white dwarf total ages spectroscopically identified as DAs, we find that 67\% of systems with precise total ages (uncertainties $<$20\%) agree within 28.28\% of the weighted mean total age of the system. This noticeably improves on samples of precisely aged systems from photometry when a DA model is assumed for all white dwarfs (57\%), and on precisely determined total ages from spectroscopy (50\%).

\item Roughly 20\% of white dwarfs in wide WD+WD binaries are inconsistent with a monotonically increasing IFMR. This discrepancy was first seen in \cite{2022ApJ...934..148H} and is still present when more spectroscopic information is available, and when the correct atmosphere models are used for SED fits to determine the total age of the white dwarf from mass and effective temperature. This inconsistency is seen in multiple samples of spectroscopically identified wide WD+WD binaries, including DA+DA, DA+DB, DA+DC, and DC+DC. These systems could not have formed through single-star evolution assuming a monotonic IFMR and, thus, could be merger products or contain close binary systems masking as single stars which contaminate the data. Roughly 20\% of the sample being merger remnants is consistent with binary population synthesis models \citep{2020A&A...636A..31T}. The implication that 20\% of the wide binary sample started off as triples is not unreasonable given the fraction of white dwarf progenitors that are expected to be born in binaries versus triples \citep{2017ApJS..230...15M, 2023ApJ...955L..14S}. 

This result could also be explained by a non-monotonic IFMR, which has gained support in recent years from white dwarfs with masses between 0.6-0.7 \msun \citep{2020NatAs...4.1102M}. We do not find that white dwarfs in inconsistent systems cluster at any mass range when compared to the full sample. 58\% of systems in the DA+DA sample that are inconsistent with a monotonic IFMR have at least one white dwarf with a mass between 0.6 \msun and 0.7\msun, but we expect most to lie in this range given the white dwarf mass distribution peaks here \citep{2019MNRAS.482.5222T}. 

\item We find that wide binaries containing DC white dwarfs are more often incompatible with a monotonically increasing IFMR than systems containing DA or DB white dwarfs. It is difficult to distinguish whether this effect is due to a larger merger population among DC white dwarfs or the incorrect application of atmospheric models. Because it is challenging to identify the atmospheric compositions of DC white dwarfs in our sample, we imposed criteria (see Section \ref{sec:inconsist}) dependent on the temperature of the DC. This method likely applies incorrect models to a subset of the DCs, which may contribute to higher inconsistency with the IFMR.

\end{itemize}

This work provides an empirical demonstration of the strengths and challenges of using white dwarfs as stellar age indicators. We find unavoidable systematic effects from past mergers among white dwarfs, though mergers are not unique to white dwarf stars and provide a challenge for any stellar age indicator. We also find that obtaining identification spectra is a crucial step for determining accurate white dwarf total ages.

The sample of spectroscopically identified white dwarfs in wide binaries will increase significantly with ongoing and upcoming all-sky, multiplexed spectroscopic surveys such as SDSS-V \citep{2023ApJS..267...44A} and DESI \citep{2023ApJ...947...37C}. These surveys will be vital in getting the largest possible sample of spectroscopically identified white dwarfs which in turn will allow for more accurate white dwarf age estimates. As the sample of white dwarfs with optical spectra increases, conclusions from tests of spectroscopic fits to Balmer lines will only strengthen. As of now, with our small sample of precise ages, we find that these spectroscopic fits do not provide more accurate total ages and, in fact, are generally as good as photometric fits to white dwarfs with unidentified spectral types.

\section{Acknowledgements}

We acknowledge helpful comments from the anonymous referee that helped to improve this work. We acknowledge support from the National Science Foundation under Grant No. AST-1908119 as well as under Grant No. PHY-1748958. PET received funding from from the European Research Council under the European Union’s Horizon 2020 research and innovation programme number 101002408 (MOS100PC). JR and BK acknowledge funding from the National Science Foundation under grant number AST-2108311 and from the NC Space Grant Graduate Research Fellowship.

We sincerely thank our excellent telescope operators and support staff Patricio Ugarte, Carlos Corco, Juan Espinoza, Rodrigo Hern\'andez, and Sergio Pizarro from SOAR and Teznie Pugh, Ana Hayslip, LaLaina Shumar, Jason Sanborn, Haylee Archer, Ishara Nisley, Sydney Perez, Ben Shafransky, Jose Fernandez, and Stephen Levine from Lowell Observatory. We also appreciate observing assistance while in training from Ryan Hegedus. These results made use of the Lowell Discovery Telescope (LDT) at Lowell Observatory. Lowell is a private, non-profit institution dedicated to astrophysical research and public appreciation of astronomy and operates the LDT in partnership with Boston University, the University of Maryland, the University of Toledo, Northern Arizona University and Yale University. The upgrade of the DeVeny optical spectrograph has been funded by a generous grant from John and Ginger Giovale and by a grant from the Mt. Cuba Astronomical Foundation. These results are based in part on observations obtained at the Southern Astrophysical Research (SOAR) telescope, which is a joint project of the Minist\'{e}rio da Ci\^{e}ncia, Tecnologia e Inova\c{c}\~{o}es (MCTI/LNA) do Brasil, the US National Science Foundation’s NOIRLab, the University of North Carolina at Chapel Hill (UNC), and Michigan State University (MSU).

\bibliography{main}{}

\begin{thebibliography}{}
\expandafter\ifx\csname natexlab\endcsname\relax\def\natexlab#1{#1}\fi
\providecommand{\url}[1]{\href{#1}{#1}}
\providecommand{\dodoi}[1]{doi:~\href{http://doi.org/#1}{\nolinkurl{#1}}}
\providecommand{\doeprint}[1]{\href{http://ascl.net/#1}{\nolinkurl{http://ascl.net/#1}}}
\providecommand{\doarXiv}[1]{\href{https://arxiv.org/abs/#1}{\nolinkurl{https://arxiv.org/abs/#1}}}

\bibitem[{{Ahumada} {et~al.}(2020){Ahumada}, {Allende Prieto}, {Almeida}, {Anders}, {Anderson}, {Andrews}, {Anguiano}, {Arcodia}, {Armengaud}, {Aubert}, {Avila}, {Avila-Reese}, {Badenes}, {Balland}, {Barger}, {Barrera-Ballesteros}, {Basu}, {Bautista}, {Beaton}, {Beers}, {Benavides}, {Bender}, {Bernardi}, {Bershady}, {Beutler}, {Bidin}, {Bird}, {Bizyaev}, {Blanc}, {Blanton}, {Boquien}, {Borissova}, {Bovy}, {Brandt}, {Brinkmann}, {Brownstein}, {Bundy}, {Bureau}, {Burgasser}, {Burtin}, {Cano-D{\'\i}az}, {Capasso}, {Cappellari}, {Carrera}, {Chabanier}, {Chaplin}, {Chapman}, {Cherinka}, {Chiappini}, {Doohyun Choi}, {Chojnowski}, {Chung}, {Clerc}, {Coffey}, {Comerford}, {Comparat}, {da Costa}, {Cousinou}, {Covey}, {Crane}, {Cunha}, {Ilha}, {Dai}, {Damsted}, {Darling}, {Davidson}, {Davies}, {Dawson}, {De}, {de la Macorra}, {De Lee}, {Queiroz}, {Deconto Machado}, {de la Torre}, {Dell'Agli}, {du Mas des Bourboux}, {Diamond-Stanic}, {Dillon}, {Donor}, {Drory}, {Duckworth}, {Dwelly}, {Ebelke}, {Eftekharzadeh}, {Davis
  Eigenbrot}, {Elsworth}, {Eracleous}, {Erfanianfar}, {Escoffier}, {Fan}, {Farr}, {Fern{\'a}ndez-Trincado}, {Feuillet}, {Finoguenov}, {Fofie}, {Fraser-McKelvie}, {Frinchaboy}, {Fromenteau}, {Fu}, {Galbany}, {Garcia}, {Garc{\'\i}a-Hern{\'a}ndez}, {Garma Oehmichen}, {Ge}, {Geimba Maia}, {Geisler}, {Gelfand}, {Goddy}, {Gonzalez-Perez}, {Grabowski}, {Green}, {Grier}, {Guo}, {Guy}, {Harding}, {Hasselquist}, {Hawken}, {Hayes}, {Hearty}, {Hekker}, {Hogg}, {Holtzman}, {Horta}, {Hou}, {Hsieh}, {Huber}, {Hunt}, {Ider Chitham}, {Imig}, {Jaber}, {Jimenez Angel}, {Johnson}, {Jones}, {J{\"o}nsson}, {Jullo}, {Kim}, {Kinemuchi}, {Kirkpatrick}, {Kite}, {Klaene}, {Kneib}, {Kollmeier}, {Kong}, {Kounkel}, {Krishnarao}, {Lacerna}, {Lan}, {Lane}, {Law}, {Le Goff}, {Leung}, {Lewis}, {Li}, {Lian}, {Lin}, {Long}, {Longa-Pe{\~n}a}, {Lundgren}, {Lyke}, {Mackereth}, {MacLeod}, {Majewski}, {Manchado}, {Maraston}, {Martini}, {Masseron}, {Masters}, {Mathur}, {McDermid}, {Merloni}, {Merrifield}, {M{\'e}sz{\'a}ros}, {Miglio}, {Minniti},
  {Minsley}, {Miyaji}, {Mohammad}, {Mosser}, {Mueller}, {Muna}, {Mu{\~n}oz-Guti{\'e}rrez}, {Myers}, {Nadathur}, {Nair}, {Nandra}, {Correa do Nascimento}, {Nevin}, {Newman}, {Nidever}, {Nitschelm}, {Noterdaeme}, {O'Connell}, {Olmstead}, {Oravetz}, {Oravetz}, {Osorio}, {Pace}, {Padilla}, {Palanque-Delabrouille}, {Palicio}, {Pan}, {Pan}, {Parker}, {Paviot}, {Peirani}, {Ram{\'r}ez}, {Penny}, {Percival}, {Perez-Fournon}, {P{\'e}rez-R{\`a}fols}, {Petitjean}, {Pieri}, {Pinsonneault}, {Poovelil}, {Povick}, {Prakash}, {Price-Whelan}, {Raddick}, {Raichoor}, {Ray}, {Rembold}, {Rezaie}, {Riffel}, {Riffel}, {Rix}, {Robin}, {Roman-Lopes}, {Rom{\'a}n-Z{\'u}{\~n}iga}, {Rose}, {Ross}, {Rossi}, {Rowlands}, {Rubin}, {Salvato}, {S{\'a}nchez}, {S{\'a}nchez-Menguiano}, {S{\'a}nchez-Gallego}, {Sayres}, {Schaefer}, {Schiavon}, {Schimoia}, {Schlafly}, {Schlegel}, {Schneider}, {Schultheis}, {Schwope}, {Seo}, {Serenelli}, {Shafieloo}, {Shamsi}, {Shao}, {Shen}, {Shetrone}, {Shirley}, {Silva Aguirre}, {Simon}, {Skrutskie}, {Slosar},
  {Smethurst}, {Sobeck}, {Sodi}, {Souto}, {Stark}, {Stassun}, {Steinmetz}, {Stello}, {Stermer}, {Storchi-Bergmann}, {Streblyanska}, {Stringfellow}, {Stutz}, {Su{\'a}rez}, {Sun}, {Taghizadeh-Popp}, {Talbot}, {Tayar}, {Thakar}, {Theriault}, {Thomas}, {Thomas}, {Tinker}, {Tojeiro}, {Toledo}, {Tremonti}, {Troup}, {Tuttle}, {Unda-Sanzana}, {Valentini}, {Vargas-Gonz{\'a}lez}, {Vargas-Maga{\~n}a}, {V{\'a}zquez-Mata}, {Vivek}, {Wake}, {Wang}, {Weaver}, {Weijmans}, {Wild}, {Wilson}, {Wilson}, {Wolthuis}, {Wood-Vasey}, {Yan}, {Yang}, {Y{\`e}che}, {Zamora}, {Zarrouk}, {Zasowski}, {Zhang}, {Zhao}, {Zhao}, {Zheng}, {Zheng}, {Zhu}, \& {Zou}}]{2020ApJS..249....3A}
{Ahumada}, R., {Allende Prieto}, C., {Almeida}, A., {et~al.} 2020, \apjs, 249, 3, \dodoi{10.3847/1538-4365/ab929e}

\bibitem[{{Almeida} {et~al.}(2023){Almeida}, {Anderson}, {Argudo-Fern{\'a}ndez}, {Badenes}, {Barger}, {Barrera-Ballesteros}, {Bender}, {Benitez}, {Besser}, {Bird}, {Bizyaev}, {Blanton}, {Bochanski}, {Bovy}, {Brandt}, {Brownstein}, {Buchner}, {Bulbul}, {Burchett}, {Cano D{\'\i}az}, {Carlberg}, {Casey}, {Chandra}, {Cherinka}, {Chiappini}, {Coker}, {Comparat}, {Conroy}, {Contardo}, {Cortes}, {Covey}, {Crane}, {Cunha}, {Dabbieri}, {Davidson}, {Davis}, {de Andrade Queiroz}, {De Lee}, {M{\'e}ndez Delgado}, {Demasi}, {Di Mille}, {Donor}, {Dow}, {Dwelly}, {Eracleous}, {Eriksen}, {Fan}, {Farr}, {Frederick}, {Fries}, {Frinchaboy}, {G{\"a}nsicke}, {Ge}, {Gonz{\'a}lez {\'A}vila}, {Grabowski}, {Grier}, {Guiglion}, {Gupta}, {Hall}, {Hawkins}, {Hayes}, {Hermes}, {Hern{\'a}ndez-Garc{\'\i}a}, {Hogg}, {Holtzman}, {Ibarra-Medel}, {Ji}, {Jofre}, {Johnson}, {Jones}, {Kinemuchi}, {Kluge}, {Koekemoer}, {Kollmeier}, {Kounkel}, {Krishnarao}, {Krumpe}, {Lacerna}, {Lago}, {Laporte}, {Liu}, {Liu}, {Liu}, {Lopes}, {Macktoobian},
  {Majewski}, {Malanushenko}, {Maoz}, {Masseron}, {Masters}, {Matijevic}, {McBride}, {Medan}, {Merloni}, {Morrison}, {Myers}, {M{\'e}sz{\'a}ros}, {Negrete}, {Nidever}, {Nitschelm}, {Oravetz}, {Oravetz}, {Pan}, {Peng}, {Pinsonneault}, {Pogge}, {Qiu}, {Ramirez}, {Rix}, {Fern{\'a}ndez Rosso}, {Runnoe}, {Salvato}, {Sanchez}, {Santana}, {Saydjari}, {Sayres}, {Schlaufman}, {Schneider}, {Schwope}, {Serna}, {Shen}, {Sobeck}, {Song}, {Souto}, {Spoo}, {Stassun}, {Steinmetz}, {Straumit}, {Stringfellow}, {S{\'a}nchez-Gallego}, {Taghizadeh-Popp}, {Tayar}, {Thakar}, {Tissera}, {Tkachenko}, {Hernandez Toledo}, {Trakhtenbrot}, {Fern{\'a}ndez-Trincado}, {Troup}, {Trump}, {Tuttle}, {Ulloa}, {Vazquez-Mata}, {Vera Alfaro}, {Villanova}, {Wachter}, {Weijmans}, {Wheeler}, {Wilson}, {Wojno}, {Wolf}, {Xue}, {Ybarra}, {Zari}, \& {Zasowski}}]{2023ApJS..267...44A}
{Almeida}, A., {Anderson}, S.~F., {Argudo-Fern{\'a}ndez}, M., {et~al.} 2023, \apjs, 267, 44, \dodoi{10.3847/1538-4365/acda98}

\bibitem[{{Andrews} {et~al.}(2015){Andrews}, {Ag{\"u}eros}, {Gianninas}, {Kilic}, {Dhital}, \& {Anderson}}]{2015ApJ...815...63A}
{Andrews}, J.~J., {Ag{\"u}eros}, M.~A., {Gianninas}, A., {et~al.} 2015, \apj, 815, 63, \dodoi{10.1088/0004-637X/815/1/63}

\bibitem[{{Bagnulo} \& {Landstreet}(2022)}]{2022ApJ...935L..12B}
{Bagnulo}, S., \& {Landstreet}, J.~D. 2022, \apjl, 935, L12, \dodoi{10.3847/2041-8213/ac84d3}

\bibitem[{{Barnes}(2007)}]{2007ApJ...669.1167B}
{Barnes}, S.~A. 2007, \apj, 669, 1167, \dodoi{10.1086/519295}

\bibitem[{{Barnett} {et~al.}(2021){Barnett}, {Williams}, {B{\'e}dard}, \& {Bolte}}]{2021AJ....162..162B}
{Barnett}, J.~W., {Williams}, K.~A., {B{\'e}dard}, A., \& {Bolte}, M. 2021, \aj, 162, 162, \dodoi{10.3847/1538-3881/ac1423}

\bibitem[{{Barrientos} \& {Chanam{\'e}}(2021)}]{2021ApJ...923..181B}
{Barrientos}, M., \& {Chanam{\'e}}, J. 2021, \apj, 923, 181, \dodoi{10.3847/1538-4357/ac2f49}

\bibitem[{{Baxter} {et~al.}(2014){Baxter}, {Dobbie}, {Parker}, {Casewell}, {Lodieu}, {Burleigh}, {Lawrie}, {K{\"u}lebi}, {Koester}, \& {Holland}}]{2014MNRAS.440.3184B}
{Baxter}, R.~B., {Dobbie}, P.~D., {Parker}, Q.~A., {et~al.} 2014, \mnras, 440, 3184, \dodoi{10.1093/mnras/stu464}

\bibitem[{{B{\'e}dard} {et~al.}(2020){B{\'e}dard}, {Bergeron}, {Brassard}, \& {Fontaine}}]{2020ApJ...901...93B}
{B{\'e}dard}, A., {Bergeron}, P., {Brassard}, P., \& {Fontaine}, G. 2020, \apj, 901, 93, \dodoi{10.3847/1538-4357/abafbe}

\bibitem[{{Berger} {et~al.}(2018){Berger}, {Huber}, {Gaidos}, \& {van Saders}}]{2018ApJ...866...99B}
{Berger}, T.~A., {Huber}, D., {Gaidos}, E., \& {van Saders}, J.~L. 2018, \apj, 866, 99, \dodoi{10.3847/1538-4357/aada83}

\bibitem[{{Bergeron} {et~al.}(2019){Bergeron}, {Dufour}, {Fontaine}, {Coutu}, {Blouin}, {Genest-Beaulieu}, {B{\'e}dard}, \& {Rolland}}]{2019ApJ...876...67B}
{Bergeron}, P., {Dufour}, P., {Fontaine}, G., {et~al.} 2019, \apj, 876, 67, \dodoi{10.3847/1538-4357/ab153a}

\bibitem[{{Bergeron} {et~al.}(2011){Bergeron}, {Wesemael}, {Dufour}, {Beauchamp}, {Hunter}, {Saffer}, {Gianninas}, {Ruiz}, {Limoges}, {Dufour}, {Fontaine}, \& {Liebert}}]{2011ApJ...737...28B}
{Bergeron}, P., {Wesemael}, F., {Dufour}, P., {et~al.} 2011, \apj, 737, 28, \dodoi{10.1088/0004-637X/737/1/28}

\bibitem[{{Bida} {et~al.}(2014){Bida}, {Dunham}, {Massey}, \& {Roe}}]{2014SPIE.9147E..2NB}
{Bida}, T.~A., {Dunham}, E.~W., {Massey}, P., \& {Roe}, H.~G. 2014, in Society of Photo-Optical Instrumentation Engineers (SPIE) Conference Series, Vol. 9147, Ground-based and Airborne Instrumentation for Astronomy V, ed. S.~K. {Ramsay}, I.~S. {McLean}, \& H.~{Takami}, 91472N, \dodoi{10.1117/12.2056872}

\bibitem[{{Brice{\~n}o} {et~al.}(2018){Brice{\~n}o}, {Heathcote}, {Cecil}, {O'Donoghue}, {Tighe}, {Schurter}, {Mart{\'\i}nez}, {Cantarruti}, \& {Estay}}]{2018SPIE10700E..3ZB}
{Brice{\~n}o}, C., {Heathcote}, S., {Cecil}, G., {et~al.} 2018, in Society of Photo-Optical Instrumentation Engineers (SPIE) Conference Series, Vol. 10700, Ground-based and Airborne Telescopes VII, ed. H.~K. {Marshall} \& J.~{Spyromilio}, 107003Z, \dodoi{10.1117/12.2314026}

\bibitem[{{Briggs} {et~al.}(2015){Briggs}, {Ferrario}, {Tout}, {Wickramasinghe}, \& {Hurley}}]{2015MNRAS.447.1713B}
{Briggs}, G.~P., {Ferrario}, L., {Tout}, C.~A., {Wickramasinghe}, D.~T., \& {Hurley}, J.~R. 2015, \mnras, 447, 1713, \dodoi{10.1093/mnras/stu2539}

\bibitem[{{Brown} {et~al.}(2007){Brown}, {Geller}, {Kenyon}, {Kurtz}, \& {Bromley}}]{2007ApJ...671.1708B}
{Brown}, W.~R., {Geller}, M.~J., {Kenyon}, S.~J., {Kurtz}, M.~J., \& {Bromley}, B.~C. 2007, \apj, 671, 1708, \dodoi{10.1086/523642}

\bibitem[{{Caiazzo} {et~al.}(2021){Caiazzo}, {Burdge}, {Fuller}, {Heyl}, {Kulkarni}, {Prince}, {Richer}, {Schwab}, {Andreoni}, {Bellm}, {Drake}, {Duev}, {Graham}, {Helou}, {Mahabal}, {Masci}, {Smith}, \& {Soumagnac}}]{2021Natur.595...39C}
{Caiazzo}, I., {Burdge}, K.~B., {Fuller}, J., {et~al.} 2021, \nat, 595, 39, \dodoi{10.1038/s41586-021-03615-y}

\bibitem[{{Catal{\'a}n} {et~al.}(2008){Catal{\'a}n}, {Isern}, {Garc{\'\i}a-Berro}, \& {Ribas}}]{2008MNRAS.387.1693C}
{Catal{\'a}n}, S., {Isern}, J., {Garc{\'\i}a-Berro}, E., \& {Ribas}, I. 2008, \mnras, 387, 1693, \dodoi{10.1111/j.1365-2966.2008.13356.x}

\bibitem[{{Chaplin} {et~al.}(2014){Chaplin}, {Basu}, {Huber}, {Serenelli}, {Casagrande}, {Silva Aguirre}, {Ball}, {Creevey}, {Gizon}, {Handberg}, {Karoff}, {Lutz}, {Marques}, {Miglio}, {Stello}, {Suran}, {Pricopi}, {Metcalfe}, {Monteiro}, {Molenda-{\.Z}akowicz}, {Appourchaux}, {Christensen-Dalsgaard}, {Elsworth}, {Garc{\'\i}a}, {Houdek}, {Kjeldsen}, {Bonanno}, {Campante}, {Corsaro}, {Gaulme}, {Hekker}, {Mathur}, {Mosser}, {R{\'e}gulo}, \& {Salabert}}]{2014ApJS..210....1C}
{Chaplin}, W.~J., {Basu}, S., {Huber}, D., {et~al.} 2014, \apjs, 210, 1, \dodoi{10.1088/0067-0049/210/1/1}

\bibitem[{{Cheng} {et~al.}(2019){Cheng}, {Cummings}, \& {M{\'e}nard}}]{2019ApJ...886..100C}
{Cheng}, S., {Cummings}, J.~D., \& {M{\'e}nard}, B. 2019, \apj, 886, 100, \dodoi{10.3847/1538-4357/ab4989}

\bibitem[{{Claytor} {et~al.}(2020){Claytor}, {van Saders}, {Santos}, {Garc{\'\i}a}, {Mathur}, {Tayar}, {Pinsonneault}, \& {Shetrone}}]{2020ApJ...888...43C}
{Claytor}, Z.~R., {van Saders}, J.~L., {Santos}, {\^A}. R.~G., {et~al.} 2020, \apj, 888, 43, \dodoi{10.3847/1538-4357/ab5c24}

\bibitem[{{Clemens} {et~al.}(2004){Clemens}, {Crain}, \& {Anderson}}]{2004SPIE.5492..331C}
{Clemens}, J.~C., {Crain}, J.~A., \& {Anderson}, R. 2004, in Society of Photo-Optical Instrumentation Engineers (SPIE) Conference Series, Vol. 5492, Ground-based Instrumentation for Astronomy, ed. A.~F.~M. {Moorwood} \& M.~{Iye}, 331--340, \dodoi{10.1117/12.550069}

\bibitem[{{Cooper} {et~al.}(2023){Cooper}, {Koposov}, {Allende Prieto}, {Manser}, {Kizhuprakkat}, {Myers}, {Dey}, {G{\"a}nsicke}, {Li}, {Rockosi}, {Valluri}, {Najita}, {Deason}, {Raichoor}, {Wang}, {Ting}, {Kim}, {Carrillo}, {Wang}, {Beraldo e Silva}, {Han}, {Ding}, {S{\'a}nchez-Conde}, {Aguilar}, {Ahlen}, {Bailey}, {Belokurov}, {Brooks}, {Cunha}, {Dawson}, {de la Macorra}, {Doel}, {Eisenstein}, {Fagrelius}, {Fanning}, {Font-Ribera}, {Forero-Romero}, {Gazta{\~n}aga}, {Gontcho a Gontcho}, {Guy}, {Honscheid}, {Kehoe}, {Kisner}, {Kremin}, {Landriau}, {Levi}, {Martini}, {Meisner}, {Miquel}, {Moustakas}, {Nie}, {Palanque-Delabrouille}, {Percival}, {Poppett}, {Prada}, {Rehemtulla}, {Schlafly}, {Schlegel}, {Schubnell}, {Sharples}, {Tarl{\'e}}, {Wechsler}, {Weinberg}, {Zhou}, \& {Zou}}]{2023ApJ...947...37C}
{Cooper}, A.~P., {Koposov}, S.~E., {Allende Prieto}, C., {et~al.} 2023, \apj, 947, 37, \dodoi{10.3847/1538-4357/acb3c0}

\bibitem[{{Coutu} {et~al.}(2019){Coutu}, {Dufour}, {Bergeron}, {Blouin}, {Loranger}, {Allard}, \& {Dunlap}}]{2019ApJ...885...74C}
{Coutu}, S., {Dufour}, P., {Bergeron}, P., {et~al.} 2019, \apj, 885, 74, \dodoi{10.3847/1538-4357/ab46b9}

\bibitem[{{Croom} {et~al.}(2004){Croom}, {Smith}, {Boyle}, {Shanks}, {Miller}, {Outram}, \& {Loaring}}]{2004MNRAS.349.1397C}
{Croom}, S.~M., {Smith}, R.~J., {Boyle}, B.~J., {et~al.} 2004, \mnras, 349, 1397, \dodoi{10.1111/j.1365-2966.2004.07619.x}

\bibitem[{{Cukanovaite} {et~al.}(2021){Cukanovaite}, {Tremblay}, {Bergeron}, {Freytag}, {Ludwig}, \& {Steffen}}]{2021MNRAS.501.5274C}
{Cukanovaite}, E., {Tremblay}, P.-E., {Bergeron}, P., {et~al.} 2021, \mnras, 501, 5274, \dodoi{10.1093/mnras/staa3684}

\bibitem[{{Cummings} \& {Kalirai}(2018)}]{2018AJ....156..165C}
{Cummings}, J.~D., \& {Kalirai}, J.~S. 2018, \aj, 156, 165, \dodoi{10.3847/1538-3881/aad5df}

\bibitem[{{Cummings} {et~al.}(2019){Cummings}, {Kalirai}, {Choi}, {Georgy}, {Tremblay}, \& {Ramirez-Ruiz}}]{2019ApJ...871L..18C}
{Cummings}, J.~D., {Kalirai}, J.~S., {Choi}, J., {et~al.} 2019, \apjl, 871, L18, \dodoi{10.3847/2041-8213/aafc2d}

\bibitem[{{Cummings} {et~al.}(2015){Cummings}, {Kalirai}, {Tremblay}, \& {Ramirez-Ruiz}}]{2015ApJ...807...90C}
{Cummings}, J.~D., {Kalirai}, J.~S., {Tremblay}, P.~E., \& {Ramirez-Ruiz}, E. 2015, \apj, 807, 90, \dodoi{10.1088/0004-637X/807/1/90}

\bibitem[{{Cummings} {et~al.}(2016){Cummings}, {Kalirai}, {Tremblay}, \& {Ramirez-Ruiz}}]{2016ApJ...818...84C}
---. 2016, \apj, 818, 84, \dodoi{10.3847/0004-637X/818/1/84}

\bibitem[{{Cummings} {et~al.}(2018){Cummings}, {Kalirai}, {Tremblay}, {Ramirez-Ruiz}, \& {Choi}}]{2018ApJ...866...21C}
{Cummings}, J.~D., {Kalirai}, J.~S., {Tremblay}, P.~E., {Ramirez-Ruiz}, E., \& {Choi}, J. 2018, \apj, 866, 21, \dodoi{10.3847/1538-4357/aadfd6}

\bibitem[{{Cunningham} {et~al.}(2024){Cunningham}, {Tremblay}, \& {W. O'Brien}}]{2024MNRAS.527.3602C}
{Cunningham}, T., {Tremblay}, P.-E., \& {W. O'Brien}, M. 2024, \mnras, 527, 3602, \dodoi{10.1093/mnras/stad3275}

\bibitem[{{Curtis} {et~al.}(2020){Curtis}, {Ag{\"u}eros}, {Matt}, {Covey}, {Douglas}, {Angus}, {Saar}, {Cody}, {Vanderburg}, {Law}, {Kraus}, {Latham}, {Baranec}, {Riddle}, {Ziegler}, {Lund}, {Torres}, {Meibom}, {Aguirre}, \& {Wright}}]{2020ApJ...904..140C}
{Curtis}, J.~L., {Ag{\"u}eros}, M.~A., {Matt}, S.~P., {et~al.} 2020, \apj, 904, 140, \dodoi{10.3847/1538-4357/abbf58}

\bibitem[{{Dobbie} {et~al.}(2012){Dobbie}, {Baxter}, {K{\"u}lebi}, {Parker}, {Koester}, {Jordan}, {Lodieu}, \& {Euchner}}]{2012MNRAS.421..202D}
{Dobbie}, P.~D., {Baxter}, R., {K{\"u}lebi}, B., {et~al.} 2012, \mnras, 421, 202, \dodoi{10.1111/j.1365-2966.2012.20291.x}

\bibitem[{{Dotter}(2016)}]{2016ApJS..222....8D}
{Dotter}, A. 2016, \apjs, 222, 8, \dodoi{10.3847/0067-0049/222/1/8}

\bibitem[{{Dunlap} \& {Clemens}(2015)}]{2015ASPC..493..547D}
{Dunlap}, B.~H., \& {Clemens}, J.~C. 2015, in Astronomical Society of the Pacific Conference Series, Vol. 493, 19th European Workshop on White Dwarfs, ed. P.~{Dufour}, P.~{Bergeron}, \& G.~{Fontaine}, 547

\bibitem[{{Eisenstein} {et~al.}(2006){Eisenstein}, {Liebert}, {Harris}, {Kleinman}, {Nitta}, {Silvestri}, {Anderson}, {Barentine}, {Brewington}, {Brinkmann}, {Harvanek}, {Krzesi{\'n}ski}, {Neilsen}, {Long}, {Schneider}, \& {Snedden}}]{2006ApJS..167...40E}
{Eisenstein}, D.~J., {Liebert}, J., {Harris}, H.~C., {et~al.} 2006, \apjs, 167, 40, \dodoi{10.1086/507110}

\bibitem[{{El-Badry} \& {Rix}(2018)}]{2018MNRAS.480.4884E}
{El-Badry}, K., \& {Rix}, H.-W. 2018, \mnras, 480, 4884, \dodoi{10.1093/mnras/sty2186}

\bibitem[{{El-Badry} {et~al.}(2021){El-Badry}, {Rix}, \& {Heintz}}]{2021MNRAS.506.2269E}
{El-Badry}, K., {Rix}, H.-W., \& {Heintz}, T.~M. 2021, \mnras, \dodoi{10.1093/mnras/stab323}

\bibitem[{{El-Badry} {et~al.}(2018){El-Badry}, {Rix}, \& {Weisz}}]{2018ApJ...860L..17E}
{El-Badry}, K., {Rix}, H.-W., \& {Weisz}, D.~R. 2018, \apjl, 860, L17, \dodoi{10.3847/2041-8213/aaca9c}

\bibitem[{{Fields} {et~al.}(2016){Fields}, {Farmer}, {Petermann}, {Iliadis}, \& {Timmes}}]{2016ApJ...823...46F}
{Fields}, C.~E., {Farmer}, R., {Petermann}, I., {Iliadis}, C., \& {Timmes}, F.~X. 2016, \apj, 823, 46, \dodoi{10.3847/0004-637X/823/1/46}

\bibitem[{{Finley} \& {Koester}(1997)}]{1997ApJ...489L..79F}
{Finley}, D.~S., \& {Koester}, D. 1997, \apjl, 489, L79, \dodoi{10.1086/310967}

\bibitem[{{Fontaine} {et~al.}(2001){Fontaine}, {Brassard}, \& {Bergeron}}]{2001PASP..113..409F}
{Fontaine}, G., {Brassard}, P., \& {Bergeron}, P. 2001, \pasp, 113, 409, \dodoi{10.1086/319535}

\bibitem[{{Fouesneau} {et~al.}(2019){Fouesneau}, {Rix}, {von Hippel}, {Hogg}, \& {Tian}}]{2019ApJ...870....9F}
{Fouesneau}, M., {Rix}, H.-W., {von Hippel}, T., {Hogg}, D.~W., \& {Tian}, H. 2019, \apj, 870, 9, \dodoi{10.3847/1538-4357/aaee74}

\bibitem[{{Gaia Collaboration} {et~al.}(2016){Gaia Collaboration}, {Prusti}, {de Bruijne}, {Brown}, {Vallenari}, {Babusiaux}, {Bailer-Jones}, {Bastian}, {Biermann}, {Evans}, {Eyer}, {Jansen}, {Jordi}, {Klioner}, {Lammers}, {Lindegren}, {Luri}, {Mignard}, {Milligan}, {Panem}, {Poinsignon}, {Pourbaix}, {Randich}, {Sarri}, {Sartoretti}, {Siddiqui}, {Soubiran}, {Valette}, {van Leeuwen}, {Walton}, {Aerts}, {Arenou}, {Cropper}, {Drimmel}, {H{\o}g}, {Katz}, {Lattanzi}, {O'Mullane}, {Grebel}, {Holland}, {Huc}, {Passot}, {Bramante}, {Cacciari}, {Casta{\~n}eda}, {Chaoul}, {Cheek}, {De Angeli}, {Fabricius}, {Guerra}, {Hern{\'a}ndez}, {Jean-Antoine-Piccolo}, {Masana}, {Messineo}, {Mowlavi}, {Nienartowicz}, {Ord{\'o}{\~n}ez-Blanco}, {Panuzzo}, {Portell}, {Richards}, {Riello}, {Seabroke}, {Tanga}, {Th{\'e}venin}, {Torra}, {Els}, {Gracia-Abril}, {Comoretto}, {Garcia-Reinaldos}, {Lock}, {Mercier}, {Altmann}, {Andrae}, {Astraatmadja}, {Bellas-Velidis}, {Benson}, {Berthier}, {Blomme}, {Busso}, {Carry}, {Cellino}, {Clementini},
  {Cowell}, {Creevey}, {Cuypers}, {Davidson}, {De Ridder}, {de Torres}, {Delchambre}, {Dell'Oro}, {Ducourant}, {Fr{\'e}mat}, {Garc{\'\i}a-Torres}, {Gosset}, {Halbwachs}, {Hambly}, {Harrison}, {Hauser}, {Hestroffer}, {Hodgkin}, {Huckle}, {Hutton}, {Jasniewicz}, {Jordan}, {Kontizas}, {Korn}, {Lanzafame}, {Manteiga}, {Moitinho}, {Muinonen}, {Osinde}, {Pancino}, {Pauwels}, {Petit}, {Recio-Blanco}, {Robin}, {Sarro}, {Siopis}, {Smith}, {Smith}, {Sozzetti}, {Thuillot}, {van Reeven}, {Viala}, {Abbas}, {Abreu Aramburu}, {Accart}, {Aguado}, {Allan}, {Allasia}, {Altavilla}, {{\'A}lvarez}, {Alves}, {Anderson}, {Andrei}, {Anglada Varela}, {Antiche}, {Antoja}, {Ant{\'o}n}, {Arcay}, {Atzei}, {Ayache}, {Bach}, {Baker}, {Balaguer-N{\'u}{\~n}ez}, {Barache}, {Barata}, {Barbier}, {Barblan}, {Baroni}, {Barrado y Navascu{\'e}s}, {Barros}, {Barstow}, {Becciani}, {Bellazzini}, {Bellei}, {Bello Garc{\'\i}a}, {Belokurov}, {Bendjoya}, {Berihuete}, {Bianchi}, {Bienaym{\'e}}, {Billebaud}, {Blagorodnova}, {Blanco-Cuaresma}, {Boch},
  {Bombrun}, {Borrachero}, {Bouquillon}, {Bourda}, {Bouy}, {Bragaglia}, {Breddels}, {Brouillet}, {Br{\"u}semeister}, {Bucciarelli}, {Budnik}, {Burgess}, {Burgon}, {Burlacu}, {Busonero}, {Buzzi}, {Caffau}, {Cambras}, {Campbell}, {Cancelliere}, {Cantat-Gaudin}, {Carlucci}, {Carrasco}, {Castellani}, {Charlot}, {Charnas}, {Charvet}, {Chassat}, {Chiavassa}, {Clotet}, {Cocozza}, {Collins}, {Collins}, {Costigan}, {Crifo}, {Cross}, {Crosta}, {Crowley}, {Dafonte}, {Damerdji}, {Dapergolas}, {David}, {David}, {De Cat}, {de Felice}, {de Laverny}, {De Luise}, {De March}, {de Martino}, {de Souza}, {Debosscher}, {del Pozo}, {Delbo}, {Delgado}, {Delgado}, {di Marco}, {Di Matteo}, {Diakite}, {Distefano}, {Dolding}, {Dos Anjos}, {Drazinos}, {Dur{\'a}n}, {Dzigan}, {Ecale}, {Edvardsson}, {Enke}, {Erdmann}, {Escolar}, {Espina}, {Evans}, {Eynard Bontemps}, {Fabre}, {Fabrizio}, {Faigler}, {Falc{\~a}o}, {Farr{\`a}s Casas}, {Faye}, {Federici}, {Fedorets}, {Fern{\'a}ndez-Hern{\'a}ndez}, {Fernique}, {Fienga}, {Figueras}, {Filippi},
  {Findeisen}, {Fonti}, {Fouesneau}, {Fraile}, {Fraser}, {Fuchs}, {Furnell}, {Gai}, {Galleti}, {Galluccio}, {Garabato}, {Garc{\'\i}a-Sedano}, {Gar{\'e}}, {Garofalo}, {Garralda}, {Gavras}, {Gerssen}, {Geyer}, {Gilmore}, {Girona}, {Giuffrida}, {Gomes}, {Gonz{\'a}lez-Marcos}, {Gonz{\'a}lez-N{\'u}{\~n}ez}, {Gonz{\'a}lez-Vidal}, {Granvik}, {Guerrier}, {Guillout}, {Guiraud}, {G{\'u}rpide}, {Guti{\'e}rrez-S{\'a}nchez}, {Guy}, {Haigron}, {Hatzidimitriou}, {Haywood}, {Heiter}, {Helmi}, {Hobbs}, {Hofmann}, {Holl}, {Holland}, {Hunt}, {Hypki}, {Icardi}, {Irwin}, {Jevardat de Fombelle}, {Jofr{\'e}}, {Jonker}, {Jorissen}, {Julbe}, {Karampelas}, {Kochoska}, {Kohley}, {Kolenberg}, {Kontizas}, {Koposov}, {Kordopatis}, {Koubsky}, {Kowalczyk}, {Krone-Martins}, {Kudryashova}, {Kull}, {Bachchan}, {Lacoste-Seris}, {Lanza}, {Lavigne}, {Le Poncin-Lafitte}, {Lebreton}, {Lebzelter}, {Leccia}, {Leclerc}, {Lecoeur-Taibi}, {Lemaitre}, {Lenhardt}, {Leroux}, {Liao}, {Licata}, {Lindstr{\o}m}, {Lister}, {Livanou}, {Lobel}, {L{\"o}ffler},
  {L{\'o}pez}, {Lopez-Lozano}, {Lorenz}, {Loureiro}, {MacDonald}, {Magalh{\~a}es Fernandes}, {Managau}, {Mann}, {Mantelet}, {Marchal}, {Marchant}, {Marconi}, {Marie}, {Marinoni}, {Marrese}, {Marschalk{\'o}}, {Marshall}, {Mart{\'\i}n-Fleitas}, {Martino}, {Mary}, {Matijevi{\v{c}}}, {Mazeh}, {McMillan}, {Messina}, {Mestre}, {Michalik}, {Millar}, {Miranda}, {Molina}, {Molinaro}, {Molinaro}, {Moln{\'a}r}, {Moniez}, {Montegriffo}, {Monteiro}, {Mor}, {Mora}, {Morbidelli}, {Morel}, {Morgenthaler}, {Morley}, {Morris}, {Mulone}, {Muraveva}, {Musella}, {Narbonne}, {Nelemans}, {Nicastro}, {Noval}, {Ord{\'e}novic}, {Ordieres-Mer{\'e}}, {Osborne}, {Pagani}, {Pagano}, {Pailler}, {Palacin}, {Palaversa}, {Parsons}, {Paulsen}, {Pecoraro}, {Pedrosa}, {Pentik{\"a}inen}, {Pereira}, {Pichon}, {Piersimoni}, {Pineau}, {Plachy}, {Plum}, {Poujoulet}, {Pr{\v{s}}a}, {Pulone}, {Ragaini}, {Rago}, {Rambaux}, {Ramos-Lerate}, {Ranalli}, {Rauw}, {Read}, {Regibo}, {Renk}, {Reyl{\'e}}, {Ribeiro}, {Rimoldini}, {Ripepi}, {Riva}, {Rixon},
  {Roelens}, {Romero-G{\'o}mez}, {Rowell}, {Royer}, {Rudolph}, {Ruiz-Dern}, {Sadowski}, {Sagrist{\`a} Sell{\'e}s}, {Sahlmann}, {Salgado}, {Salguero}, {Sarasso}, {Savietto}, {Schnorhk}, {Schultheis}, {Sciacca}, {Segol}, {Segovia}, {Segransan}, {Serpell}, {Shih}, {Smareglia}, {Smart}, {Smith}, {Solano}, {Solitro}, {Sordo}, {Soria Nieto}, {Souchay}, {Spagna}, {Spoto}, {Stampa}, {Steele}, {Steidelm{\"u}ller}, {Stephenson}, {Stoev}, {Suess}, {S{\"u}veges}, {Surdej}, {Szabados}, {Szegedi-Elek}, {Tapiador}, {Taris}, {Tauran}, {Taylor}, {Teixeira}, {Terrett}, {Tingley}, {Trager}, {Turon}, {Ulla}, {Utrilla}, {Valentini}, {van Elteren}, {Van Hemelryck}, {van Leeuwen}, {Varadi}, {Vecchiato}, {Veljanoski}, {Via}, {Vicente}, {Vogt}, {Voss}, {Votruba}, {Voutsinas}, {Walmsley}, {Weiler}, {Weingrill}, {Werner}, {Wevers}, {Whitehead}, {Wyrzykowski}, {Yoldas}, {{\v{Z}}erjal}, {Zucker}, {Zurbach}, {Zwitter}, {Alecu}, {Allen}, {Allende Prieto}, {Amorim}, {Anglada-Escud{\'e}}, {Arsenijevic}, {Azaz}, {Balm}, {Beck}, {Bernstein},
  {Bigot}, {Bijaoui}, {Blasco}, {Bonfigli}, {Bono}, {Boudreault}, {Bressan}, {Brown}, {Brunet}, {Bunclark}, {Buonanno}, {Butkevich}, {Carret}, {Carrion}, {Chemin}, {Ch{\'e}reau}, {Corcione}, {Darmigny}, {de Boer}, {de Teodoro}, {de Zeeuw}, {Delle Luche}, {Domingues}, {Dubath}, {Fodor}, {Fr{\'e}zouls}, {Fries}, {Fustes}, {Fyfe}, {Gallardo}, {Gallegos}, {Gardiol}, {Gebran}, {Gomboc}, {G{\'o}mez}, {Grux}, {Gueguen}, {Heyrovsky}, {Hoar}, {Iannicola}, {Isasi Parache}, {Janotto}, {Joliet}, {Jonckheere}, {Keil}, {Kim}, {Klagyivik}, {Klar}, {Knude}, {Kochukhov}, {Kolka}, {Kos}, {Kutka}, {Lainey}, {LeBouquin}, {Liu}, {Loreggia}, {Makarov}, {Marseille}, {Martayan}, {Martinez-Rubi}, {Massart}, {Meynadier}, {Mignot}, {Munari}, {Nguyen}, {Nordlander}, {Ocvirk}, {O'Flaherty}, {Olias Sanz}, {Ortiz}, {Osorio}, {Oszkiewicz}, {Ouzounis}, {Palmer}, {Park}, {Pasquato}, {Peltzer}, {Peralta}, {P{\'e}turaud}, {Pieniluoma}, {Pigozzi}, {Poels}, {Prat}, {Prod'homme}, {Raison}, {Rebordao}, {Risquez}, {Rocca-Volmerange}, {Rosen},
  {Ruiz-Fuertes}, {Russo}, {Sembay}, {Serraller Vizcaino}, {Short}, {Siebert}, {Silva}, {Sinachopoulos}, {Slezak}, {Soffel}, {Sosnowska}, {Strai{\v{z}}ys}, {ter Linden}, {Terrell}, {Theil}, {Tiede}, {Troisi}, {Tsalmantza}, {Tur}, {Vaccari}, {Vachier}, {Valles}, {Van Hamme}, {Veltz}, {Virtanen}, {Wallut}, {Wichmann}, {Wilkinson}, {Ziaeepour}, \& {Zschocke}}]{2016A&A...595A...1G}
{Gaia Collaboration}, {Prusti}, T., {de Bruijne}, J.~H.~J., {et~al.} 2016, \aap, 595, A1, \dodoi{10.1051/0004-6361/201629272}

\bibitem[{{Gaia Collaboration} {et~al.}(2018){Gaia Collaboration}, {Brown}, {Vallenari}, {Prusti}, {de Bruijne}, {Babusiaux}, {Bailer-Jones}, {Biermann}, {Evans}, {Eyer}, {Jansen}, {Jordi}, {Klioner}, {Lammers}, {Lindegren}, {Luri}, {Mignard}, {Panem}, {Pourbaix}, {Randich}, {Sartoretti}, {Siddiqui}, {Soubiran}, {van Leeuwen}, {Walton}, {Arenou}, {Bastian}, {Cropper}, {Drimmel}, {Katz}, {Lattanzi}, {Bakker}, {Cacciari}, {Casta{\~n}eda}, {Chaoul}, {Cheek}, {De Angeli}, {Fabricius}, {Guerra}, {Holl}, {Masana}, {Messineo}, {Mowlavi}, {Nienartowicz}, {Panuzzo}, {Portell}, {Riello}, {Seabroke}, {Tanga}, {Th{\'e}venin}, {Gracia-Abril}, {Comoretto}, {Garcia-Reinaldos}, {Teyssier}, {Altmann}, {Andrae}, {Audard}, {Bellas-Velidis}, {Benson}, {Berthier}, {Blomme}, {Burgess}, {Busso}, {Carry}, {Cellino}, {Clementini}, {Clotet}, {Creevey}, {Davidson}, {De Ridder}, {Delchambre}, {Dell'Oro}, {Ducourant}, {Fern{\'a}ndez-Hern{\'a}ndez}, {Fouesneau}, {Fr{\'e}mat}, {Galluccio}, {Garc{\'\i}a-Torres},
  {Gonz{\'a}lez-N{\'u}{\~n}ez}, {Gonz{\'a}lez-Vidal}, {Gosset}, {Guy}, {Halbwachs}, {Hambly}, {Harrison}, {Hern{\'a}ndez}, {Hestroffer}, {Hodgkin}, {Hutton}, {Jasniewicz}, {Jean-Antoine-Piccolo}, {Jordan}, {Korn}, {Krone-Martins}, {Lanzafame}, {Lebzelter}, {L{\"o}ffler}, {Manteiga}, {Marrese}, {Mart{\'\i}n-Fleitas}, {Moitinho}, {Mora}, {Muinonen}, {Osinde}, {Pancino}, {Pauwels}, {Petit}, {Recio-Blanco}, {Richards}, {Rimoldini}, {Robin}, {Sarro}, {Siopis}, {Smith}, {Sozzetti}, {S{\"u}veges}, {Torra}, {van Reeven}, {Abbas}, {Abreu Aramburu}, {Accart}, {Aerts}, {Altavilla}, {{\'A}lvarez}, {Alvarez}, {Alves}, {Anderson}, {Andrei}, {Anglada Varela}, {Antiche}, {Antoja}, {Arcay}, {Astraatmadja}, {Bach}, {Baker}, {Balaguer-N{\'u}{\~n}ez}, {Balm}, {Barache}, {Barata}, {Barbato}, {Barblan}, {Barklem}, {Barrado}, {Barros}, {Barstow}, {Bartholom{\'e} Mu{\~n}oz}, {Bassilana}, {Becciani}, {Bellazzini}, {Berihuete}, {Bertone}, {Bianchi}, {Bienaym{\'e}}, {Blanco-Cuaresma}, {Boch}, {Boeche}, {Bombrun}, {Borrachero},
  {Bossini}, {Bouquillon}, {Bourda}, {Bragaglia}, {Bramante}, {Breddels}, {Bressan}, {Brouillet}, {Br{\"u}semeister}, {Brugaletta}, {Bucciarelli}, {Burlacu}, {Busonero}, {Butkevich}, {Buzzi}, {Caffau}, {Cancelliere}, {Cannizzaro}, {Cantat-Gaudin}, {Carballo}, {Carlucci}, {Carrasco}, {Casamiquela}, {Castellani}, {Castro-Ginard}, {Charlot}, {Chemin}, {Chiavassa}, {Cocozza}, {Costigan}, {Cowell}, {Crifo}, {Crosta}, {Crowley}, {Cuypers}, {Dafonte}, {Damerdji}, {Dapergolas}, {David}, {David}, {de Laverny}, {De Luise}, {De March}, {de Martino}, {de Souza}, {de Torres}, {Debosscher}, {del Pozo}, {Delbo}, {Delgado}, {Delgado}, {Di Matteo}, {Diakite}, {Diener}, {Distefano}, {Dolding}, {Drazinos}, {Dur{\'a}n}, {Edvardsson}, {Enke}, {Eriksson}, {Esquej}, {Eynard Bontemps}, {Fabre}, {Fabrizio}, {Faigler}, {Falc{\~a}o}, {Farr{\`a}s Casas}, {Federici}, {Fedorets}, {Fernique}, {Figueras}, {Filippi}, {Findeisen}, {Fonti}, {Fraile}, {Fraser}, {Fr{\'e}zouls}, {Gai}, {Galleti}, {Garabato}, {Garc{\'\i}a-Sedano}, {Garofalo},
  {Garralda}, {Gavel}, {Gavras}, {Gerssen}, {Geyer}, {Giacobbe}, {Gilmore}, {Girona}, {Giuffrida}, {Glass}, {Gomes}, {Granvik}, {Gueguen}, {Guerrier}, {Guiraud}, {Guti{\'e}rrez-S{\'a}nchez}, {Haigron}, {Hatzidimitriou}, {Hauser}, {Haywood}, {Heiter}, {Helmi}, {Heu}, {Hilger}, {Hobbs}, {Hofmann}, {Holland}, {Huckle}, {Hypki}, {Icardi}, {Jan{\ss}en}, {Jevardat de Fombelle}, {Jonker}, {Juh{\'a}sz}, {Julbe}, {Karampelas}, {Kewley}, {Klar}, {Kochoska}, {Kohley}, {Kolenberg}, {Kontizas}, {Kontizas}, {Koposov}, {Kordopatis}, {Kostrzewa-Rutkowska}, {Koubsky}, {Lambert}, {Lanza}, {Lasne}, {Lavigne}, {Le Fustec}, {Le Poncin-Lafitte}, {Lebreton}, {Leccia}, {Leclerc}, {Lecoeur-Taibi}, {Lenhardt}, {Leroux}, {Liao}, {Licata}, {Lindstr{\o}m}, {Lister}, {Livanou}, {Lobel}, {L{\'o}pez}, {Managau}, {Mann}, {Mantelet}, {Marchal}, {Marchant}, {Marconi}, {Marinoni}, {Marschalk{\'o}}, {Marshall}, {Martino}, {Marton}, {Mary}, {Massari}, {Matijevi{\v{c}}}, {Mazeh}, {McMillan}, {Messina}, {Michalik}, {Millar}, {Molina}, {Molinaro},
  {Moln{\'a}r}, {Montegriffo}, {Mor}, {Morbidelli}, {Morel}, {Morris}, {Mulone}, {Muraveva}, {Musella}, {Nelemans}, {Nicastro}, {Noval}, {O'Mullane}, {Ord{\'e}novic}, {Ord{\'o}{\~n}ez-Blanco}, {Osborne}, {Pagani}, {Pagano}, {Pailler}, {Palacin}, {Palaversa}, {Panahi}, {Pawlak}, {Piersimoni}, {Pineau}, {Plachy}, {Plum}, {Poggio}, {Poujoulet}, {Pr{\v{s}}a}, {Pulone}, {Racero}, {Ragaini}, {Rambaux}, {Ramos-Lerate}, {Regibo}, {Reyl{\'e}}, {Riclet}, {Ripepi}, {Riva}, {Rivard}, {Rixon}, {Roegiers}, {Roelens}, {Romero-G{\'o}mez}, {Rowell}, {Royer}, {Ruiz-Dern}, {Sadowski}, {Sagrist{\`a} Sell{\'e}s}, {Sahlmann}, {Salgado}, {Salguero}, {Sanna}, {Santana-Ros}, {Sarasso}, {Savietto}, {Schultheis}, {Sciacca}, {Segol}, {Segovia}, {S{\'e}gransan}, {Shih}, {Siltala}, {Silva}, {Smart}, {Smith}, {Solano}, {Solitro}, {Sordo}, {Soria Nieto}, {Souchay}, {Spagna}, {Spoto}, {Stampa}, {Steele}, {Steidelm{\"u}ller}, {Stephenson}, {Stoev}, {Suess}, {Surdej}, {Szabados}, {Szegedi-Elek}, {Tapiador}, {Taris}, {Tauran}, {Taylor},
  {Teixeira}, {Terrett}, {Teyssandier}, {Thuillot}, {Titarenko}, {Torra Clotet}, {Turon}, {Ulla}, {Utrilla}, {Uzzi}, {Vaillant}, {Valentini}, {Valette}, {van Elteren}, {Van Hemelryck}, {van Leeuwen}, {Vaschetto}, {Vecchiato}, {Veljanoski}, {Viala}, {Vicente}, {Vogt}, {von Essen}, {Voss}, {Votruba}, {Voutsinas}, {Walmsley}, {Weiler}, {Wertz}, {Wevers}, {Wyrzykowski}, {Yoldas}, {{\v{Z}}erjal}, {Ziaeepour}, {Zorec}, {Zschocke}, {Zucker}, {Zurbach}, \& {Zwitter}}]{2018A&A...616A...1G}
{Gaia Collaboration}, {Brown}, A.~G.~A., {Vallenari}, A., {et~al.} 2018, \aap, 616, A1, \dodoi{10.1051/0004-6361/201833051}

\bibitem[{{Gaia Collaboration} {et~al.}(2021){Gaia Collaboration}, {Brown}, {Vallenari}, {Prusti}, {de Bruijne}, {Babusiaux}, {Biermann}, {Creevey}, {Evans}, {Eyer}, {Hutton}, {Jansen}, {Jordi}, {Klioner}, {Lammers}, {Lindegren}, {Luri}, {Mignard}, {Panem}, {Pourbaix}, {Randich}, {Sartoretti}, {Soubiran}, {Walton}, {Arenou}, {Bailer-Jones}, {Bastian}, {Cropper}, {Drimmel}, {Katz}, {Lattanzi}, {van Leeuwen}, {Bakker}, {Cacciari}, {Casta{\~n}eda}, {De Angeli}, {Ducourant}, {Fabricius}, {Fouesneau}, {Fr{\'e}mat}, {Guerra}, {Guerrier}, {Guiraud}, {Jean-Antoine Piccolo}, {Masana}, {Messineo}, {Mowlavi}, {Nicolas}, {Nienartowicz}, {Pailler}, {Panuzzo}, {Riclet}, {Roux}, {Seabroke}, {Sordo}, {Tanga}, {Th{\'e}venin}, {Gracia-Abril}, {Portell}, {Teyssier}, {Altmann}, {Andrae}, {Bellas-Velidis}, {Benson}, {Berthier}, {Blomme}, {Brugaletta}, {Burgess}, {Busso}, {Carry}, {Cellino}, {Cheek}, {Clementini}, {Damerdji}, {Davidson}, {Delchambre}, {Dell'Oro}, {Fern{\'a}ndez-Hern{\'a}ndez}, {Galluccio}, {Garc{\'\i}a-Lario},
  {Garcia-Reinaldos}, {Gonz{\'a}lez-N{\'u}{\~n}ez}, {Gosset}, {Haigron}, {Halbwachs}, {Hambly}, {Harrison}, {Hatzidimitriou}, {Heiter}, {Hern{\'a}ndez}, {Hestroffer}, {Hodgkin}, {Holl}, {Jan{\ss}en}, {Jevardat de Fombelle}, {Jordan}, {Krone-Martins}, {Lanzafame}, {L{\"o}ffler}, {Lorca}, {Manteiga}, {Marchal}, {Marrese}, {Moitinho}, {Mora}, {Muinonen}, {Osborne}, {Pancino}, {Pauwels}, {Petit}, {Recio-Blanco}, {Richards}, {Riello}, {Rimoldini}, {Robin}, {Roegiers}, {Rybizki}, {Sarro}, {Siopis}, {Smith}, {Sozzetti}, {Ulla}, {Utrilla}, {van Leeuwen}, {van Reeven}, {Abbas}, {Abreu Aramburu}, {Accart}, {Aerts}, {Aguado}, {Ajaj}, {Altavilla}, {{\'A}lvarez}, {{\'A}lvarez Cid-Fuentes}, {Alves}, {Anderson}, {Anglada Varela}, {Antoja}, {Audard}, {Baines}, {Baker}, {Balaguer-N{\'u}{\~n}ez}, {Balbinot}, {Balog}, {Barache}, {Barbato}, {Barros}, {Barstow}, {Bartolom{\'e}}, {Bassilana}, {Bauchet}, {Baudesson-Stella}, {Becciani}, {Bellazzini}, {Bernet}, {Bertone}, {Bianchi}, {Blanco-Cuaresma}, {Boch}, {Bombrun}, {Bossini},
  {Bouquillon}, {Bragaglia}, {Bramante}, {Breedt}, {Bressan}, {Brouillet}, {Bucciarelli}, {Burlacu}, {Busonero}, {Butkevich}, {Buzzi}, {Caffau}, {Cancelliere}, {C{\'a}novas}, {Cantat-Gaudin}, {Carballo}, {Carlucci}, {Carnerero}, {Carrasco}, {Casamiquela}, {Castellani}, {Castro-Ginard}, {Castro Sampol}, {Chaoul}, {Charlot}, {Chemin}, {Chiavassa}, {Cioni}, {Comoretto}, {Cooper}, {Cornez}, {Cowell}, {Crifo}, {Crosta}, {Crowley}, {Dafonte}, {Dapergolas}, {David}, {David}, {de Laverny}, {De Luise}, {De March}, {De Ridder}, {de Souza}, {de Teodoro}, {de Torres}, {del Peloso}, {del Pozo}, {Delbo}, {Delgado}, {Delgado}, {Delisle}, {Di Matteo}, {Diakite}, {Diener}, {Distefano}, {Dolding}, {Eappachen}, {Edvardsson}, {Enke}, {Esquej}, {Fabre}, {Fabrizio}, {Faigler}, {Fedorets}, {Fernique}, {Fienga}, {Figueras}, {Fouron}, {Fragkoudi}, {Fraile}, {Franke}, {Gai}, {Garabato}, {Garcia-Gutierrez}, {Garc{\'\i}a-Torres}, {Garofalo}, {Gavras}, {Gerlach}, {Geyer}, {Giacobbe}, {Gilmore}, {Girona}, {Giuffrida}, {Gomel}, {Gomez},
  {Gonzalez-Santamaria}, {Gonz{\'a}lez-Vidal}, {Granvik}, {Guti{\'e}rrez-S{\'a}nchez}, {Guy}, {Hauser}, {Haywood}, {Helmi}, {Hidalgo}, {Hilger}, {H{\l}adczuk}, {Hobbs}, {Holland}, {Huckle}, {Jasniewicz}, {Jonker}, {Juaristi Campillo}, {Julbe}, {Karbevska}, {Kervella}, {Khanna}, {Kochoska}, {Kontizas}, {Kordopatis}, {Korn}, {Kostrzewa-Rutkowska}, {Kruszy{\'n}ska}, {Lambert}, {Lanza}, {Lasne}, {Le Campion}, {Le Fustec}, {Lebreton}, {Lebzelter}, {Leccia}, {Leclerc}, {Lecoeur-Taibi}, {Liao}, {Licata}, {Lindstr{\o}m}, {Lister}, {Livanou}, {Lobel}, {Madrero Pardo}, {Managau}, {Mann}, {Marchant}, {Marconi}, {Marcos Santos}, {Marinoni}, {Marocco}, {Marshall}, {Martin Polo}, {Mart{\'\i}n-Fleitas}, {Masip}, {Massari}, {Mastrobuono-Battisti}, {Mazeh}, {McMillan}, {Messina}, {Michalik}, {Millar}, {Mints}, {Molina}, {Molinaro}, {Moln{\'a}r}, {Montegriffo}, {Mor}, {Morbidelli}, {Morel}, {Morris}, {Mulone}, {Munoz}, {Muraveva}, {Murphy}, {Musella}, {Noval}, {Ord{\'e}novic}, {Orr{\`u}}, {Osinde}, {Pagani}, {Pagano},
  {Palaversa}, {Palicio}, {Panahi}, {Pawlak}, {Pe{\~n}alosa Esteller}, {Penttil{\"a}}, {Piersimoni}, {Pineau}, {Plachy}, {Plum}, {Poggio}, {Poretti}, {Poujoulet}, {Pr{\v{s}}a}, {Pulone}, {Racero}, {Ragaini}, {Rainer}, {Raiteri}, {Rambaux}, {Ramos}, {Ramos-Lerate}, {Re Fiorentin}, {Regibo}, {Reyl{\'e}}, {Ripepi}, {Riva}, {Rixon}, {Robichon}, {Robin}, {Roelens}, {Rohrbasser}, {Romero-G{\'o}mez}, {Rowell}, {Royer}, {Rybicki}, {Sadowski}, {Sagrist{\`a} Sell{\'e}s}, {Sahlmann}, {Salgado}, {Salguero}, {Samaras}, {Sanchez Gimenez}, {Sanna}, {Santove{\~n}a}, {Sarasso}, {Schultheis}, {Sciacca}, {Segol}, {Segovia}, {S{\'e}gransan}, {Semeux}, {Shahaf}, {Siddiqui}, {Siebert}, {Siltala}, {Slezak}, {Smart}, {Solano}, {Solitro}, {Souami}, {Souchay}, {Spagna}, {Spoto}, {Steele}, {Steidelm{\"u}ller}, {Stephenson}, {S{\"u}veges}, {Szabados}, {Szegedi-Elek}, {Taris}, {Tauran}, {Taylor}, {Teixeira}, {Thuillot}, {Tonello}, {Torra}, {Torra}, {Turon}, {Unger}, {Vaillant}, {van Dillen}, {Vanel}, {Vecchiato}, {Viala}, {Vicente},
  {Voutsinas}, {Weiler}, {Wevers}, {Wyrzykowski}, {Yoldas}, {Yvard}, {Zhao}, {Zorec}, {Zucker}, {Zurbach}, \& {Zwitter}}]{2021A&A...649A...1G}
---. 2021, \aap, 649, A1, \dodoi{10.1051/0004-6361/202039657}

\bibitem[{{Gentile Fusillo} {et~al.}(2019){Gentile Fusillo}, {Tremblay}, {G{\"a}nsicke}, {Manser}, {Cunningham}, {Cukanovaite}, {Hollands}, {Marsh}, {Raddi}, {Jordan}, {Toonen}, {Geier}, {Barstow}, \& {Cummings}}]{2019MNRAS.482.4570G}
{Gentile Fusillo}, N.~P., {Tremblay}, P.-E., {G{\"a}nsicke}, B.~T., {et~al.} 2019, \mnras, 482, 4570, \dodoi{10.1093/mnras/sty3016}

\bibitem[{{Gentile Fusillo} {et~al.}(2021){Gentile Fusillo}, {Tremblay}, {Cukanovaite}, {Vorontseva}, {Lallement}, {Hollands}, {G{\"a}nsicke}, {Burdge}, {McCleery}, \& {Jordan}}]{2021MNRAS.508.3877G}
{Gentile Fusillo}, N.~P., {Tremblay}, P.~E., {Cukanovaite}, E., {et~al.} 2021, \mnras, 508, 3877, \dodoi{10.1093/mnras/stab2672}

\bibitem[{{Giammichele} {et~al.}(2012){Giammichele}, {Bergeron}, \& {Dufour}}]{2012ApJS..199...29G}
{Giammichele}, N., {Bergeron}, P., \& {Dufour}, P. 2012, \apjs, 199, 29, \dodoi{10.1088/0067-0049/199/2/29}

\bibitem[{{Gianninas} {et~al.}(2011){Gianninas}, {Bergeron}, \& {Ruiz}}]{2011ApJ...743..138G}
{Gianninas}, A., {Bergeron}, P., \& {Ruiz}, M.~T. 2011, \apj, 743, 138, \dodoi{10.1088/0004-637X/743/2/138}

\bibitem[{{Hansen} {et~al.}(2004){Hansen}, {Richer}, {Fahlman}, {Stetson}, {Brewer}, {Currie}, {Gibson}, {Ibata}, {Rich}, \& {Shara}}]{2004ApJS..155..551H}
{Hansen}, B. M.~S., {Richer}, H.~B., {Fahlman}, G.~G., {et~al.} 2004, \apjs, 155, 551, \dodoi{10.1086/424832}

\bibitem[{{Heintz} {et~al.}(2022){Heintz}, {Hermes}, {El-Badry}, {Walsh}, {van Saders}, {Fields}, \& {Koester}}]{2022ApJ...934..148H}
{Heintz}, T.~M., {Hermes}, J.~J., {El-Badry}, K., {et~al.} 2022, \apj, 934, 148, \dodoi{10.3847/1538-4357/ac78d9}

\bibitem[{{Hollands} {et~al.}(2024){Hollands}, {Littlefair}, \& {Parsons}}]{2024MNRAS.527.9061H}
{Hollands}, M.~A., {Littlefair}, S.~P., \& {Parsons}, S.~G. 2024, \mnras, 527, 9061, \dodoi{10.1093/mnras/stad3729}

\bibitem[{{Holmberg} {et~al.}(2009){Holmberg}, {Nordstr{\"o}m}, \& {Andersen}}]{2009A&A...501..941H}
{Holmberg}, J., {Nordstr{\"o}m}, B., \& {Andersen}, J. 2009, \aap, 501, 941, \dodoi{10.1051/0004-6361/200811191}

\bibitem[{{Kalirai}(2012)}]{2012Natur.486...90K}
{Kalirai}, J.~S. 2012, \nat, 486, 90, \dodoi{10.1038/nature11062}

\bibitem[{{Kalirai} {et~al.}(2008){Kalirai}, {Hansen}, {Kelson}, {Reitzel}, {Rich}, \& {Richer}}]{2008ApJ...676..594K}
{Kalirai}, J.~S., {Hansen}, B. M.~S., {Kelson}, D.~D., {et~al.} 2008, \apj, 676, 594, \dodoi{10.1086/527028}

\bibitem[{{Kawka}(2020)}]{2020IAUS..357...60K}
{Kawka}, A. 2020, in White Dwarfs as Probes of Fundamental Physics: Tracers of Planetary, Stellar and Galactic Evolution, ed. M.~A. {Barstow}, S.~J. {Kleinman}, J.~L. {Provencal}, \& L.~{Ferrario}, Vol. 357, 60--74, \dodoi{10.1017/S1743921320000745}

\bibitem[{{Kawka} {et~al.}(2020){Kawka}, {Vennes}, \& {Ferrario}}]{2020MNRAS.491L..40K}
{Kawka}, A., {Vennes}, S., \& {Ferrario}, L. 2020, \mnras, 491, L40, \dodoi{10.1093/mnrasl/slz165}

\bibitem[{{Kepler} {et~al.}(2015){Kepler}, {Pelisoli}, {Koester}, {Ourique}, {Kleinman}, {Romero}, {Nitta}, {Eisenstein}, {Costa}, {K{\"u}lebi}, {Jordan}, {Dufour}, {Giommi}, \& {Rebassa-Mansergas}}]{2015MNRAS.446.4078K}
{Kepler}, S.~O., {Pelisoli}, I., {Koester}, D., {et~al.} 2015, \mnras, 446, 4078, \dodoi{10.1093/mnras/stu2388}

\bibitem[{{Kepler} {et~al.}(2016){Kepler}, {Pelisoli}, {Koester}, {Ourique}, {Romero}, {Reindl}, {Kleinman}, {Eisenstein}, {Valois}, \& {Amaral}}]{2016MNRAS.455.3413K}
---. 2016, \mnras, 455, 3413, \dodoi{10.1093/mnras/stv2526}

\bibitem[{{Kilic} {et~al.}(2020){Kilic}, {Bergeron}, {Kosakowski}, {Brown}, {Ag{\"u}eros}, \& {Blouin}}]{2020ApJ...898...84K}
{Kilic}, M., {Bergeron}, P., {Kosakowski}, A., {et~al.} 2020, \apj, 898, 84, \dodoi{10.3847/1538-4357/ab9b8d}

\bibitem[{{Kilic} {et~al.}(2023){Kilic}, {Moss}, {Kosakowski}, {Bergeron}, {Conly}, {Brown}, {Toonen}, {Williams}, \& {Dufour}}]{2023MNRAS.518.2341K}
{Kilic}, M., {Moss}, A.~G., {Kosakowski}, A., {et~al.} 2023, \mnras, 518, 2341, \dodoi{10.1093/mnras/stac3182}

\bibitem[{{Kilkenny} {et~al.}(2015){Kilkenny}, {O'Donoghue}, {Worters}, {Koen}, {Hambly}, \& {MacGillivray}}]{2015MNRAS.453.1879K}
{Kilkenny}, D., {O'Donoghue}, D., {Worters}, H.~L., {et~al.} 2015, \mnras, 453, 1879, \dodoi{10.1093/mnras/stv1771}

\bibitem[{{Kleinman} {et~al.}(2013){Kleinman}, {Kepler}, {Koester}, {Pelisoli}, {Pe{\c{c}}anha}, {Nitta}, {Costa}, {Krzesinski}, {Dufour}, {Lachapelle}, {Bergeron}, {Yip}, {Harris}, {Eisenstein}, {Althaus}, \& {C{\'o}rsico}}]{2013ApJS..204....5K}
{Kleinman}, S.~J., {Kepler}, S.~O., {Koester}, D., {et~al.} 2013, \apjs, 204, 5, \dodoi{10.1088/0067-0049/204/1/5}

\bibitem[{{Koester}(2010)}]{2010MmSAI..81..921K}
{Koester}, D. 2010, \memsai, 81, 921

\bibitem[{{Koester} {et~al.}(1998){Koester}, {Dreizler}, {Weidemann}, \& {Allard}}]{1998AA...338..612K}
{Koester}, D., {Dreizler}, S., {Weidemann}, V., \& {Allard}, N.~F. 1998, \aap, 338, 612

\bibitem[{{Koester} \& {Kepler}(2019)}]{2019A&A...628A.102K}
{Koester}, D., \& {Kepler}, S.~O. 2019, \aap, 628, A102, \dodoi{10.1051/0004-6361/201935946}

\bibitem[{{Koester} {et~al.}(2009){Koester}, {Voss}, {Napiwotzki}, {Christlieb}, {Homeier}, {Lisker}, {Reimers}, \& {Heber}}]{2009ANA...505..441K}
{Koester}, D., {Voss}, B., {Napiwotzki}, R., {et~al.} 2009, \aap, 505, 441, \dodoi{10.1051/0004-6361/200912531}

\bibitem[{{Kraft}(1967)}]{1967ApJ...150..551K}
{Kraft}, R.~P. 1967, \apj, 150, 551, \dodoi{10.1086/149359}

\bibitem[{{K{\"u}lebi} {et~al.}(2010){K{\"u}lebi}, {Jordan}, {Nelan}, {Bastian}, \& {Altmann}}]{2010ANA...524A..36K}
{K{\"u}lebi}, B., {Jordan}, S., {Nelan}, E., {Bastian}, U., \& {Altmann}, M. 2010, \aap, 524, A36, \dodoi{10.1051/0004-6361/201015237}

\bibitem[{{Lanzafame} {et~al.}(2019){Lanzafame}, {Distefano}, {Barnes}, \& {Spada}}]{2019ApJ...877..157L}
{Lanzafame}, A.~C., {Distefano}, E., {Barnes}, S.~A., \& {Spada}, F. 2019, \apj, 877, 157, \dodoi{10.3847/1538-4357/ab1aa2}

\bibitem[{{Liebert} {et~al.}(2005){Liebert}, {Bergeron}, \& {Holberg}}]{2005ApJS..156...47L}
{Liebert}, J., {Bergeron}, P., \& {Holberg}, J.~B. 2005, \apjs, 156, 47, \dodoi{10.1086/425738}

\bibitem[{{Limoges} {et~al.}(2013){Limoges}, {L{\'e}pine}, \& {Bergeron}}]{2013AJ....145..136L}
{Limoges}, M.~M., {L{\'e}pine}, S., \& {Bergeron}, P. 2013, \aj, 145, 136, \dodoi{10.1088/0004-6256/145/5/136}

\bibitem[{{Lu} {et~al.}(2021){Lu}, {Angus}, {Curtis}, {David}, \& {Kiman}}]{2021AJ....161..189L}
{Lu}, Y.~L., {Angus}, R., {Curtis}, J.~L., {David}, T.~J., \& {Kiman}, R. 2021, \aj, 161, 189, \dodoi{10.3847/1538-3881/abe4d6}

\bibitem[{{Mamajek} \& {Hillenbrand}(2008)}]{2008ApJ...687.1264M}
{Mamajek}, E.~E., \& {Hillenbrand}, L.~A. 2008, \apj, 687, 1264, \dodoi{10.1086/591785}

\bibitem[{{Marigo} {et~al.}(2020){Marigo}, {Cummings}, {Curtis}, {Kalirai}, {Chen}, {Tremblay}, {Ramirez-Ruiz}, {Bergeron}, {Bladh}, {Bressan}, {Girardi}, {Pastorelli}, {Trabucchi}, {Cheng}, {Aringer}, \& {Tio}}]{2020NatAs...4.1102M}
{Marigo}, P., {Cummings}, J.~D., {Curtis}, J.~L., {et~al.} 2020, Nature Astronomy, 4, 1102, \dodoi{10.1038/s41550-020-1132-1}

\bibitem[{{McCleery} {et~al.}(2020){McCleery}, {Tremblay}, {Gentile Fusillo}, {Hollands}, {G{\"a}nsicke}, {Izquierdo}, {Toonen}, {Cunningham}, \& {Rebassa-Mansergas}}]{2020MNRAS.499.1890M}
{McCleery}, J., {Tremblay}, P.-E., {Gentile Fusillo}, N.~P., {et~al.} 2020, \mnras, 499, 1890, \dodoi{10.1093/mnras/staa2030}

\bibitem[{{McCook} \& {Sion}(1999)}]{1999ApJS..121....1M}
{McCook}, G.~P., \& {Sion}, E.~M. 1999, \apjs, 121, 1, \dodoi{10.1086/313186}

\bibitem[{{Metcalfe} \& {Egeland}(2019)}]{2019ApJ...871...39M}
{Metcalfe}, T.~S., \& {Egeland}, R. 2019, \apj, 871, 39, \dodoi{10.3847/1538-4357/aaf575}

\bibitem[{{Moe} \& {Di Stefano}(2017)}]{2017ApJS..230...15M}
{Moe}, M., \& {Di Stefano}, R. 2017, \apjs, 230, 15, \dodoi{10.3847/1538-4365/aa6fb6}

\bibitem[{{O'Brien} {et~al.}(2023){O'Brien}, {Tremblay}, {Gentile Fusillo}, {Hollands}, {G{\"a}nsicke}, {Koester}, {Pelisoli}, {Cukanovaite}, {Cunningham}, {Doyle}, {Elms}, {Farihi}, {Hermes}, {Holberg}, {Jordan}, {Klein}, {Kleinman}, {Manser}, {De Martino}, {Marsh}, {McCleery}, {Melis}, {Nitta}, {Parsons}, {Raddi}, {Rebassa-Mansergas}, {Schreiber}, {Silvotti}, {Steeghs}, {Toloza}, {Toonen}, {Torres}, {Weinberger}, \& {Zuckerman}}]{2023MNRAS.518.3055O}
{O'Brien}, M.~W., {Tremblay}, P.~E., {Gentile Fusillo}, N.~P., {et~al.} 2023, \mnras, 518, 3055, \dodoi{10.1093/mnras/stac3303}

\bibitem[{{O'Brien} {et~al.}(2024){O'Brien}, {Tremblay}, {Klein}, {Koester}, {Melis}, {B{\'e}dard}, {Cukanovaite}, {Cunningham}, {Doyle}, {G{\"a}nsicke}, {Gentile Fusillo}, {Hollands}, {McCleery}, {Pelisoli}, {Toonen}, {Weinberger}, \& {Zuckerman}}]{2024MNRAS.527.8687O}
{O'Brien}, M.~W., {Tremblay}, P.~E., {Klein}, B.~L., {et~al.} 2024, \mnras, 527, 8687, \dodoi{10.1093/mnras/stad3773}

\bibitem[{{O'Donoghue} {et~al.}(2013){O'Donoghue}, {Kilkenny}, {Koen}, {Hambly}, {MacGillivray}, \& {Stobie}}]{2013MNRAS.431..240O}
{O'Donoghue}, D., {Kilkenny}, D., {Koen}, C., {et~al.} 2013, \mnras, 431, 240, \dodoi{10.1093/mnras/stt158}

\bibitem[{{Paxton} {et~al.}(2011){Paxton}, {Bildsten}, {Dotter}, {Herwig}, {Lesaffre}, \& {Timmes}}]{2011ApJS..192....3P}
{Paxton}, B., {Bildsten}, L., {Dotter}, A., {et~al.} 2011, \apjs, 192, 3, \dodoi{10.1088/0067-0049/192/1/3}

\bibitem[{{Paxton} {et~al.}(2013){Paxton}, {Cantiello}, {Arras}, {Bildsten}, {Brown}, {Dotter}, {Mankovich}, {Montgomery}, {Stello}, {Timmes}, \& {Townsend}}]{2013ApJS..208....4P}
{Paxton}, B., {Cantiello}, M., {Arras}, P., {et~al.} 2013, \apjs, 208, 4, \dodoi{10.1088/0067-0049/208/1/4}

\bibitem[{{Paxton} {et~al.}(2015){Paxton}, {Marchant}, {Schwab}, {Bauer}, {Bildsten}, {Cantiello}, {Dessart}, {Farmer}, {Hu}, {Langer}, {Townsend}, {Townsley}, \& {Timmes}}]{2015ApJS..220...15P}
{Paxton}, B., {Marchant}, P., {Schwab}, J., {et~al.} 2015, \apjs, 220, 15, \dodoi{10.1088/0067-0049/220/1/15}

\bibitem[{Prochaska {et~al.}(2020)Prochaska, Hennawi, Westfall, Cooke, Wang, Hsyu, Davies, Farina, \& Pelliccia}]{2020JOSS....5.2308P}
Prochaska, J.~X., Hennawi, J.~F., Westfall, K.~B., {et~al.} 2020, Journal of Open Source Software, 5, 2308, \dodoi{10.21105/joss.02308}

\bibitem[{{Qiu} {et~al.}(2021){Qiu}, {Tian}, {Wang}, {Nie}, {von Hippel}, {Liu}, {Fouesneau}, \& {Rix}}]{2021ApJS..253...58Q}
{Qiu}, D., {Tian}, H.-J., {Wang}, X.-D., {et~al.} 2021, \apjs, 253, 58, \dodoi{10.3847/1538-4365/abe468}

\bibitem[{{Rebassa-Mansergas} {et~al.}(2023){Rebassa-Mansergas}, {Maldonado}, {Raddi}, {Torres}, {Hoskin}, {Cunningham}, {Hollands}, {Ren}, {G{\"a}nsicke}, {Tremblay}, \& {Camisassa}}]{2023MNRAS.526.4787R}
{Rebassa-Mansergas}, A., {Maldonado}, J., {Raddi}, R., {et~al.} 2023, \mnras, 526, 4787, \dodoi{10.1093/mnras/stad3050}

\bibitem[{{Sahu} {et~al.}(2023){Sahu}, {G{\"a}nsicke}, {Tremblay}, {Koester}, {Hermes}, {Wilson}, {Toloza}, {Hoskin}, {Farihi}, {Manser}, \& {Redfield}}]{2023MNRAS.526.5800S}
{Sahu}, S., {G{\"a}nsicke}, B.~T., {Tremblay}, P.-E., {et~al.} 2023, \mnras, 526, 5800, \dodoi{10.1093/mnras/stad2663}

\bibitem[{{Salaris} {et~al.}(2009){Salaris}, {Serenelli}, {Weiss}, \& {Miller Bertolami}}]{2009ApJ...692.1013S}
{Salaris}, M., {Serenelli}, A., {Weiss}, A., \& {Miller Bertolami}, M. 2009, \apj, 692, 1013, \dodoi{10.1088/0004-637X/692/2/1013}

\bibitem[{{Salpeter}(1955)}]{1955ApJ...121..161S}
{Salpeter}, E.~E. 1955, \apj, 121, 161, \dodoi{10.1086/145971}

\bibitem[{{Shariat} {et~al.}(2023){Shariat}, {Naoz}, {Hansen}, {Angelo}, {Michaely}, \& {Stephan}}]{2023ApJ...955L..14S}
{Shariat}, C., {Naoz}, S., {Hansen}, B. M.~S., {et~al.} 2023, \apjl, 955, L14, \dodoi{10.3847/2041-8213/acf76b}

\bibitem[{{Sion} {et~al.}(1991){Sion}, {Oswalt}, {Liebert}, \& {Hintzen}}]{1991AJ....101.1476S}
{Sion}, E.~M., {Oswalt}, T.~D., {Liebert}, J., \& {Hintzen}, P. 1991, \aj, 101, 1476, \dodoi{10.1086/115779}

\bibitem[{{Skumanich}(1972)}]{1972ApJ...171..565S}
{Skumanich}, A. 1972, \apj, 171, 565, \dodoi{10.1086/151310}

\bibitem[{{Temmink} {et~al.}(2020){Temmink}, {Toonen}, {Zapartas}, {Justham}, \& {G{\"a}nsicke}}]{2020A&A...636A..31T}
{Temmink}, K.~D., {Toonen}, S., {Zapartas}, E., {Justham}, S., \& {G{\"a}nsicke}, B.~T. 2020, \aap, 636, A31, \dodoi{10.1051/0004-6361/201936889}

\bibitem[{{Tian} {et~al.}(2020){Tian}, {El-Badry}, {Rix}, \& {Gould}}]{Tian}
{Tian}, H.-J., {El-Badry}, K., {Rix}, H.-W., \& {Gould}, A. 2020, \apjs, 246, 4, \dodoi{10.3847/1538-4365/ab54c4}

\bibitem[{{Tremblay} {et~al.}(2011){Tremblay}, {Bergeron}, \& {Gianninas}}]{2011ApJ...730..128T}
{Tremblay}, P.~E., {Bergeron}, P., \& {Gianninas}, A. 2011, \apj, 730, 128, \dodoi{10.1088/0004-637X/730/2/128}

\bibitem[{{Tremblay} {et~al.}(2019){Tremblay}, {Cukanovaite}, {Gentile Fusillo}, {Cunningham}, \& {Hollands}}]{2019MNRAS.482.5222T}
{Tremblay}, P.~E., {Cukanovaite}, E., {Gentile Fusillo}, N.~P., {Cunningham}, T., \& {Hollands}, M.~A. 2019, \mnras, 482, 5222, \dodoi{10.1093/mnras/sty3067}

\bibitem[{{Tremblay} {et~al.}(2013){Tremblay}, {Ludwig}, {Steffen}, \& {Freytag}}]{2013A&A...559A.104T}
{Tremblay}, P.~E., {Ludwig}, H.~G., {Steffen}, M., \& {Freytag}, B. 2013, \aap, 559, A104, \dodoi{10.1051/0004-6361/201322318}

\bibitem[{{Tremblay} {et~al.}(2020){Tremblay}, {Hollands}, {Gentile Fusillo}, {McCleery}, {Izquierdo}, {G{\"a}nsicke}, {Cukanovaite}, {Koester}, {Brown}, {Charpinet}, {Cunningham}, {Farihi}, {Giammichele}, {van Grootel}, {Hermes}, {Hoskin}, {Jordan}, {Kepler}, {Kleinman}, {Manser}, {Marsh}, {de Martino}, {Nitta}, {Parsons}, {Pelisoli}, {Raddi}, {Rebassa-Mansergas}, {Ren}, {Schreiber}, {Silvotti}, {Toloza}, {Toonen}, \& {Torres}}]{2020MNRAS.497..130T}
{Tremblay}, P.~E., {Hollands}, M.~A., {Gentile Fusillo}, N.~P., {et~al.} 2020, \mnras, 497, 130, \dodoi{10.1093/mnras/staa1892}

\bibitem[{{van Saders} {et~al.}(2016){van Saders}, {Ceillier}, {Metcalfe}, {Silva Aguirre}, {Pinsonneault}, {Garc{\'\i}a}, {Mathur}, \& {Davies}}]{2016Natur.529..181V}
{van Saders}, J.~L., {Ceillier}, T., {Metcalfe}, T.~S., {et~al.} 2016, \nat, 529, 181, \dodoi{10.1038/nature16168}

\bibitem[{{Vincent} {et~al.}(2024){Vincent}, {Barstow}, {Jordan}, {Mander}, {Bergeron}, \& {Dufour}}]{2024A&A...682A...5V}
{Vincent}, O., {Barstow}, M.~A., {Jordan}, S., {et~al.} 2024, \aap, 682, A5, \dodoi{10.1051/0004-6361/202347694}

\bibitem[{{Weidemann} \& {Koester}(1983)}]{1983A&A...121...77W}
{Weidemann}, V., \& {Koester}, D. 1983, \aap, 121, 77

\bibitem[{{Williams} {et~al.}(2009){Williams}, {Bolte}, \& {Koester}}]{2009ApJ...693..355W}
{Williams}, K.~A., {Bolte}, M., \& {Koester}, D. 2009, \apj, 693, 355, \dodoi{10.1088/0004-637X/693/1/355}

\bibitem[{{Williams} {et~al.}(2016){Williams}, {Montgomery}, {Winget}, {Falcon}, \& {Bierwagen}}]{2016ApJ...817...27W}
{Williams}, K.~A., {Montgomery}, M.~H., {Winget}, D.~E., {Falcon}, R.~E., \& {Bierwagen}, M. 2016, \apj, 817, 27, \dodoi{10.3847/0004-637X/817/1/27}

\bibitem[{{Xiang} \& {Rix}(2022)}]{2022Natur.603..599X}
{Xiang}, M., \& {Rix}, H.-W. 2022, \nat, 603, 599, \dodoi{10.1038/s41586-022-04496-5}

\bibitem[{{Zhao} {et~al.}(2012){Zhao}, {Oswalt}, {Willson}, {Wang}, \& {Zhao}}]{2012ApJ...746..144Z}
{Zhao}, J.~K., {Oswalt}, T.~D., {Willson}, L.~A., {Wang}, Q., \& {Zhao}, G. 2012, \apj, 746, 144, \dodoi{10.1088/0004-637X/746/2/144}

\end{thebibliography}
\bibliographystyle{aasjournal}

\appendix
\restartappendixnumbering

\section{Systematic Surface Gravity Differences between Spectroscopy and Photometry}\label{sec:appendix}
We find a systematic trend that the lowest surface gravities determined from spectroscopy deviate more strongly from those determined from photometry (see bottom panel of Figure~\ref{fig:phot_vs_spec}, where there appears to be a correlated trend). Ignoring typical 3-D corrections for spectroscopic surface gravities \citep{2013A&A...559A.104T} does not exaggerate this trend, and instead pushes the spectroscopic vs. photometric surface gravities into better agreement at lower spectroscopic surface gravities. The systematic offset also cannot be explained by the photometry being crowded or blended. Blended photometry would push the photometric surface gravities lower, not higher, thus bringing the spectroscopic and photometric surface gravities into better agreement. Double-degenerate binaries also cannot be invoked since they would broaden the lines, making the derived spectroscopic \logg\ higher.

\begin{figure}[h]
    \centering
    \includegraphics[width=0.975\textwidth]{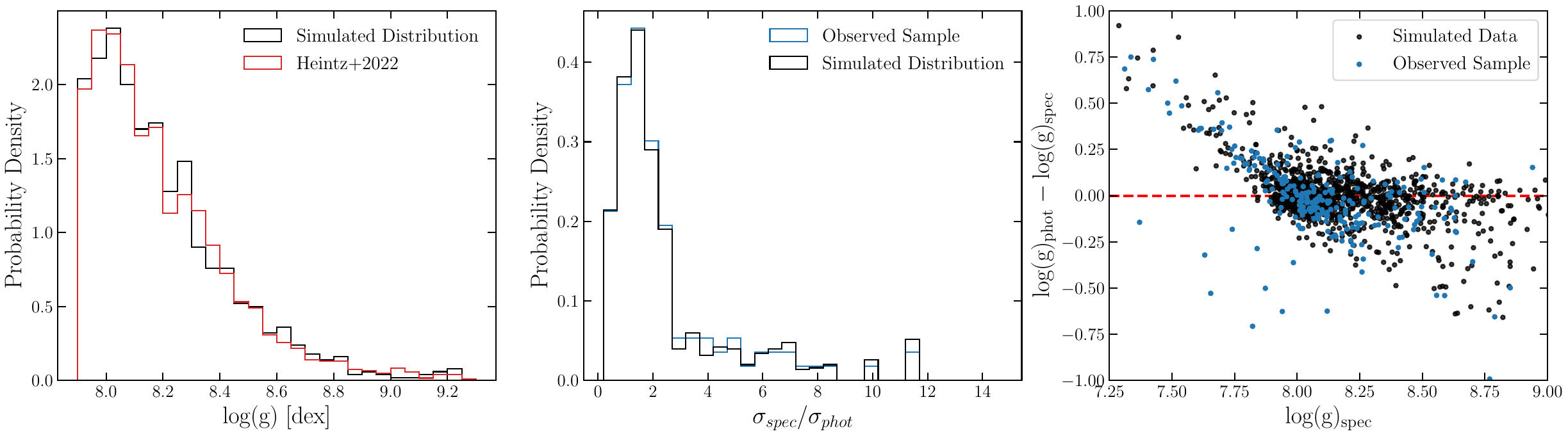}
    \caption{A simulated sample of 1000 white dwarfs to help illustrate the systematic trend seen in Figure~\ref{fig:phot_vs_spec}. The left panel shows the distribution of the true \logg\ values for our simulated sample (see text in Section~\ref{sec:appendix}) and the sample of photometric \logg\ values from \cite{2022ApJ...934..148H}, with white dwarfs with \logg~$<$~7.9 removed. The middle panel shows the ratio of spectroscopic to photometric uncertainties for both the simulated sample and reported \logg\ values from this work. The right panel shows the differences between the observed \logg\ measurements from photometry and spectroscopy from this work on top of simulated measurements described in Section~\ref{sec:appendix}.}
    \label{fig:app_logg}
\end{figure}

Instead, we find the systematic shift between surface gravities derived from photometry and spectroscopy at \logg~$<$~7.9 can be explained statistically. Our formal spectroscopic fit uncertainties are 1.6 times larger on average than the uncertainties reported from photometry (see middle panel of Figure~\ref{fig:app_logg}). We also expect the number of true \logg~$<$~8.0 isolated white dwarfs should be small (e.g., \citealt{2024MNRAS.527.8687O}). Thus, we do not see white dwarfs where photometry and spectroscopy both return a low \logg; instead, white dwarfs with low spectroscopic surface gravities are, in fact, closer to \logg~=~8.0.

To illustrate this, we generate a mock sample of 1000 white dwarfs with simulated true \logg\ values similarly distributed to those from observed photometry in our wide WD+WD sample, with a strict cut at \logg~=~7.9 (see first panel of Figure~\ref{fig:app_logg}). Without the strict cut around \logg~=~7.9, we are not able to reproduce the observed trend. We then introduce random errors into the measured photometric surface gravities ranging from $0.03-0.5$\,dex, which are drawn from the distribution of uncertainties on the photometric surface gravities reported in \cite{2022ApJ...934..148H}. We then require that the random errors introduced to the measured spectroscopic \logg\ values reflect the relative empirical uncertainties (see Figure~\ref{fig:app_logg}, middle panel). The resulting simulated distribution of \logg\ differences between photometry and spectroscopy is compared to the observed sample in the right panel of Figure~\ref{fig:app_logg}. 

The simulated distribution captures the observed trend in the spectroscopic sample very well, and implies that most of the white dwarfs in our sample with spectroscopic \logg~$<$~7.9 are, in fact, normal mass white dwarfs with surface gravities closer to \logg~=~8.0. The larger scatter in spectroscopic surface gravities and the larger uncertainties are a consequence of the Balmer lines becoming weaker at lower effective temperatures (especially $<8000$~K), and we recommend using the photometric \logg\ in these cases, which should be more representative of the intrinsic surface gravity of the white dwarf. We note that there are some white dwarfs that have \logg~$<$~7.9 from spectroscopy, but even smaller surface gravities from photometry (blue points in the bottom-left of the rightmost panel of Figure~\ref{fig:app_logg}). These white dwarfs are probably close binaries that appear over-luminous in the Gaia CMD, and therefore have apparently lower surface gravities. For these objects, the spectroscopic surface gravities are a better estimate of the true value.

\section{LDT and SOAR Spectra with DA Fits}
Here we tabulate all the spectroscopic parameters and spectral types determined through our LDT and SOAR follow-up observations. We also include a figure to the fits to each DA in our sample.

\startlongtable
\begin{deluxetable*}{llccccccc}
\tabletypesize{\scriptsize}
\tablecolumns{8}
\tablewidth{0pc}
\tablecaption{Observed Spectral Types and Fitted Atmospheric Parameters from SOAR and LDT \label{tab:new_obs}}
\tablehead{
        \colhead{Gaia DR3 ID}&
        \colhead{WDJ Name}&
        \colhead{$G$}&
        \colhead{Spectral Type}&
        \colhead{Instrument and Facility}&
        \colhead{SNR}&
        \colhead{UTC Date}&
        \colhead{$T_{\mathrm{eff}}$}&
        \colhead{\logg}\\
  &  & (mag) &  &  &  &  & (K) & (dex) }
\startdata
1100267237077449728 & WDJ065350.43+635558.04 & 16.27 & DC & LDT DeVeny & 10 & 2020-01-20 &  &  \\
1141846540594232064 & WDJ073615.22+794043.26 & 18.55 & DC & LDT DeVeny & 23 & 2021-04-14 &  &  \\
1141846540594232192 & WDJ073615.19+794035.14 & 15.79 & DA & LDT DeVeny & 180 & 2021-04-14 & $20350\pm240$ & $8.01\pm0.03$ \\
1165926910393567872 & WDJ153401.77+101012.04 & 17.75 & DA & LDT DeVeny & 44 & 2020-08-11 & $7070\pm80$ & $7.75\pm0.14$ \\
1165926910393568384 & WDJ153402.44+101024.24 & 18.09 & DA & LDT DeVeny & 32 & 2020-08-11 & $6410\pm110$ & $7.49\pm0.30$ \\
1246692121126938240 & WDJ135704.68+192133.34 & 16.85 & DA & LDT DeVeny & 98 & 2022-05-05 & $13820\pm280$ & $7.97\pm0.03$ \\
1282448170543051520 & WDJ144528.12+292124.29 & 17.36 & DA & LDT DeVeny & 36 & 2019-08-25 &  &  \\
1282448170543051648 & WDJ144528.47+292132.07 & 14.51 & DA & LDT DeVeny & 245 & 2019-08-25 & $13040\pm160$ & $8.07\pm0.03$ \\
1643897255075822976 & WDJ151950.19+641540.48 & 17.79 & DA & LDT DeVeny & 43 & 2021-04-14 & $6160\pm120$ & $7.31\pm0.31$ \\
1643897255075823104 & WDJ151949.71+641534.74 & 18.42 & DA & LDT DeVeny & 25 & 2021-04-14 &  &  \\
1680759791146338048 & WDJ122741.06+661224.33 & 18.08 & DA & LDT DeVeny & 34 & 2020-01-19 & $13890\pm520$ & $8.18\pm0.07$ \\
1680759859865814656 & WDJ122739.17+661224.39 & 17.97 & DAH & LDT DeVeny & 35 & 2020-01-19 &  &  \\
1682410707854900352 & WDJ123030.22+675259.56 & 19.01 & DC & LDT DeVeny & 12 & 2020-01-19 &  &  \\
1682410776574377600 & WDJ123029.22+675310.96 & 19.36 & DC & LDT DeVeny & 7 & 2020-01-19 &  &  \\
1829120747684237312 & WDJ201836.79+212924.44 & 17.31 & DA & LDT DeVeny & 30 & 2019-08-25 &  &  \\
1829120747684239104 & WDJ201836.72+212931.06 & 16.52 & DC & LDT DeVeny & 54 & 2019-08-25 &  &  \\
1840810854079992832 & WDJ211531.38+253452.74 & 17.60 & DA & LDT DeVeny & 41 & 2019-10-02 & $8800\pm110$ & $8.13\pm0.07$ \\
1840810858375964544 & WDJ211531.58+253457.52 & 17.25 & DC & LDT DeVeny & 52 & 2019-10-02 &  &  \\
1874954641491354624 & WDJ222301.60+220131.69 & 15.77 & DA & LDT DeVeny & 85 & 2020-08-11 & $19680\pm240$ & $8.45\pm0.03$ \\
1874954645786146304 & WDJ222301.71+220125.28 & 16.11 & DA & LDT DeVeny & 70 & 2020-08-11 & $14400\pm440$ & $8.27\pm0.04$ \\
1911420636118031744 & WDJ231941.38+342614.52 & 17.34 & DA & LDT DeVeny & 54 & 2019-08-25 & $16960\pm200$ & $8.13\pm0.03$ \\
1911420636119326976 & WDJ231941.48+342609.63 & 17.64 & DA & LDT DeVeny & 46 & 2019-08-25 & $14430\pm470$ & $8.17\pm0.04$ \\
1932585268583421312 & WDJ224133.59+412052.51 & 18.05 & DA & LDT DeVeny & 41 & 2019-10-02 & $9750\pm120$ & $8.11\pm0.06$ \\
1932585302943159808 & WDJ224134.34+412054.34 & 18.58 & DA & LDT DeVeny & 30 & 2019-10-02 & $11580\pm210$ & $8.67\pm0.07$ \\
2080526555267049984 & WDJ194530.33+465006.84 & 17.17 & DC & LDT DeVeny & 38 & 2019-10-02 &  &  \\
2080526555267050496 & WDJ194530.35+465015.52 & 16.97 & DA & LDT DeVeny & 50 & 2019-10-02 &  &  \\
2142145179748734976 & WDJ193124.32+570413.55 & 16.27 & DA & LDT DeVeny & 68 & 2019-08-25 & $16070\pm290$ & $8.59\pm0.03$ \\
2142145385907166208 & WDJ193124.43+570419.66 & 14.98 & DA & LDT DeVeny & 130 & 2019-08-25 & $24590\pm300$ & $8.24\pm0.03$ \\
2150591795574568192 & WDJ182519.30+553849.16 & 18.74 & DC & LDT DeVeny & 13 & 2021-06-12 &  &  \\
2150591799870095872 & WDJ182518.67+553847.70 & 18.49 & DC & LDT DeVeny & 16 & 2021-06-12 &  &  \\
2214973561500636544 & WDJ232117.26+692553.73 & 17.45 & DC & LDT DeVeny & 34 & 2020-09-22 &  &  \\
2214973771957744128 & WDJ232117.29+692622.44 & 14.73 & DA & LDT DeVeny & 200 & 2020-09-22 & $20840\pm250$ & $8.03\pm0.01$ \\
2238516205691628544 & WDJ193744.39+590010.43 & 16.53 & DA & LDT DeVeny & 53 & 2019-08-25 & $9230\pm110$ & $8.12\pm0.05$ \\
2238516240051369984 & WDJ193745.94+590032.16 & 15.36 & DA & LDT DeVeny & 110 & 2019-08-25 & $12770\pm150$ & $8.01\pm0.03$ \\
2310995767378550016 & WDJ235343.05-362050.73 & 16.95 & DC & SOAR Goodman & 61 & 2018-08-08 &  &  \\
2310996145335674240 & WDJ235344.03-362042.50 & 17.60 & DA & SOAR Goodman & 35 & 2018-08-08 & $6810\pm80$ & $8.27\pm0.13$ \\
2329764705742441216 & WDJ232322.50-292921.55 & 19.14 & DA & SOAR Goodman & 20 & 2018-09-10 & $6740\pm130$ & $8.38\pm0.23$ \\
2329764710037398528 & WDJ232322.85-292916.58 & 18.47 & DA & SOAR Goodman & 36 & 2018-09-10 & $9190\pm110$ & $8.61\pm0.04$ \\
2359214952893455872 & WDJ010131.48-162909.12 & 18.44 & DA & SOAR Goodman & 49 & 2019-09-29 & $13320\pm220$ & $8.48\pm0.03$ \\
2359214952893456000 & WDJ010131.06-162908.14 & 18.68 & DA & SOAR Goodman & 33 & 2019-09-29 & $10100\pm120$ & $8.12\pm0.04$ \\
2454507598448711296 & WDJ012059.34-162248.18 & 18.80 & DA & SOAR Goodman & 25 & 2019-09-29 & $6280\pm190$ & $8.54\pm0.33$ \\
2454507602744210432 & WDJ012058.81-162246.57 & 15.84 & DA & SOAR Goodman & 206 & 2019-09-29 & $14930\pm180$ & $8.10\pm0.01$ \\
2543653978300417024 & WDJ002925.63+001552.72 & 18.47 & DA & SOAR Goodman & 29 & 2018-09-08 & $10190\pm120$ & $8.16\pm0.06$ \\
2543654008364462848 & WDJ002925.30+001559.78 & 19.54 & DC & SOAR Goodman & 13 & 2018-09-08 &  &  \\
2564851394251630848 & WDJ013213.18+052633.00 & 16.30 & DA & LDT DeVeny & 72 & 2020-08-11 & $14570\pm200$ & $7.99\pm0.03$ \\
2564945432560219008 & WDJ013222.84+052925.09 & 18.21 & DAZ & LDT DeVeny & 33 & 2020-08-11 &  &  \\
2607475916712247168 & WDJ230327.91-075504.82 & 18.26 & DA & LDT DeVeny & 41 & 2020-08-11 & $13690\pm340$ & $8.12\pm0.05$ \\
2607475916712248064 & WDJ230328.14-075457.97 & 18.26 & DA & LDT DeVeny & 41 & 2020-08-11 & $13140\pm270$ & $8.14\pm0.06$ \\
2616210918121365760 & WDJ222236.56-082806.01 & 17.19 & DA & SOAR Goodman & 29 & 2018-08-31 & $11840\pm200$ & $8.24\pm0.06$ \\
2616210922414728960 & WDJ222236.30-082807.97 & 16.54 & DA & SOAR Goodman & 41 & 2018-08-31 & $15640\pm190$ & $8.07\pm0.03$ \\
2634608741244388096 & WDJ230419.49-070149.84 & 19.03 & DC & LDT DeVeny & 16 & 2020-11-17 &  &  \\
2634608741244966016 & WDJ230418.93-070124.51 & 17.07 & DC & LDT DeVeny & 120 & 2020-11-17 &  &  \\
2664365301168573824 & WDJ231650.13+064123.48 & 18.61 & DA & SOAR Goodman & 18 & 2018-09-10 & $7970\pm100$ & $8.37\pm0.13$ \\
2664365374183501312 & WDJ231650.24+064128.42 & 16.08 & DA & SOAR Goodman & 100 & 2018-09-10 & $18150\pm220$ & $8.04\pm0.01$ \\
2676567307551465088 & WDJ215839.14-023916.44 & 17.09 & DC & LDT DeVeny & 9 & 2020-09-22 &  &  \\
2679752588442464768 & WDJ215602.44-013829.07 & 17.63 & DA & LDT DeVeny & 42 & 2019-08-24 & $9140\pm110$ & $8.06\pm0.07$ \\
2679765511998843520 & WDJ215545.48-013455.69 & 17.73 & DA & LDT DeVeny & 36 & 2019-08-24 & $8690\pm100$ & $8.24\pm0.08$ \\
2689069472018944256 & WDJ210824.62-004133.72 & 19.65 & DA & SOAR Goodman & 10 & 2018-09-08 & $7360\pm180$ & $8.14\pm0.28$ \\
2689069472018944896 & WDJ210823.65-004130.83 & 18.29 & DA & SOAR Goodman & 37 & 2018-09-08 & $10730\pm130$ & $7.99\pm0.03$ \\
2731501687319341568 & WDJ224231.15+125004.99 & 16.35 & DA & LDT DeVeny & 85 & 2020-09-22 & $16240\pm190$ & $8.10\pm0.03$ \\
2731502443233585280 & WDJ224230.34+125002.42 & 16.61 & DA & LDT DeVeny & 74 & 2020-09-22 & $14380\pm360$ & $8.19\pm0.03$ \\
2747684849212364928 & WDJ002335.53+064325.89 & 18.45 & DA & LDT DeVeny & 16 & 2020-01-20 & $7620\pm150$ & $7.48\pm0.29$ \\
2747684853507815552 & WDJ002336.21+064320.90 & 17.42 & DA & LDT DeVeny & 45 & 2020-01-20 & $10410\pm120$ & $8.07\pm0.06$ \\
2790494815860044544 & WDJ010456.47+211958.87 & 17.70 & DA & LDT DeVeny & 26 & 2019-08-24 &  &  \\
2790494850219788160 & WDJ010457.96+212017.54 & 17.80 & DC & LDT DeVeny & 21 & 2019-08-24 &  &  \\
282878984339974528 & WDJ054015.76+610112.41 & 17.64 & DA & LDT DeVeny & 34 & 2021-03-23 & $12140\pm240$ & $8.22\pm0.06$ \\
282878984339976320 & WDJ054015.70+610123.34 & 18.32 & DA & LDT DeVeny & 19 & 2021-03-23 & $11300\pm330$ & $8.40\pm0.12$ \\
2841680934336158976 & WDJ232755.40+263824.46 & 17.58 & DA & LDT DeVeny & 59 & 2020-11-17 & $6620\pm80$ & $8.14\pm0.15$ \\
2841680934336159104 & WDJ232755.27+263821.32 & 18.30 & DC & LDT DeVeny & 22 & 2020-11-17 &  &  \\
2843077206728009472 & WDJ230249.40+243028.69 & 17.95 & DQ & LDT DeVeny & 31 & 2020-11-17 &  &  \\
2843077211025217408 & WDJ230250.40+243014.16 & 18.18 & DA & LDT DeVeny & 25 & 2020-11-17 & $6310\pm140$ & $7.54\pm0.39$ \\
31047184711544832 & WDJ030953.89+150511.53 & 18.43 & DA & LDT DeVeny & 30 & 2020-01-20 & $6560\pm180$ & $8.32\pm0.38$ \\
31047257726638592 & WDJ030953.95+150521.83 & 15.21 & DA & LDT DeVeny & 260 & 2020-01-20 & $22110\pm270$ & $8.07\pm0.01$ \\
3284951791556594816 & WDJ041840.44+055115.05 & 18.55 & DB & LDT DeVeny & 28 & 2020-01-20 &  &  \\
3284951791557832064 & WDJ041840.69+055113.90 & 19.23 & DA & LDT DeVeny & 17 & 2020-01-20 & $11770\pm400$ & $8.26\pm0.12$ \\
3330855616042214912 & WDJ062210.56+110747.91 & 17.15 & DA & SOAR Goodman & 35 & 2018-12-30 & $13930\pm440$ & $8.63\pm0.03$ \\
3330855616042218112 & WDJ062208.96+110747.74 & 17.25 & DA & SOAR Goodman & 24 & 2018-12-30 & $16400\pm360$ & $8.49\pm0.05$ \\
3334379344646313088 & WDJ053724.80+073112.00 & 17.41 & DA & SOAR Goodman & 36 & 2018-12-30 & $6590\pm80$ & $8.46\pm0.14$ \\
3334380100562482432 & WDJ053720.93+073232.37 & 16.96 & DC & LDT DeVeny & 42 & 2020-01-19 &  &  \\
3404213863611804672 & WDJ053834.31+223918.74 & 17.12 & DC & LDT DeVeny & 65 & 2020-11-17 &  &  \\
3404213863614488192 & WDJ053834.52+223919.57 & 17.28 & DA & LDT DeVeny & 62 & 2020-11-17 &  &  \\
3562650151984291840 & WDJ110021.77-160014.95 & 17.91 & DC & LDT DeVeny & 20 & 2021-04-14 &  &  \\
3562650151984292224 & WDJ110022.76-160011.79 & 17.62 & DA & LDT DeVeny & 28 & 2021-04-14 & $6270\pm260$ & $8.17\pm0.59$ \\
3603920492731480960 & WDJ133651.76-162020.10 & 17.60 & DA & LDT DeVeny & 44 & 2021-04-14 & $25010\pm390$ & $8.19\pm0.06$ \\
3603920595812980864 & WDJ133652.21-162010.94 & 18.48 & DA & LDT DeVeny & 23 & 2021-04-14 & $11890\pm440$ & $8.70\pm0.10$ \\
377231139432432384 & WDJ005657.17+441018.64 & 15.94 & DA & LDT DeVeny & 91 & 2020-08-11 & $13260\pm160$ & $8.05\pm0.03$ \\
377231345590861824 & WDJ005656.68+441029.62 & 16.33 & DA & LDT DeVeny & 71 & 2020-08-11 & $11130\pm130$ & $8.05\pm0.03$ \\
3823125109439975936 & WDJ095618.10-035211.48 & 18.12 & DA & LDT DeVeny & 30 & 2020-01-19 & $9880\pm120$ & $8.32\pm0.09$ \\
3823125109439978624 & WDJ095617.48-035153.93 & 17.89 & DC & LDT DeVeny & 34 & 2020-01-19 &  &  \\
3875651975353757440 & WDJ101501.75+080611.24 & 17.49 & DC & SOAR Goodman & 26 & 2019-01-07 &  &  \\
3875652014008894720 & WDJ101502.61+080635.70 & 15.92 & DA & SOAR Goodman & 107 & 2019-01-07 & $6730\pm80$ & $8.22\pm0.04$ \\
3883615188318129536 & WDJ101955.91+121631.53 & 15.76 & DA & SOAR Goodman & 67 & 2019-01-07 & $22790\pm270$ & $8.05\pm0.03$ \\
3886617649631192320 & WDJ101954.61+121718.03 & 16.87 & DA & SOAR Goodman & 34 & 2019-01-07 & $13020\pm170$ & $8.22\pm0.04$ \\
3905533368502667520 & WDJ121510.11+094836.37 & 17.77 & DA & LDT DeVeny & 34 & 2020-01-19 & $8510\pm100$ & $7.95\pm0.10$ \\
3905534918987016576 & WDJ121509.68+094847.14 & 17.92 & DA & LDT DeVeny & 34 & 2020-01-19 & $9960\pm120$ & $8.43\pm0.07$ \\
4190500054845023488 & WDJ195333.12-101954.85 & 17.26 & DA & SOAR Goodman & 38 & 2018-10-27 & $13700\pm310$ & $8.03\pm0.04$ \\
428567401760258560 & WDJ003230.15+600616.03 & 18.18 & DC & LDT DeVeny & 19 & 2020-01-19 &  &  \\
428567406061520896 & WDJ003230.51+600620.08 & 18.88 & DA & LDT DeVeny & 10 & 2020-01-19 & $8160\pm220$ & $7.34\pm0.38$ \\
4297600901227590272 & WDJ194738.02+062510.15 & 18.87 & DA & SOAR Goodman & 17 & 2018-09-08 & $7990\pm110$ & $8.22\pm0.15$ \\
4297600905539762432 & WDJ194737.61+062508.48 & 18.63 & DC & SOAR Goodman & 21 & 2018-09-08 &  & \\
4311973098757316608 & WDJ185400.72+104854.69 & 18.17 & DA & LDT DeVeny & 43 & 2021-04-14 &  &  \\
4311973103092890752 & WDJ185400.68+104851.56 & 18.01 & DA & LDT DeVeny & 45 & 2021-04-14 & $10210\pm120$ & $8.21\pm0.07$ \\
4471738941493617408 & WDJ181332.20+060412.61 & 17.26 & DZ & LDT DeVeny & 50 & 2019-08-24 &  &  \\
4471739010213094272 & WDJ181330.47+060412.51 & 17.69 & DA & LDT DeVeny & 34 & 2019-08-24 & $6220\pm130$ & $7.70\pm0.36$ \\
4500927474812609536 & WDJ175137.69+151250.96 & 18.53 & DA & SOAR Goodman & 20 & 2018-09-10 & $7230\pm120$ & $8.51\pm0.19$ \\
4500927504877462016 & WDJ175137.51+151304.83 & 18.87 & DC & SOAR Goodman & 10 & 2018-09-10 &  &  \\
4848810205061914112 & WDJ034011.79-420701.07 & 19.39 & DC & SOAR Goodman & 23 & 2019-01-09 &  &  \\
4848810205061914496 & WDJ034011.52-420702.38 & 19.32 & DA & SOAR Goodman & 25 & 2019-01-09 & $6780\pm140$ & $8.77\pm0.21$ \\
4855870439806768384 & WDJ035012.19-383051.98 & 19.58 & DC & SOAR Goodman & 20 & 2018-12-31 &  &  \\
4855870439806768512 & WDJ035012.13-383057.28 & 19.05 & DA & SOAR Goodman & 27 & 2018-12-31 &  &  \\
4940155500795290624 & WDJ022440.28-461133.91 & 18.09 & DAH & SOAR Goodman & 32 & 2019-07-04 &  &  \\
4940155500795290880 & WDJ022440.65-461140.65 & 17.41 & DA & SOAR Goodman & 52 & 2019-07-04 & $14010\pm290$ & $8.12\pm0.03$ \\
5062948237833763840 & WDJ024051.59-324837.30 & 17.52 & DA & SOAR Goodman & 42 & 2018-08-08 & $8380\pm100$ & $8.50\pm0.05$ \\
5062948340912201088 & WDJ024051.94-324814.09 & 17.59 & DAH & SOAR Goodman & 39 & 2018-08-08 &  &  \\
5131731035268115584 & WDJ022556.26-175608.23 & 18.16 & DA & SOAR Goodman & 21 & 2019-07-05 & $6230\pm160$ & $8.19\pm0.32$ \\
5131731039563630720 & WDJ022556.45-175614.79 & 17.36 & DZ & SOAR Goodman & 44 & 2019-07-05 &  &  \\
5146493662498666624 & WDJ022031.42-153248.98 & 18.45 & DA & SOAR Goodman & 32 & 2018-09-03 & $6620\pm80$ & $8.01\pm0.13$ \\
5146493662498666752 & WDJ022031.01-153248.08 & 18.17 & DC & SOAR Goodman & 42 & 2018-09-03 &  &  \\
5281331583776780416 & WDJ071256.85-664407.33 & 19.24 & DA & SOAR Goodman & 24 & 2018-12-31 & $8560\pm100$ & $8.48\pm0.07$ \\
5281331588075295232 & WDJ071254.12-664409.71 & 19.20 & DA & SOAR Goodman & 25 & 2018-12-31 & $8830\pm110$ & $8.47\pm0.07$ \\
5427657584803951232 & WDJ092016.10-412709.51 & 17.58 & DA & SOAR Goodman & 33 & 2018-12-31 & $8370\pm100$ & $8.06\pm0.06$ \\
5427657619159602944 & WDJ092016.04-412702.01 & 17.59 & DA & SOAR Goodman & 32 & 2018-12-31 & $8320\pm100$ & $7.99\pm0.06$ \\
5490676707920375936 & WDJ072441.18-541935.42 & 19.45 & DA & SOAR Goodman & 19 & 2019-01-09 & $9640\pm120$ & $8.63\pm0.09$ \\
5490676742283052800 & WDJ072441.73-541935.33 & 19.20 & DA & SOAR Goodman & 23 & 2019-01-09 & $9060\pm110$ & $8.41\pm0.07$ \\
5551794951535526144 & WDJ064716.66-470744.24 & 18.98 & DAH & SOAR Goodman & 29 & 2018-12-31 &  &  \\
5551794951535528448 & WDJ064715.56-470733.36 & 18.90 & DAH & SOAR Goodman & 30 & 2018-12-31 &  &  \\
5564028981196462336 & WDJ065335.34-395533.29 & 15.35 & DA & SOAR Goodman & 56 & 2019-01-07 & $7220\pm90$ & $8.26\pm0.05$ \\
5564029702750970112 & WDJ065330.21-395429.12 & 15.92 & DA & SOAR Goodman & 38 & 2019-01-07 & $6400\pm110$ & $8.56\pm0.18$ \\
5780478664144123264 & WDJ162541.00-772112.71 & 18.13 & DA & SOAR Goodman & 44 & 2019-05-05 & $9350\pm110$ & $8.13\pm0.04$ \\
5780478664144129152 & WDJ162540.54-772104.20 & 18.55 & DA & SOAR Goodman & 31 & 2019-05-05 & $8510\pm100$ & $8.24\pm0.06$ \\
6002657567182879104 & WDJ152915.54-405524.74 & 18.52 & DC & SOAR Goodman & 28 & 2019-09-28 &  &  \\
6002657567182879616 & WDJ152915.14-405526.00 & 18.08 & DA & SOAR Goodman & 43 & 2019-09-28 & $7980\pm100$ & $8.51\pm0.05$ \\
6093462311215003776 & WDJ135006.24-502534.22 & 18.07 & DAH & SOAR Goodman & 34 & 2019-07-04 &  &  \\
6093462315507278208 & WDJ135006.04-502540.07 & 18.57 & DA & SOAR Goodman & 20 & 2019-07-04 & $6320\pm160$ & $8.23\pm0.30$ \\
6458700145311223936 & WDJ213406.66-575113.38 & 17.94 & DA & SOAR Goodman & 26 & 2018-09-09 & $7090\pm90$ & $7.97\pm0.13$ \\
6458700149606803456 & WDJ213406.86-575117.46 & 17.63 & DC & SOAR Goodman & 38 & 2018-09-09 &  &  \\
6590545195937635456 & WDJ213136.44-345905.01 & 19.46 & DA & SOAR Goodman & 12 & 2018-09-09 & $7180\pm190$ & $8.64\pm0.28$ \\
6590545195937636992 & WDJ213136.47-345858.34 & 18.41 & DA & SOAR Goodman & 22 & 2018-09-09 & $7950\pm100$ & $8.08\pm0.09$ \\
6642323156995855232 & WDJ194406.24-534220.66 & 17.91 & DA & SOAR Goodman & 40 & 2019-05-05 &  &  \\
6642323156995855360 & WDJ194406.22-534225.12 & 17.63 & DA & SOAR Goodman & 45 & 2019-05-05 & $13220\pm260$ & $8.55\pm0.03$ \\
6898489884295407488 & WDJ211507.40-074151.48 & 17.35 & DA & SOAR Goodman & 25 & 2018-10-27 & $8140\pm100$ & $8.21\pm0.08$ \\
6898489884295412352 & WDJ211507.42-074134.46 & 16.82 & DA & SOAR Goodman & 40 & 2018-10-27 & $10580\pm130$ & $8.41\pm0.03$ \\
733402121474544896 & WDJ105707.88+301625.43 & 17.79 & DC & LDT DeVeny & 23 & 2021-04-14 &  &  \\
733402636870625152 & WDJ105718.15+301823.17 & 18.56 & DC & LDT DeVeny & 13 & 2021-04-14 &  &  \\
867739936760402560 & WDJ074723.48+243823.89 & 17.94 & DA & LDT DeVeny & 49 & 2020-11-17 &  &  \\
867740005479885184 & WDJ074721.53+243847.89 & 18.57 & DC & LDT DeVeny & 24 & 2020-11-17 &  &  \\
961949666442453248 & WDJ061725.77+433850.02 & 19.34 & DC & LDT DeVeny & 11 & 2020-01-19 &  &  \\
961949735161929728 & WDJ061723.93+433846.21 & 19.19 & DA & LDT DeVeny & 13 & 2020-01-19 & $8240\pm200$ & $8.62\pm0.27$
\enddata
\tablecomments{The S/N ratios tabulated here are calculated by taking the mean spectrum flux over spectrum error in a 200~\AA window around 4600~\AA}
\end{deluxetable*}

\pagebreak

\begin{figure}[h]
    \centering
    \includegraphics[width=0.9\columnwidth]{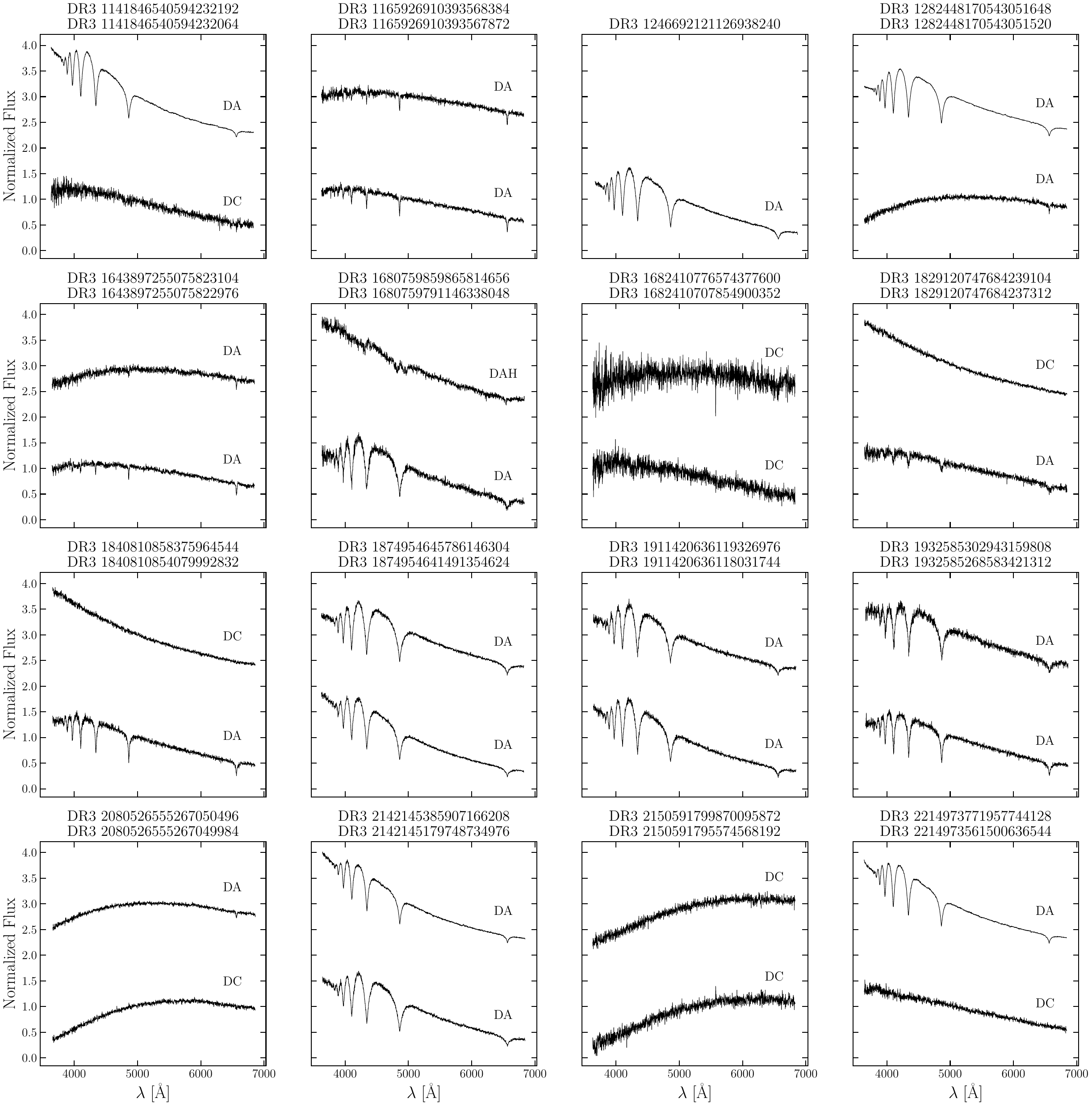}
    \caption{All newly observed white dwarf optical spectra with the DeVeny Spectrograph on the LDT.}
    \label{fig:LDT_spectra}    
\end{figure}

\pagebreak

\begin{figure}[h]
    \centering
    \includegraphics[width=0.9\columnwidth]{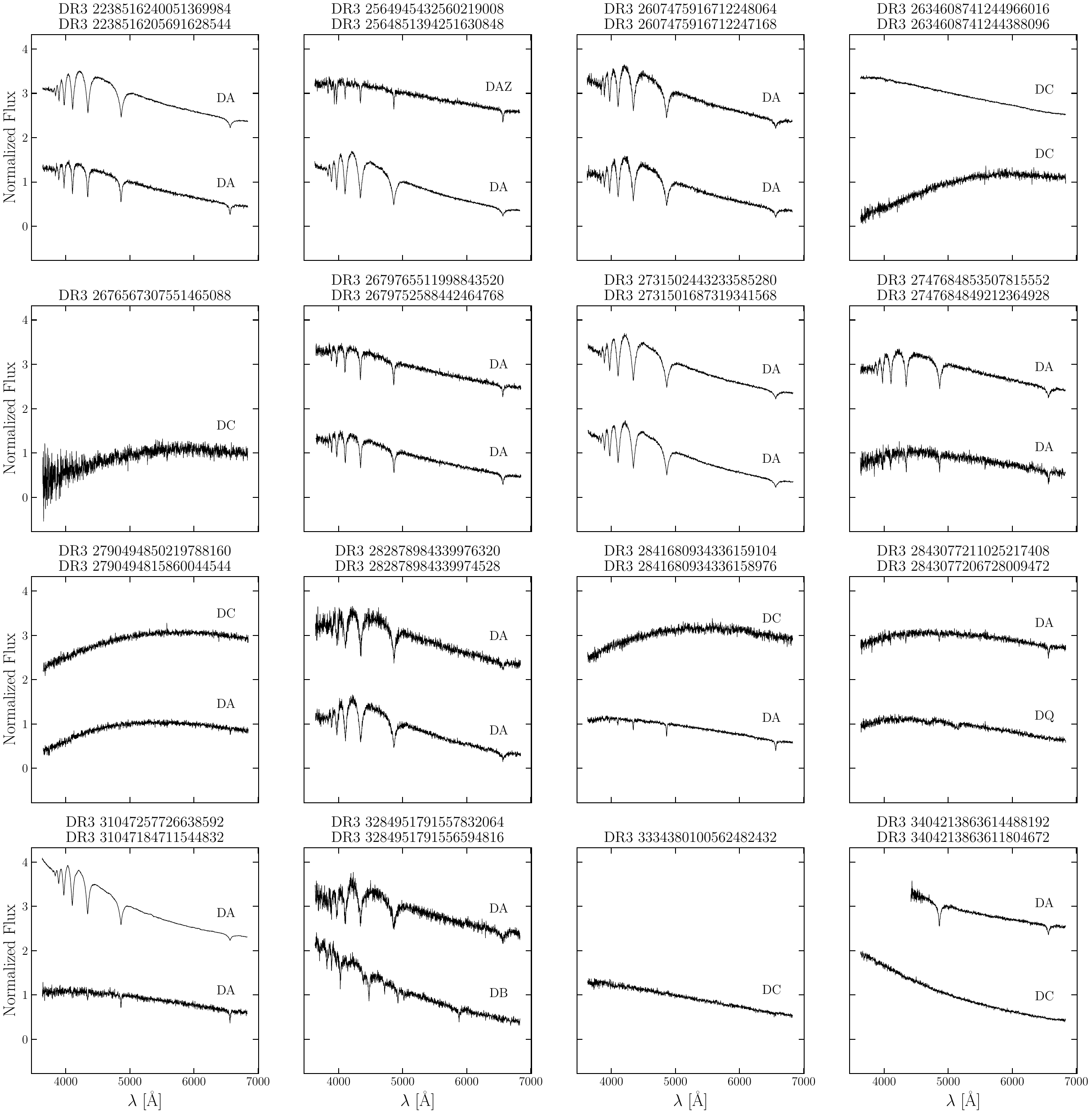}
    \caption{Same as Figure~\ref{fig:LDT_spectra}, continued.}
    \label{fig:LDT_spectra2}    
\end{figure}

\pagebreak

\begin{figure}[h]
    \centering
    \includegraphics[width=0.9\columnwidth]{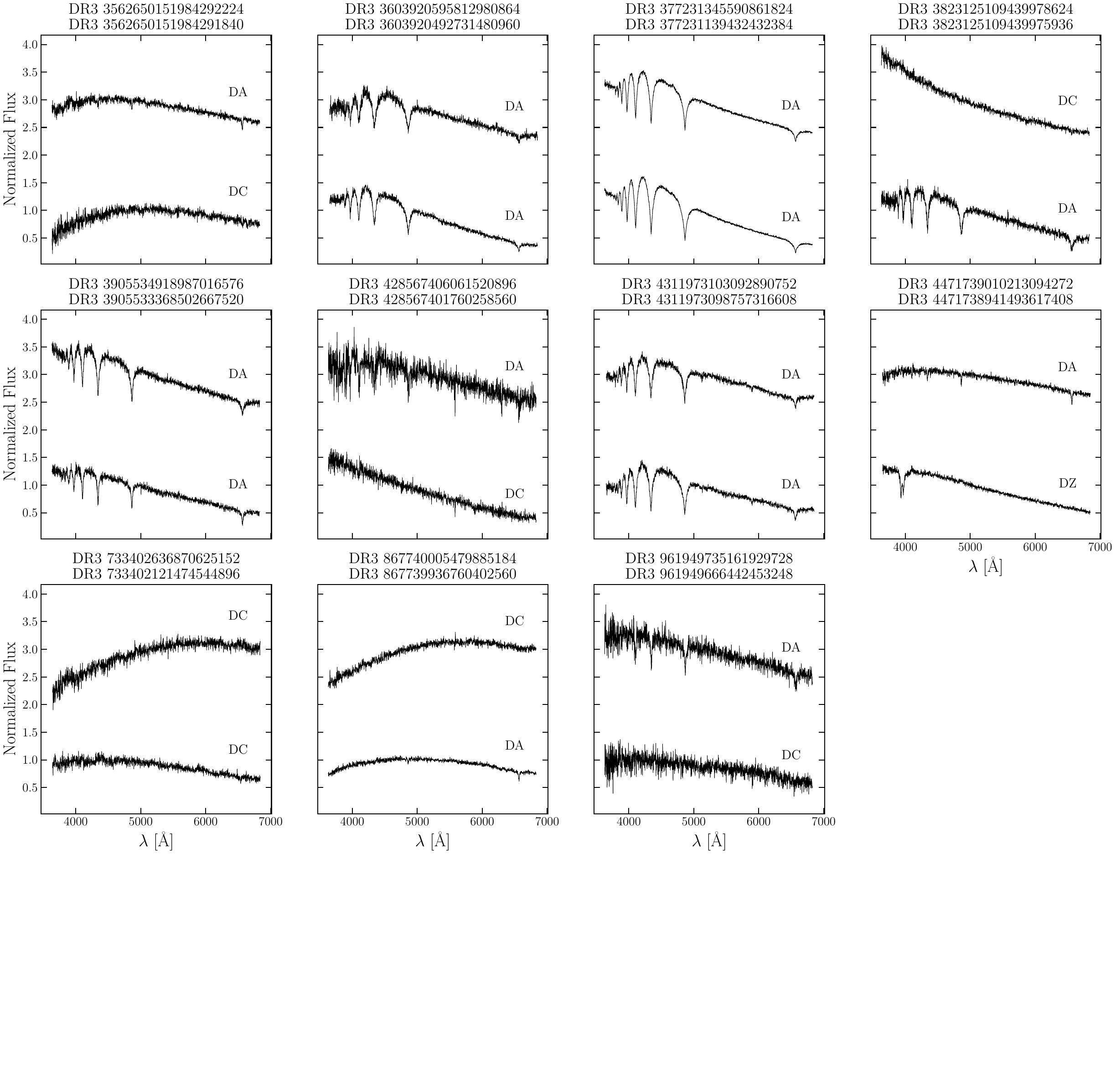}
    \caption{Same as Figure~\ref{fig:LDT_spectra}, continued.}
    \label{fig:LDT_spectra3}    
\end{figure}

\pagebreak

\begin{figure}[h]
    \centering
    \includegraphics[width=0.9\columnwidth]{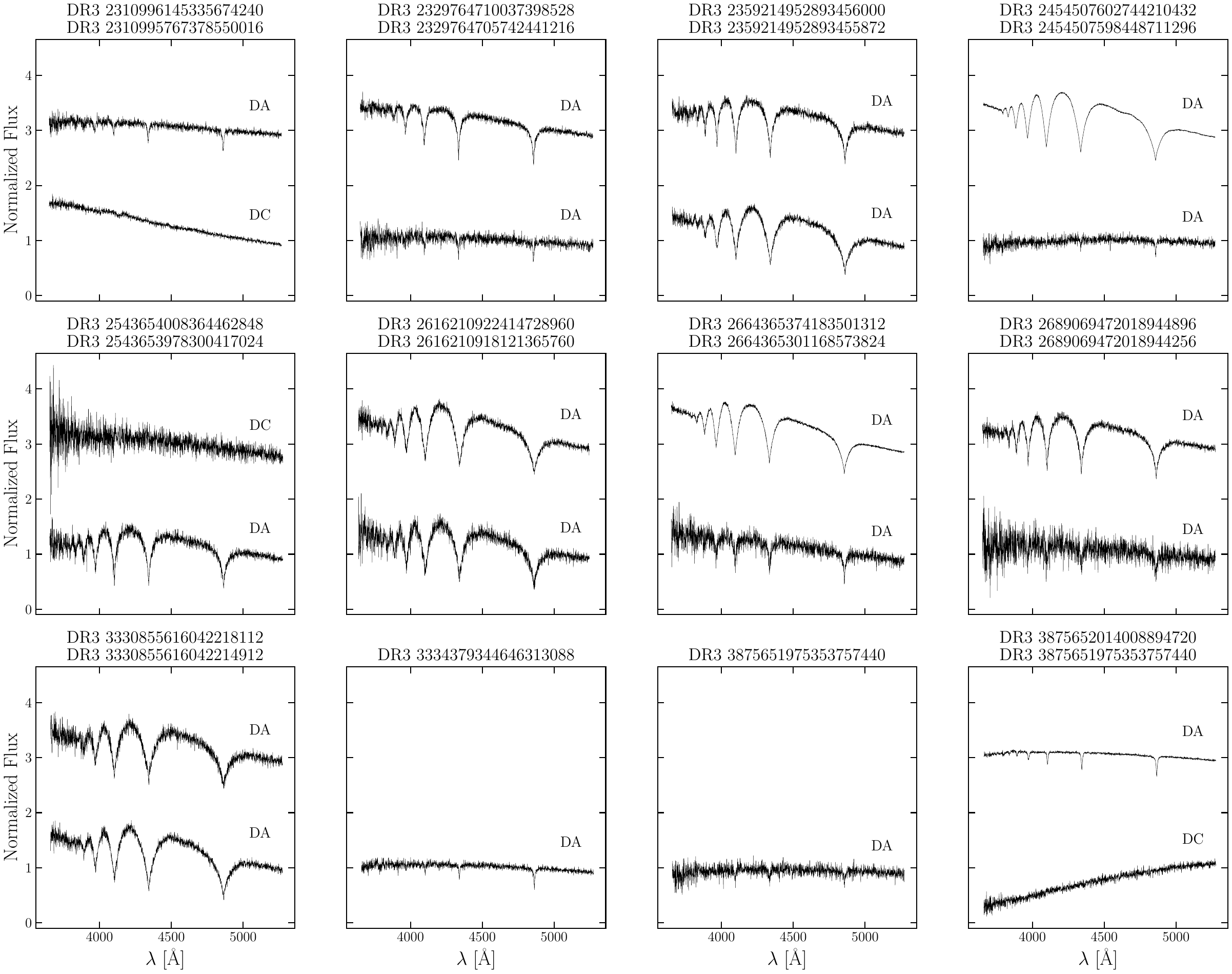}
    \caption{All newly observed white dwarf optical spectra with the Goodman Spectrograph on the SOAR telescope.}
    \label{fig:SOAR_spectra}    
\end{figure}

\pagebreak

\begin{figure}[h]
    \centering
    \includegraphics[width=0.9\columnwidth]{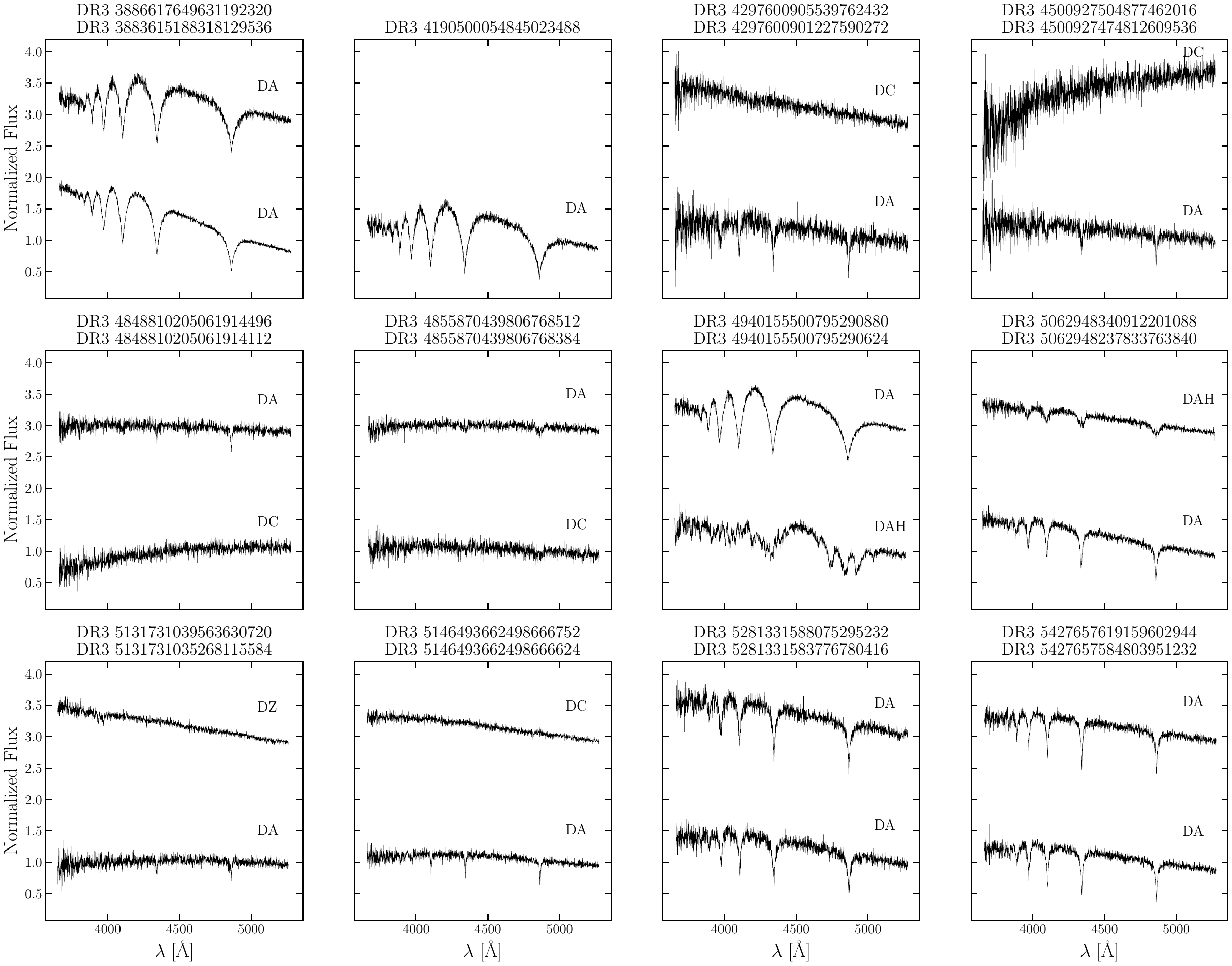}
    \caption{Same as Figure~\ref{fig:SOAR_spectra}, continued.}
    \label{fig:SOAR_spectra2}    
\end{figure}

\pagebreak

\begin{figure}[h]
    \centering
    \includegraphics[width=0.9\columnwidth]{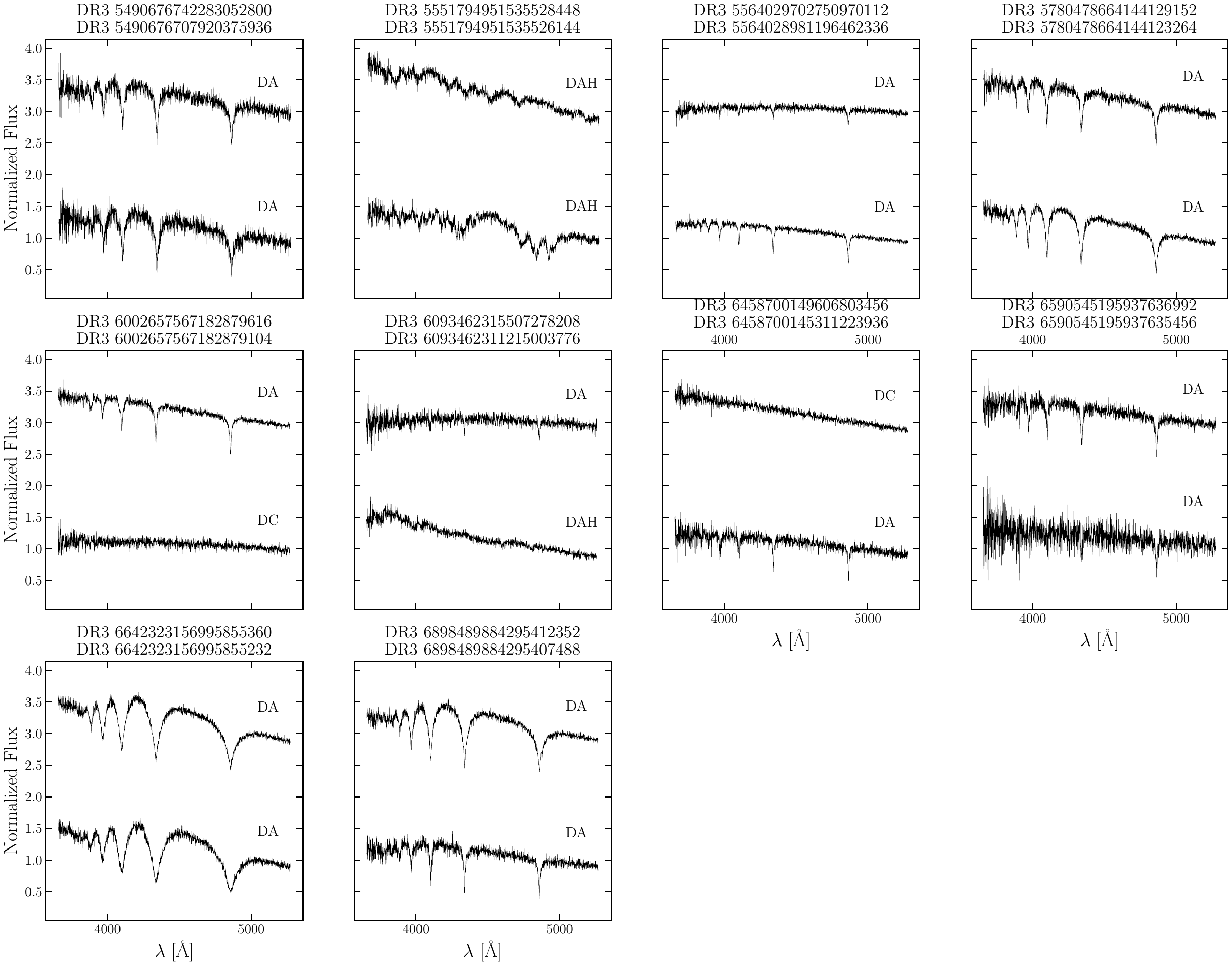}
    \caption{Same as Figure~\ref{fig:SOAR_spectra}, continued.}
    \label{fig:SOAR_spectra3}    
\end{figure}

\pagebreak

\begin{figure}[h]
    \centering
    \includegraphics[width=0.9\columnwidth]{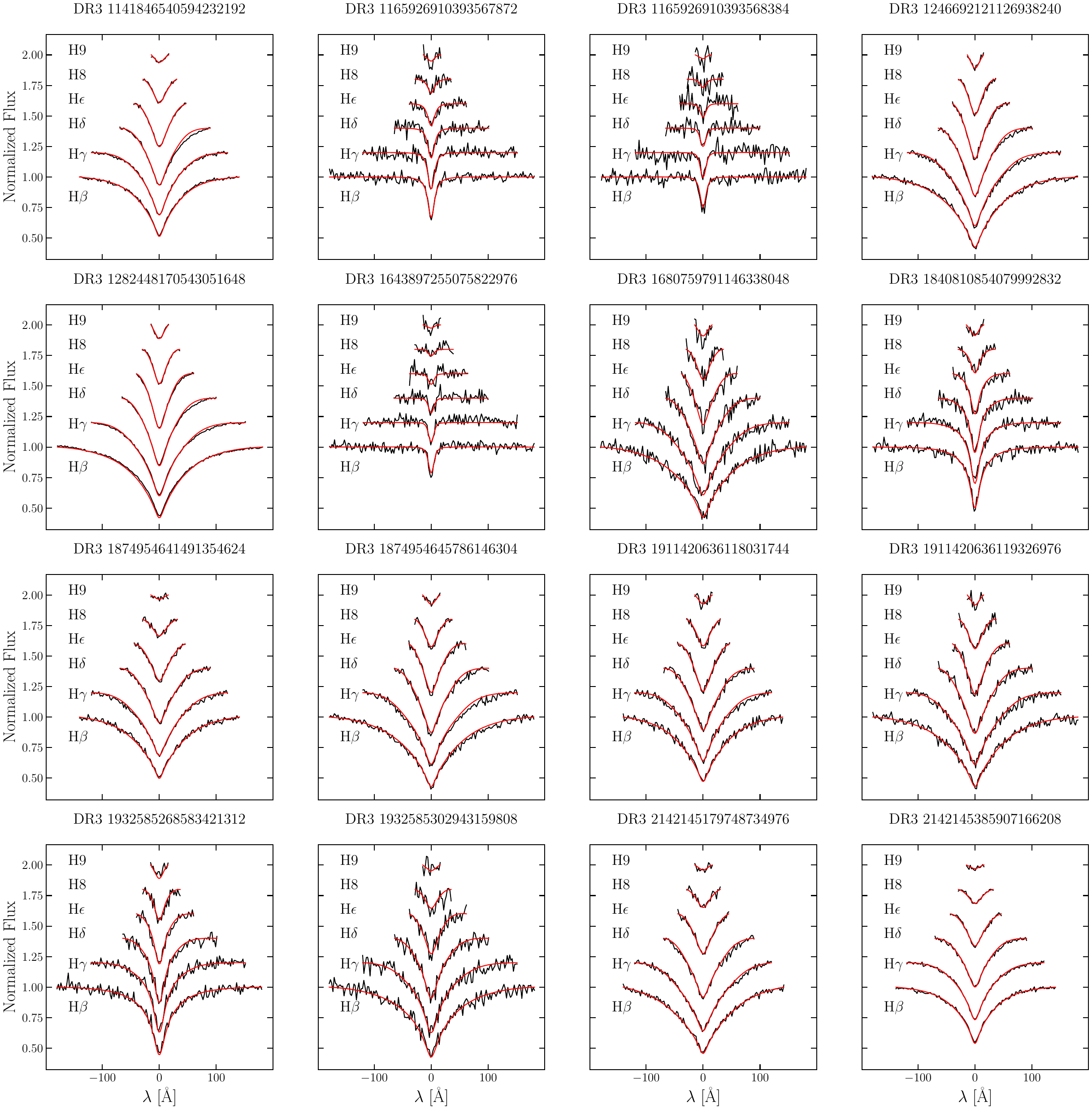}
    \caption{Spectral fits to the new DAs observed in this work (see Section~\ref{sec:specfits}), with observations in black and best-fit models in red.}
    \label{fig:DA_fits}    
\end{figure}

\pagebreak

\begin{figure}[h]
    \centering
    \includegraphics[width=0.9\columnwidth]{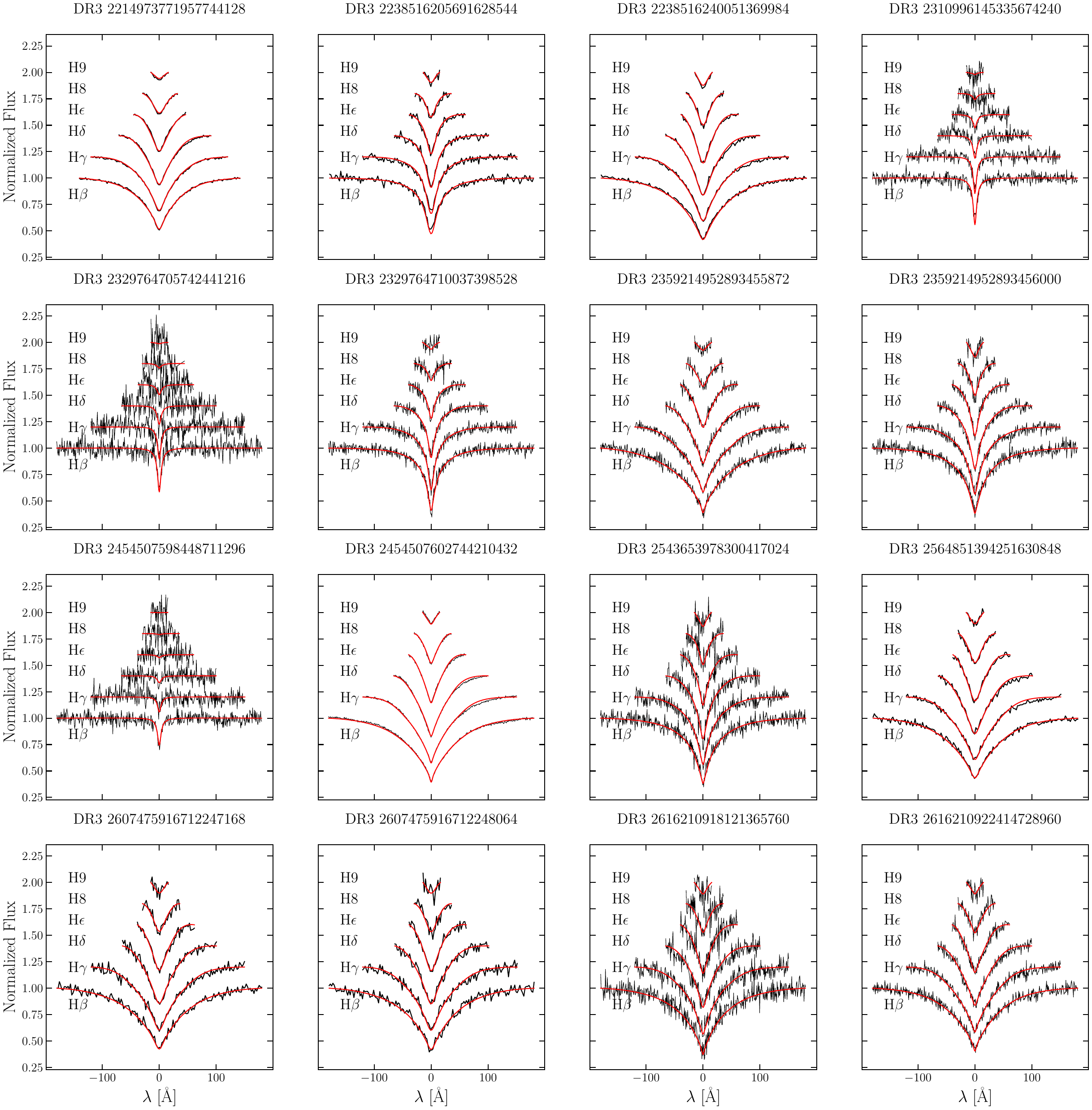}
    \caption{Same as Figure~\ref{fig:DA_fits}, continued.}
    \label{fig:DA_fits2}    
\end{figure}

\pagebreak

\begin{figure}[h]
    \centering
    \includegraphics[width=0.9\columnwidth]{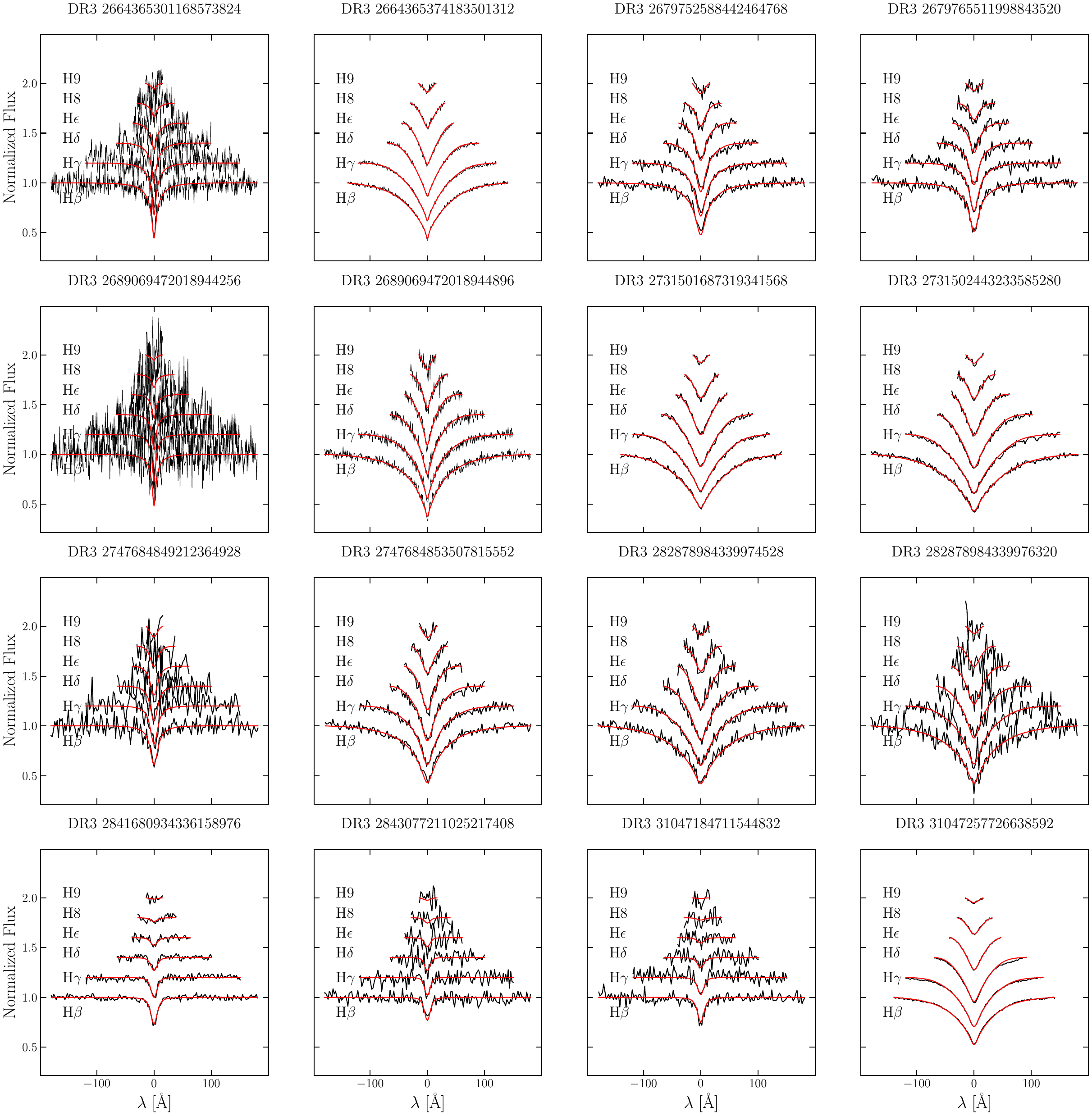}
    \caption{Same as Figure~\ref{fig:DA_fits}, continued.}
    \label{fig:DA_fits3}    
\end{figure}

\pagebreak

\begin{figure}[h]
    \centering
    \includegraphics[width=0.9\columnwidth]{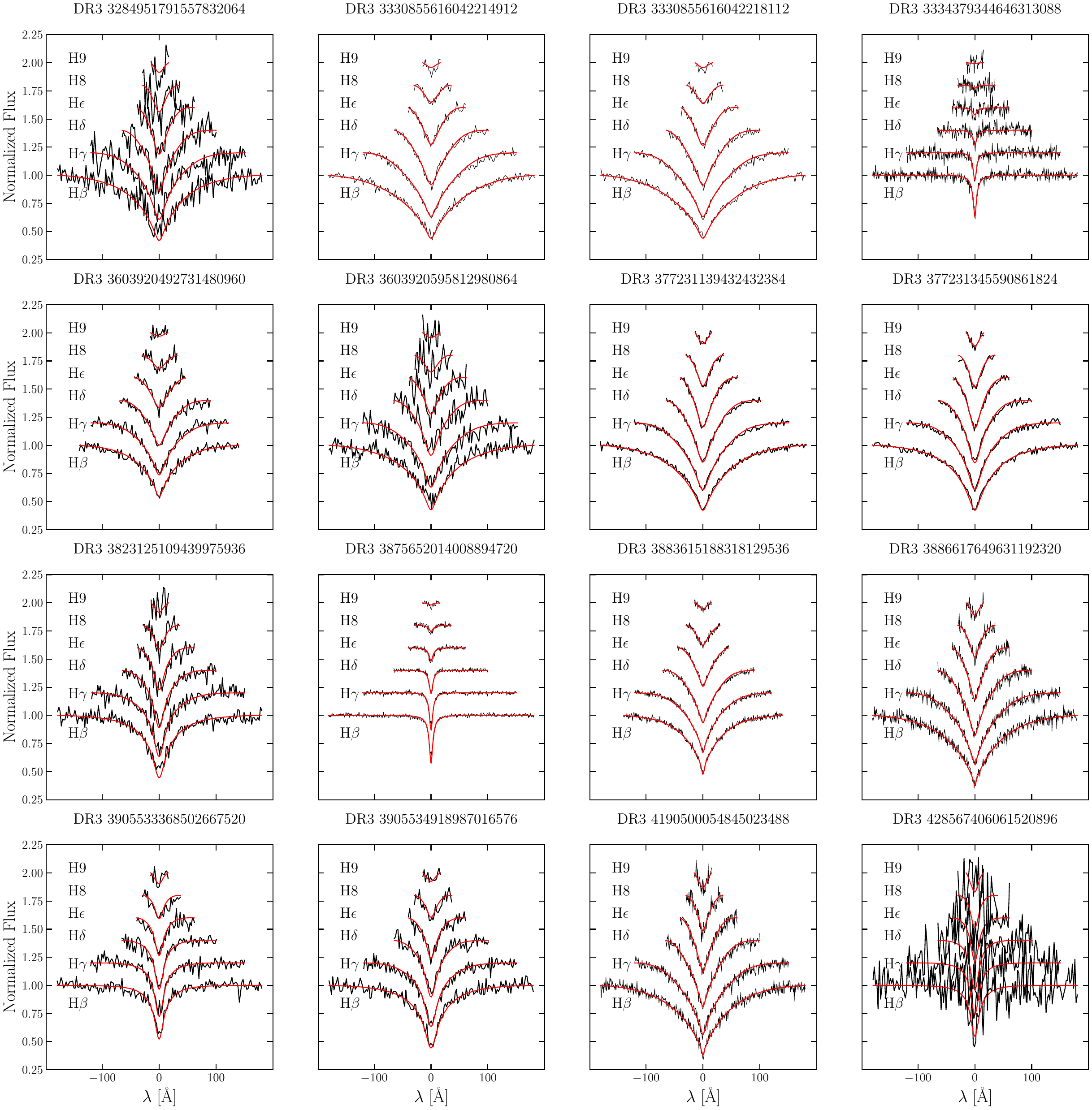}
    \caption{Same as Figure~\ref{fig:DA_fits}, continued.}
    \label{fig:DA_fits4}    
\end{figure}

\pagebreak

\begin{figure}[h]
    \centering
    \includegraphics[width=0.9\columnwidth]{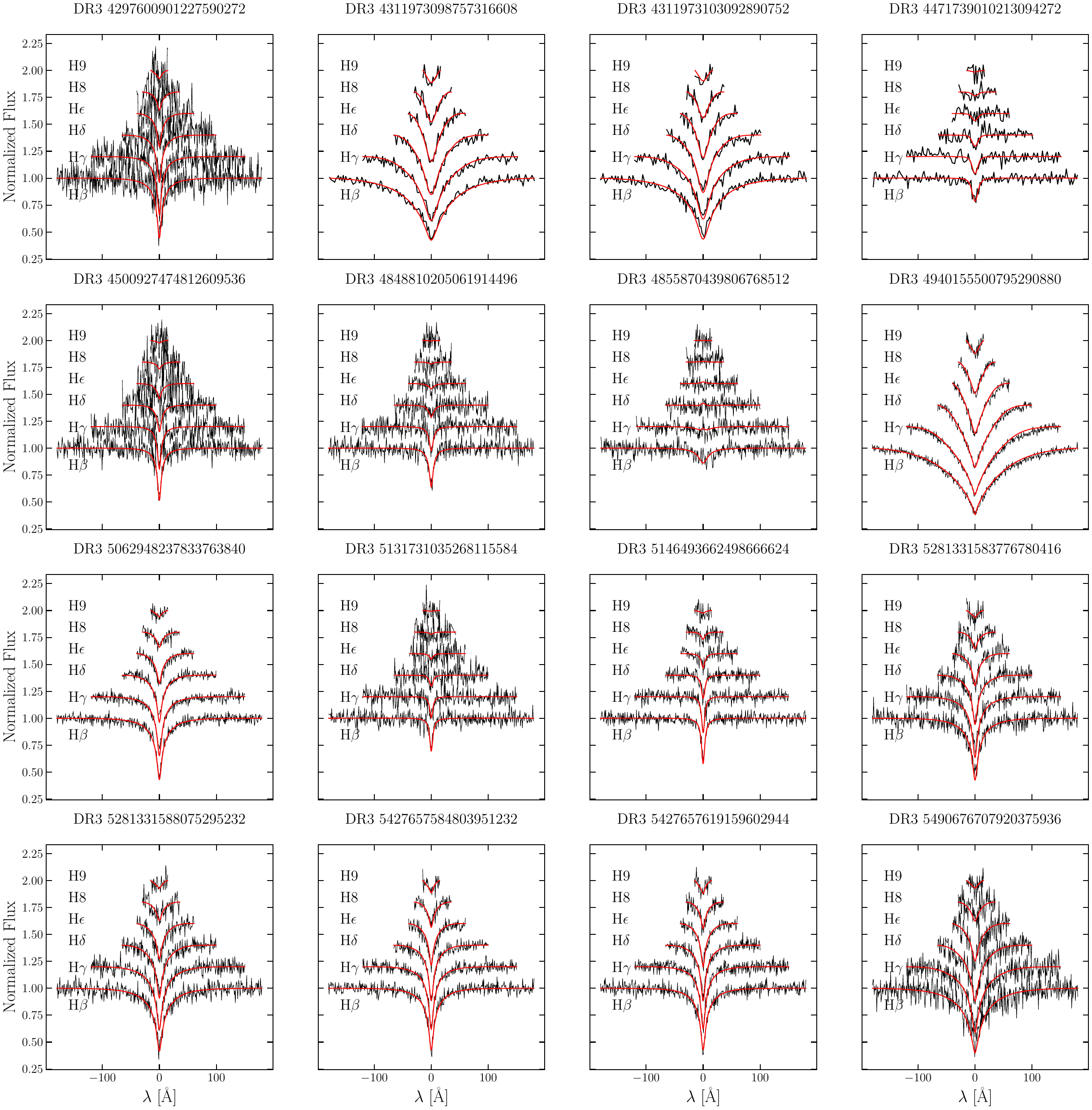}
    \caption{Same as Figure~\ref{fig:DA_fits}, continued.}
    \label{fig:DA_fits5}    
\end{figure}

\pagebreak

\begin{figure}[h]
    \centering
    \includegraphics[width=0.9\columnwidth]{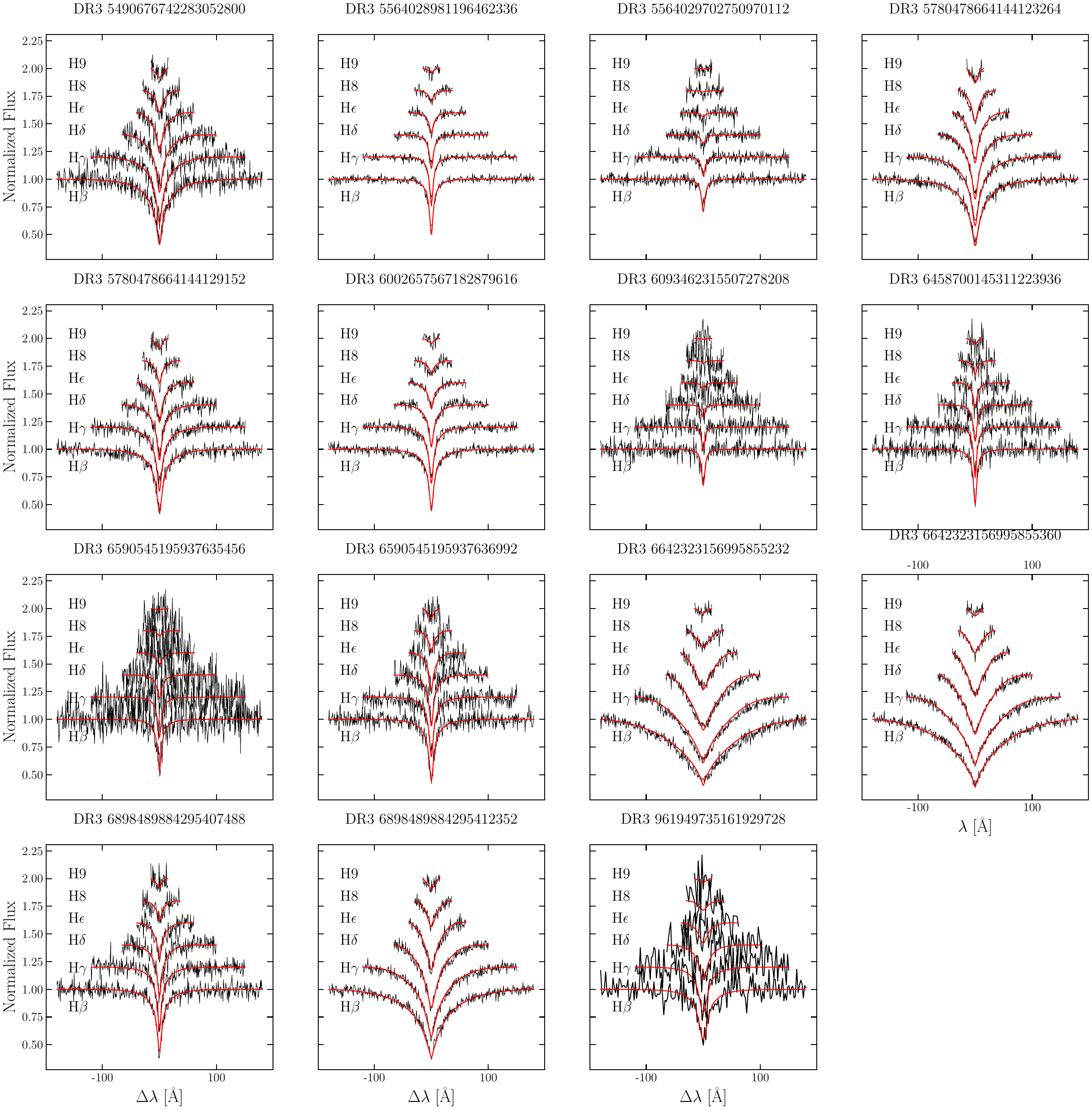}
    \caption{Same as Figure~\ref{fig:DA_fits}, continued.}
    \label{fig:DA_fits6}    
\end{figure}

\section{Literature Spectral Types and Spectroscopic Atmospheric Parameters}
Here we tabulate all the spectroscopic parameters and spectral types taken from previous works.

\startlongtable
\begin{deluxetable*}{llccccc}
\tablecolumns{6}
\tablewidth{0pc}
\tablecaption{Literature Spectral Types and Fitted Atmospheric Parameters \label{tab:lit_obs}}
\tablehead{
        \colhead{Gaia DR3 ID}&
        \colhead{WDJ Name}&
        \colhead{$G$}&
        \colhead{Spectral Type}&
        \colhead{Source}&
        \colhead{$T_{\mathrm{eff}}$}&
        \colhead{\logg}\\
  &  & (mag) &  &  & (K) & (dex) }
\startdata
1008929564913828224 & WDJ085917.36+425031.67 & 18.93 & DA & \citealt{2014MNRAS.440.3184B} & $10{,}020\pm230$ & $8.18\pm0.13$ \\
1008929569208837376 & WDJ085917.24+425027.51 & 18.45 & DA & \citealt{2014MNRAS.440.3184B} & $11{,}150\pm250$ & $8.01\pm0.08$ \\
1013776353903292928 & WDJ084952.88+471249.58 & 16.87 & DB & \citealt{2014MNRAS.440.3184B} & $16{,}990\pm390$ & $7.84\pm0.07$ \\
1013776353903293056 & WDJ084952.49+471247.88 & 17.72 & DA & \citealt{2014MNRAS.440.3184B} & $11{,}230\pm260$ & $7.88\pm0.07$ \\
1017810595208747776 & WDJ085909.61+540953.17 & 19.92 & DA & \citealt{2019MNRAS.482.4570G} &  &  \\
1042071701528617856 & WDJ084515.55+611705.79 & 17.17 & DA & \citealt{2020MNRAS.497..130T} & $5470\pm70$ & $8.02\pm0.02$ \\
1068909504756042112 & WDJ091130.39+664502.03 & 19.06 & DA & \citealt{2019MNRAS.482.5222T} & $6900\pm240$ & $8.04\pm0.43$ \\
1099325161772414208 & WDJ070326.09+622251.24 & 15.65 & DA & \citealt{2011ApJ...743..138G} & $17{,}140\pm260$ & $7.98\pm0.05$ \\
1100655330324351616 & WDJ065326.68+640342.05 & 16.49 & DA & \citealt{2011ApJ...743..138G} & $6000\pm200$ & $8.25\pm0.43$ \\
1211378350265079808 & WDJ153554.07+212506.58 & 17.22 & DA & \citealt{2013AJ....145..136L} & $6560\pm110$ & $8.59\pm0.22$ \\
1218155666922068864 & WDJ154403.20+234459.09 & 18.40 & DA & \citealt{2019MNRAS.482.5222T} & $9550\pm110$ & $7.95\pm0.11$ \\
1218155666922071552 & WDJ154403.27+234516.79 & 19.48 & DC & \citealt{2015ApJ...815...63A} &  &  \\
1224799122336440576 & WDJ154017.90+284457.36 & 20.12 & DA & \citealt{2015MNRAS.446.4078K} & $7220\pm110$ & $7.70\pm0.20$ \\
1238352149340737536 & WDJ144921.38+205421.86 & 19.02 & DAH: & \citealt{2020ApJ...898...84K} &  &  \\
1246997338682879104 & WDJ140457.31+201246.15 & 18.74 & DC & \citealt{2015MNRAS.446.4078K} &  &  \\
1247034339825405056 & WDJ140237.20+201525.87 & 19.84 & DC & \citealt{2015MNRAS.446.4078K} &  &  \\
1248152852388496896 & WDJ134426.44+184931.13 & 18.03 & DQ & \citealt{2013ApJS..204....5K} &  &  \\
1302684712116023552 & WDJ161229.09+233422.61 & 19.37 & DA & \citealt{2019MNRAS.482.5222T} & $10{,}470\pm260$ & $8.20\pm0.18$ \\
1332001785217988992 & WDJ162555.28+375920.90 & 18.17 & DA & \citealt{2019MNRAS.482.5222T} & $7080\pm90$ & $8.27\pm0.15$ \\
1337576648471644288 & WDJ170356.78+330435.85 & 18.21 & DA & \citealt{2014MNRAS.440.3184B} & $11{,}310\pm260$ & $8.01\pm0.07$ \\
1337576648471644800 & WDJ170355.91+330438.49 & 18.78 & DA & \citealt{2019MNRAS.482.5222T} & $9500\pm110$ & $7.68\pm0.13$ \\
1377401818623167104 & WDJ153953.06+385520.65 & 19.01 & DBZ: & \citealt{2015MNRAS.446.4078K} &  &  \\
1385162244707678720 & WDJ160521.16+430435.93 & 14.80 & DA & \citealt{2011ApJ...743..138G} & $37{,}550\pm560$ & $8.02\pm0.05$ \\
1386046148977928832 & WDJ160641.92+442027.32 & 20.14 & DA: & \citealt{2015MNRAS.446.4078K} &  &  \\
1398291165161647872 & WDJ155244.41+473124.05 & 19.11 & DA & \citealt{2014MNRAS.440.3184B} & $16{,}540\pm380$ & $7.91\pm0.07$ \\
1398291165161649280 & WDJ155245.20+473129.53 & 18.88 & DA & \citealt{2014MNRAS.440.3184B} & $18{,}770\pm430$ & $7.94\pm0.07$ \\
1408135749896103936 & WDJ170530.97+480310.27 & 14.53 & DA & \citealt{2011ApJ...743..138G} & $14{,}170\pm290$ & $8.08\pm0.05$ \\
1408135749896104192 & WDJ170530.44+480312.36 & 14.39 & DA & \citealt{2011ApJ...743..138G} & $9250\pm140$ & $7.65\pm0.06$ \\
1408293285000502400 & WDJ165659.52+475648.83 & 20.07 & DA & \citealt{2016MNRAS.455.3413K} & $8990\pm110$ & $8.02\pm0.12$ \\
1448097259489385216 & WDJ133441.32+255646.53 & 16.84 & DA & \citealt{2007ApJ...671.1708B} &  &  \\
1448906637486139392 & WDJ133611.40+273748.57 & 18.92 & DC & \citealt{2013ApJS..204....5K} &  &  \\
1450779346306149760 & WDJ135834.67+263343.19 & 17.07 & DA & \citealt{2019MNRAS.482.4570G} &  &  \\
1459032658542428416 & WDJ135228.45+342906.73 & 17.27 & DB & \citealt{2015MNRAS.446.4078K} & $11{,}600\pm140$ &  \\
1459033345737195008 & WDJ135219.74+342853.09 & 19.42 & DA & \citealt{2015MNRAS.446.4078K} & $6860\pm80$ & $8.21\pm0.10$ \\
1463032819282938112 & WDJ131421.52+305050.49 & 17.79 & DA & \citealt{2019MNRAS.482.5222T} & $10{,}250\pm120$ & $8.02\pm0.05$ \\
1463032819282938240 & WDJ131421.73+305051.40 & 18.13 & DA & \citealt{2014MNRAS.440.3184B} & $9170\pm210$ & $7.87\pm0.07$ \\
1473687769854312192 & WDJ132334.63+360912.03 & 18.64 & DA & \citealt{2019MNRAS.482.5222T} & $10{,}040\pm120$ & $8.02\pm0.09$ \\
1478574038183169024 & WDJ141132.59+330836.81 & 20.00 & DC: & \citealt{2015MNRAS.446.4078K} &  &  \\
1498271376679465600 & WDJ141207.61+421627.40 & 16.01 & DA & \citealt{2011ApJ...743..138G} & $16{,}020\pm280$ & $8.01\pm0.05$ \\
1498271406743008640 & WDJ141208.82+421624.85 & 18.28 & DAH & \citealt{2019MNRAS.482.5222T} &  &  \\
1503769858234576128 & WDJ134530.33+461741.85 & 19.34 & DC & \citealt{2016MNRAS.455.3413K} &  &  \\
150577636388844160 & WDJ042029.21+261755.39 & 18.38 & DA & \citealt{2020ApJ...898...84K} &  &  \\
1521307274856035968 & WDJ124038.89+383111.06 & 19.32 & DA & \citealt{2015MNRAS.446.4078K} & $7370\pm90$ & $8.09\pm0.12$ \\
1528594636063458688 & WDJ125748.69+431913.77 & 19.62 & DA: & \citealt{2016MNRAS.455.3413K} &  &  \\
1535762279348704000 & WDJ121124.22+392427.63 & 19.47 & DA & \citealt{2019MNRAS.482.5222T} & $10{,}840\pm160$ & $8.21\pm0.08$ \\
1538128771968706560 & WDJ122138.25+425933.21 & 18.84 & DA & \citealt{2019MNRAS.482.5222T} & $8260\pm110$ & $8.18\pm0.15$ \\
1538128771968879616 & WDJ122138.63+425934.13 & 19.88 & DA & \citealt{2019MNRAS.482.4570G} &  &  \\
1546811340415484800 & WDJ120311.49+494832.30 & 18.82 & DA & \citealt{2019MNRAS.482.5222T} & $7100\pm100$ & $8.09\pm0.17$ \\
1546811344714710912 & WDJ120310.96+494850.91 & 17.35 & DA & \citealt{2015ApJ...815...63A} & $11{,}410\pm220$ & $8.12\pm0.07$ \\
1552114525514355712 & WDJ133820.22+473122.85 & 19.64 & DA: & \citealt{2016MNRAS.455.3413K} &  &  \\
1564527844289256448 & WDJ130955.74+550338.46 & 17.94 & DA & \citealt{2019MNRAS.482.5222T} & $8070\pm100$ & $8.12\pm0.08$ \\
1564527878649314944 & WDJ130954.23+550339.17 & 18.52 & DA & \citealt{2019MNRAS.482.5222T} & $8050\pm100$ & $7.89\pm0.15$ \\
1568524088021418624 & WDJ123940.38+503821.76 & 20.19 & DA & \citealt{2016MNRAS.455.3413K} & $8430\pm100$ & $8.16\pm0.13$ \\
1570271658673389568 & WDJ125733.65+542850.45 & 16.69 & DA & \citealt{2019MNRAS.482.5222T} &  &  \\
1574653689250360576 & WDJ122717.44+563825.70 & 18.30 & DA & \citealt{2016MNRAS.455.3413K} & $15{,}680\pm190$ & $8.07\pm0.02$ \\
1578748824604827648 & WDJ130033.46+590406.83 & 17.93 & DAH & \citealt{2019MNRAS.482.5222T} &  &  \\
1578748858964566400 & WDJ130035.20+590415.59 & 15.38 & DA & \citealt{2011ApJ...743..138G} & $15{,}160\pm240$ & $8.00\pm0.05$ \\
1579105719206817536 & WDJ130606.76+592610.28 & 17.56 & DC: & \citealt{2016MNRAS.455.3413K} &  &  \\
1580126370938935424 & WDJ124238.82+612118.94 & 18.91 & DA & \citealt{2019MNRAS.482.5222T} & $7420\pm160$ & $7.87\pm0.26$ \\
1580767149995407616 & WDJ123156.08+573609.85 & 18.57 & DA & \citealt{2015ApJ...815...63A} & $11{,}190\pm230$ & $7.92\pm0.08$ \\
1580767149995407744 & WDJ123155.74+573611.47 & 17.91 & DA & \citealt{2015ApJ...815...63A} & $15{,}360\pm290$ & $8.01\pm0.05$ \\
1593068623526189696 & WDJ150746.82+520957.95 & 17.92 & DAH & \citealt{2012MNRAS.421..202D} &  &  \\
1593068623526189824 & WDJ150746.49+521002.04 & 17.11 & DA & \citealt{2012MNRAS.421..202D} & $17{,}600\pm210$ & $8.13\pm0.02$ \\
1597058850006489856 & WDJ154359.44+534456.73 & 20.46 & DC: & \citealt{2016MNRAS.455.3413K} &  &  \\
1602106540386957184 & WDJ153904.23+581115.83 & 17.66 & DA & \citealt{2019MNRAS.482.5222T} & $6930\pm80$ & $7.77\pm0.13$ \\
16166875377576064 & WDJ032026.89+111327.36 & 18.88 & DA & \citealt{2020ApJ...898...84K} &  &  \\
16166875378033664 & WDJ032026.96+111324.62 & 18.08 & DA & \citealt{2020ApJ...898...84K} &  &  \\
1648851933643976448 & WDJ164023.74+685424.81 & 16.49 & DA & \citealt{2011ApJ...743..138G} & $18{,}770\pm300$ & $8.05\pm0.05$ \\
1662446600351645184 & WDJ134246.13+593830.72 & 19.81 & DA & \citealt{2016MNRAS.455.3413K} & $10{,}090\pm120$ & $8.79\pm0.12$ \\
1675399225287958272 & WDJ131022.04+611804.34 & 19.12 & DC & \citealt{1991AJ....101.1476S} &  &  \\
1678050701869897088 & WDJ131520.69+652828.51 & 19.85 & DA & \citealt{2019MNRAS.482.5222T} &  &  \\
1678050873667756032 & WDJ131520.60+652828.09 & 19.54 & DA & \citealt{2019MNRAS.482.5222T} &  &  \\
1682554091043762560 & WDJ123647.43+682502.95 & 18.58 & DC & \citealt{2015MNRAS.446.4078K} &  &  \\
1765847182089067008 & WDJ214538.58+110619.69 & 19.93 & DC & \citealt{1999ApJS..121....1M} &  &  \\
1767494804558717824 & WDJ214208.53+132844.23 & 16.49 & DA & \citealt{2011ApJ...743..138G} & $7660\pm130$ & $8.09\pm0.13$ \\
1781347890859073024 & WDJ221019.65+203906.94 & 19.64 & DA & \citealt{2019MNRAS.482.5222T} & $10{,}200\pm200$ & $7.41\pm0.19$ \\
1875185779452781952 & WDJ222426.91+231536.18 & 17.73 & DA & \citealt{2014MNRAS.440.3184B} & $10{,}740\pm250$ & $8.01\pm0.07$ \\
1875185886825938816 & WDJ222427.07+231537.56 & 17.29 & DA & \citealt{2014MNRAS.440.3184B} & $13{,}200\pm290$ & $7.95\pm0.07$ \\
1942154318286035328 & WDJ233246.21+491709.15 & 18.88 & DA & \citealt{2019MNRAS.482.5222T} & $18{,}200\pm340$ & $7.92\pm0.06$ \\
2066035777984385664 & WDJ204401.02+403014.24 & 15.44 & DA & \citealt{2015ApJ...815...63A} & $13{,}590\pm240$ & $8.00\pm0.05$ \\
2066035777984385792 & WDJ204400.72+403005.88 & 17.64 & DAH & \citealt{2015ApJ...815...63A} &  &  \\
2274076297221555968 & WDJ212657.66+733844.66 & 12.89 & DA & \citealt{2011ApJ...743..138G} & $16{,}110\pm240$ & $7.97\pm0.04$ \\
2317685436639994368 & WDJ003918.81-310306.82 & 17.91 & DA & \citealt{2004MNRAS.349.1397C} &  &  \\
2364272573237917952 & WDJ003230.41-175322.62 & 16.58 & DA & \citealt{2011ApJ...743..138G} & $14{,}270\pm390$ & $7.94\pm0.06$ \\
2416481783371550976 & WDJ000734.82-160531.61 & 16.15 & DA & \citealt{2011ApJ...743..138G} & $14{,}920\pm250$ & $7.93\pm0.05$ \\
2422606780396753024 & WDJ000022.54-105142.20 & 18.77 & DA & \citealt{2011ApJ...730..128T} & $8620\pm110$ & $8.06\pm0.15$ \\
2429392661221943040 & WDJ001440.69-075856.97 & 19.26 & DC & \citealt{2020ApJ...898...84K} &  &  \\
2448158399834236928 & WDJ000012.01-030831.10 & 19.87 & DA & \citealt{2019MNRAS.482.4570G} &  &  \\
2448482961923062400 & WDJ000720.13-023456.24 & 18.79 & DA & \citealt{2016MNRAS.455.3413K} & $6760\pm80$ & $7.43\pm0.12$ \\
2469900005324049280 & WDJ010904.22-104215.39 & 18.73 & DA & \citealt{2020ApJ...898...84K} &  &  \\
2469900009618822656 & WDJ010903.42-104214.15 & 16.60 & DA & \citealt{2020ApJ...898...84K} &  &  \\
2477317211979980544 & WDJ012824.94-082253.44 & 18.61 & DA & \citealt{2013ApJS..204....5K} & $6680\pm80$ & $7.79\pm0.07$ \\
2477317211980446208 & WDJ012824.99-082252.53 & 18.82 & DA & \citealt{2013ApJS..204....5K} & $6680\pm80$ & $7.79\pm0.07$ \\
2482813425794003200 & WDJ011728.64-043939.68 & 17.95 & DC & \citealt{1999ApJS..121....1M} &  &  \\
2482813430089468160 & WDJ011728.83-043938.41 & 17.87 & DC & \citealt{1999ApJS..121....1M} &  &  \\
2490455242060072064 & WDJ022443.44-024257.96 & 19.03 & DA & \citealt{2020ApJ...898...84K} &  &  \\
2490455242060751360 & WDJ022443.13-024255.30 & 19.09 & DA & \citealt{2020ApJ...898...84K} &  &  \\
2499301496005547264 & WDJ024443.74+011312.39 & 19.86 & DA & \citealt{2019MNRAS.482.5222T} & $8600\pm240$ & $7.99\pm0.32$ \\
2501256839996606464 & WDJ022733.10+005200.36 & 19.64 & DA & \citealt{2019MNRAS.482.5222T} & $10{,}190\pm260$ & $8.40\pm0.21$ \\
2501256839996606592 & WDJ022733.16+005153.73 & 19.83 & DA & \citealt{2019MNRAS.482.5222T} & $9560\pm110$ & $8.07\pm0.12$ \\
2512937609148915840 & WDJ015343.13+042533.07 & 20.31 & DA & \citealt{2019MNRAS.482.4570G} &  &  \\
2588702614661788544 & WDJ013910.70+144726.68 & 19.45 & DA & \citealt{2019MNRAS.482.5222T} & $9240\pm240$ & $8.85\pm0.25$ \\
2629899631727265280 & WDJ222542.67-011405.31 & 18.13 & DA & \citealt{2020ApJ...898...84K} &  &  \\
2629899631727265920 & WDJ222543.52-011359.51 & 19.21 & DA & \citealt{2020ApJ...898...84K} &  &  \\
2644149993913188224 & WDJ232658.90-002339.82 & 19.21 & DA & \citealt{2013ApJS..204....5K} & $7530\pm90$ & $8.13\pm0.20$ \\
2644149993913188352 & WDJ232659.23-002347.96 & 17.52 & DA & \citealt{2011ApJ...730..128T} & $10{,}470\pm130$ & $8.08\pm0.04$ \\
2645295921252503424 & WDJ232115.34+010211.96 & 18.84 & DA & \citealt{2020ApJ...898...84K} &  &  \\
2645295955612242688 & WDJ232115.70+010224.59 & 18.99 & DA & \citealt{2020ApJ...898...84K} &  &  \\
2653032978419246208 & WDJ223730.57-004754.70 & 19.67 & DA & \citealt{2013ApJS..204....5K} &  & \\
2676566272464334720 & WDJ215847.13-024024.42 & 17.15 & DC & \citealt{2020MNRAS.497..130T} &  &  \\
2689364416012661760 & WDJ212057.73-001810.62 & 18.27 & DA & \citealt{2011ApJ...730..128T} & $8630\pm100$ & $7.96\pm0.10$ \\
2701318787466116352 & WDJ213648.79+064320.22 & 18.10 & DA & \citealt{2015MNRAS.446.4078K} & $19{,}270\pm230$ & $8.02\pm0.02$ \\
2703590241050519680 & WDJ222017.81+020342.88 & 18.77 & DA: & \citealt{2015MNRAS.446.4078K} &  &  \\
2731184546933975040 & WDJ222650.27+130832.75 & 17.71 & DA & \citealt{2019MNRAS.482.4570G} &  &  \\
2745919102257342976 & WDJ000215.38+073359.56 & 17.72 & DA & \citealt{2015ApJ...815...63A} &  &  \\
2745919106553695616 & WDJ000216.18+073350.30 & 17.95 & DAH & \citealt{2015ApJ...815...63A} &  &  \\
2749530521214502144 & WDJ003310.14+074816.35 & 20.04 & DA & \citealt{2019MNRAS.482.4570G} &  &  \\
2749533025179314816 & WDJ003247.88+074934.00 & 20.09 & DA & \citealt{2019MNRAS.482.4570G} &  &  \\
2751252493861856000 & WDJ003424.36+102641.24 & 18.26 & DA & \citealt{2015MNRAS.446.4078K} & $7110\pm90$ & $7.85\pm0.06$ \\
2757867396332920576 & WDJ233232.23+080230.00 & 18.19 & DBA & \citealt{2016MNRAS.455.3413K} & $36{,}500\pm440$ & $7.82\pm0.04$ \\
2773334978019335040 & WDJ235515.27+170806.73 & 18.23 & DA & \citealt{2015ApJ...815...63A} & $10{,}160\pm160$ & $8.31\pm0.07$ \\
2773334982315549824 & WDJ235515.44+170830.99 & 16.41 & DB & \citealt{2015ApJ...815...63A} & $21{,}470\pm440$ & $8.12\pm0.03$ \\
2777645067895669376 & WDJ005212.26+135302.04 & 17.81 & DA & \citealt{2014MNRAS.440.3184B} & $19{,}120\pm440$ & $7.94\pm0.07$ \\
2777645067895669504 & WDJ005212.72+135301.13 & 18.84 & DA & \citealt{2014MNRAS.440.3184B} & $10{,}880\pm250$ & $7.93\pm0.07$ \\
2794800056333855232 & WDJ003051.77+181053.88 & 18.82 & DA & \citealt{2015ApJ...815...63A} & $14{,}070\pm1050$ & $8.42\pm0.18$ \\
2794800060629297152 & WDJ003051.81+181046.16 & 18.85 & DA & \citealt{2015ApJ...815...63A} & $15{,}620\pm2930$ & $8.43\pm0.35$ \\
2815944352131513088 & WDJ225932.21+140439.28 & 16.59 & DA & \citealt{2011ApJ...743..138G} & $28{,}340\pm410$ & $8.39\pm0.05$ \\
2815944352131513472 & WDJ225932.74+140444.25 & 18.63 & DAH & \citealt{2015ApJ...815...63A} &  &  \\
2839231634746334848 & WDJ231814.91+234509.51 & 19.44 & DA & \citealt{2020ApJ...898...84K} &  &  \\
2839231634746334976 & WDJ231815.26+234511.42 & 18.99 & DC & \citealt{2020ApJ...898...84K} &  &  \\
2853079051690361088 & WDJ000142.79+251504.18 & 18.86 & DA & \citealt{2019MNRAS.482.5222T} & $18{,}470\pm220$ & $8.01\pm0.03$ \\
2853985019206468352 & WDJ000223.03+272358.26 & 17.68 & DB & \citealt{2013ApJS..204....5K} & $16{,}760\pm200$ & $7.96\pm0.06$ \\
2876148729784835584 & WDJ001840.46+344147.72 & 19.43 & DA & \citealt{2020ApJ...898...84K} &  &  \\
2876148734080688256 & WDJ001839.94+344154.35 & 18.08 & DA & \citealt{2019MNRAS.482.5222T} & $7470\pm90$ & $7.94\pm0.07$ \\
2879072503002076672 & WDJ234526.50+350754.23 & 19.98 & DA & \citealt{2019MNRAS.482.4570G} &  &  \\
294062563782633216 & WDJ011714.49+244021.58 & 19.70 & DA & \citealt{2019MNRAS.482.5222T} & $17{,}670\pm590$ & $8.26\pm0.10$ \\
3072961070640767488 & WDJ082730.59-021620.14 & 17.99 & DA & \citealt{2015ApJ...815...63A} & $27{,}310\pm450$ & $8.49\pm0.06$ \\
3072961074934467200 & WDJ082730.73-021618.52 & 18.26 & DA & \citealt{2015ApJ...815...63A} & $27{,}860\pm490$ & $8.58\pm0.07$ \\
3152007091864270336 & WDJ075410.54+123947.46 & 18.85 & DA & \citealt{2015ApJ...815...63A} & $14{,}190\pm1070$ & $8.24\pm0.12$ \\
3152007091867073152 & WDJ075410.59+123945.62 & 18.90 & DA & \citealt{2015ApJ...815...63A} & $13{,}690\pm630$ & $8.31\pm0.11$ \\
323375784298256896 & WDJ012535.00+385047.07 & 19.11 & DA & \citealt{2019MNRAS.482.5222T} & $13{,}570\pm410$ & $8.02\pm0.06$ \\
3238868098140387840 & WDJ051013.94+043838.43 & 14.23 & DA & \citealt{2011ApJ...743..138G} & $21{,}550\pm320$ & $8.08\pm0.04$ \\
3238868171156736768 & WDJ051013.52+043855.13 & 15.38 & DA & \citealt{2011ApJ...743..138G} & $12{,}010\pm190$ & $8.19\pm0.05$ \\
3243842361760113408 & WDJ033826.84-073155.40 & 16.49 & DA & \citealt{2013ApJS..204....5K} & $15{,}640\pm190$ & $8.07\pm0.02$ \\
3247469062209123072 & WDJ032913.02-055520.75 & 18.51 & DA & \citealt{2020ApJ...898...84K} &  &  \\
3247469130928600960 & WDJ032911.76-055501.34 & 19.36 & DA & \citealt{2020ApJ...898...84K} &  &  \\
3263351198434024576 & WDJ033236.61-004918.41 & 18.21 & DA & \citealt{2014MNRAS.440.3184B} & $10{,}400\pm240$ & $8.03\pm0.07$ \\
3263351202729621376 & WDJ033236.87-004936.91 & 15.84 & DA & \citealt{2011ApJ...743..138G} & $34{,}390\pm410$ & $7.80\pm0.02$ \\
3301217592917972864 & WDJ035454.20+074608.59 & 16.54 & DA & \citealt{2011ApJ...743..138G} & $16{,}380\pm310$ & $7.87\pm0.06$ \\
3310871201928485248 & WDJ042641.73+141215.87 & 17.85 & DA & \citealt{2011ApJ...743..138G} & $19{,}550\pm420$ & $8.88\pm0.07$ \\
339492017717622016 & WDJ022551.96+422803.52 & 17.31 & DA & \citealt{2011ApJ...743..138G} & $6320\pm280$ & $8.15\pm0.56$ \\
3566532561902107648 & WDJ111622.76-103516.10 & 18.99 & DA & \citealt{2020ApJ...898...84K} &  &  \\
3566532561902107904 & WDJ111623.74-103516.63 & 18.30 & DA & \citealt{2020ApJ...898...84K} &  &  \\
3647484552174618368 & WDJ141419.61-012217.76 & 18.62 & DA & \citealt{2015MNRAS.446.4078K} &  &  \\
3647484556468720256 & WDJ141418.87-012214.94 & 19.11 & DA & \citealt{2015MNRAS.446.4078K} & $7900\pm90$ & $7.84\pm0.05$ \\
3682458814461982592 & WDJ125458.08-021837.75 & 16.64 & DA & \citealt{2009ANA...505..441K} & $15{,}930\pm280$ & $7.81\pm0.07$ \\
3683519503881169920 & WDJ124428.57-011857.85 & 13.99 & DA & \citealt{2011ApJ...743..138G} & $24{,}480\pm370$ & $7.37\pm0.05$ \\
3706830967160380416 & WDJ123007.22+033856.12 & 18.91 & DA & \citealt{2019MNRAS.482.5222T} & $8630\pm100$ & $8.18\pm0.12$ \\
3739031165907785216 & WDJ133552.03+123709.07 & 16.30 & DA & \citealt{2020ApJ...898...84K} &  &  \\
3739031268987206784 & WDJ133552.15+123716.40 & 19.62 & DA & \citealt{2020ApJ...898...84K} &  &  \\
3745268523572692480 & WDJ132814.32+163151.39 & 16.47 & DA & \citealt{2014MNRAS.440.3184B} & $20{,}000\pm460$ & $8.19\pm0.07$ \\
3745268523573825792 & WDJ132814.40+163150.79 & 17.42 & DA & \citealt{2011ApJ...743..138G} & $18{,}800\pm320$ & $8.25\pm0.05$ \\
3796545519645331584 & WDJ112721.31-020837.66 & 16.39 & DA & \citealt{2011ApJ...743..138G} & $25{,}180\pm400$ & $7.74\pm0.05$ \\
3798238454018973056 & WDJ112630.75+001958.86 & 19.84 & DA: & \citealt{2015MNRAS.446.4078K} &  &  \\
3809511712378896384 & WDJ105306.81+025027.98 & 19.07 & DA & \citealt{2019MNRAS.482.5222T} & $12{,}330\pm410$ & $8.29\pm0.10$ \\
3826588571067552256 & WDJ095204.00-023044.14 & 18.08 & DA: & \citealt{2015MNRAS.446.4078K} &  &  \\
3828044221382810752 & WDJ094416.63-001855.53 & 19.59 & DAH: & \citealt{2015MNRAS.446.4078K} &  &  \\
3835861439819152128 & WDJ101359.85+030553.90 & 18.09 & DA & \citealt{2020ApJ...898...84K} &  &  \\
3835866563715176192 & WDJ101401.60+030550.42 & 18.03 & DA & \citealt{2020ApJ...898...84K} &  &  \\
3845506012220187008 & WDJ091659.86+032638.80 & 19.87 & DA & \citealt{2015MNRAS.446.4078K} & $8420\pm100$ & $8.63\pm0.13$ \\
3857475197016034432 & WDJ104346.72+030318.45 & 19.14 & DQ:PEC: & \citealt{2013ApJS..204....5K} &  &  \\
3874412413432643328 & WDJ100623.17+071154.30 & 18.64 & DC & \citealt{2020ApJ...898...84K} &  &  \\
3874412413432647680 & WDJ100623.08+071212.70 & 15.98 & DA & \citealt{2019MNRAS.482.5222T} & $9550\pm110$ & $7.96\pm0.03$ \\
3900127108484055552 & WDJ122319.59+050121.26 & 18.71 & DA & \citealt{2019MNRAS.482.5222T} & $8820\pm110$ & $8.25\pm0.11$ \\
3902698243410506112 & WDJ123313.51+082402.65 & 18.36 & DQPEC & \citealt{2013ApJS..204....5K} &  &  \\
3906688955223470080 & WDJ120715.25+105337.88 & 18.42 & DA & \citealt{2015MNRAS.446.4078K} & $19{,}550\pm230$ & $8.14\pm0.02$ \\
3906688955223834368 & WDJ120715.45+105339.24 & 20.16 & DA & \citealt{2019MNRAS.482.4570G} &  &  \\
3920275276810355072 & WDJ115937.82+134414.09 & 18.02 & DA & \citealt{2019MNRAS.482.5222T} & $9490\pm110$ & $7.99\pm0.07$ \\
3920275276810355200 & WDJ115937.83+134408.91 & 18.37 & DA & \citealt{2014MNRAS.440.3184B} & $16{,}400\pm380$ & $8.94\pm0.07$ \\
3930720057454344960 & WDJ125438.31+143228.61 & 16.93 & DC & \citealt{2020ApJ...898...84K} &  &  \\
3930720263611949056 & WDJ125437.34+143242.21 & 20.06 & DC & \citealt{2020ApJ...898...84K} &  &  \\
3937174942327932544 & WDJ131426.39+173228.40 & 18.43 & DAH & \citealt{2019MNRAS.482.5222T} &  &  \\
3937174946624964224 & WDJ131426.84+173209.50 & 16.28 & DA & \citealt{2015ApJ...815...63A} & $12{,}530\pm210$ & $8.04\pm0.05$ \\
3940068410255312768 & WDJ131332.15+203039.39 & 17.86 & DA & \citealt{2014MNRAS.440.3184B} & $13{,}440\pm300$ & $8.30\pm0.07$ \\
3940068414551140608 & WDJ131332.57+203039.19 & 17.57 & DA & \citealt{2014MNRAS.440.3184B} & $13{,}210\pm290$ & $8.08\pm0.07$ \\
3966161676608942976 & WDJ113617.93+141812.47 & 18.63 & DA & \citealt{2019MNRAS.482.5222T} & $9360\pm110$ & $7.88\pm0.10$ \\
3970693313784409344 & WDJ112442.94+170513.95 & 18.54 & DA & \citealt{2019MNRAS.482.5222T} & $8100\pm100$ & $8.36\pm0.11$ \\
3987906992950122112 & WDJ104051.61+213033.75 & 20.11 & DA & \citealt{2015MNRAS.446.4078K} & $7880\pm100$ & $8.12\pm0.15$ \\
3989385561210153344 & WDJ104459.57+215059.22 & 18.45 & DA & \citealt{2019MNRAS.482.5222T} & $7610\pm90$ & $7.82\pm0.14$ \\
3993035733656088320 & WDJ113136.33+234913.00 & 19.26 & DAZ & \citealt{2016MNRAS.455.3413K} & $7440\pm90$ & $8.26\pm0.06$ \\
41518658576469760 & WDJ033912.60+143120.70 & 18.96 & DA & \citealt{2020ApJ...898...84K} &  &  \\
42174998299513728 & WDJ032538.35+150900.26 & 18.07 & DC & \citealt{1999ApJS..121....1M} &  &  \\
4228388602763827584 & WDJ204711.40+002127.73 & 17.17 & DA & \citealt{2019MNRAS.482.5222T} & $14{,}390\pm180$ & $8.04\pm0.02$ \\
4228388774562523264 & WDJ204713.68+002203.96 & 18.24 & DQ:M: & \citealt{2013ApJS..204....5K} &  &  \\
4284122782775063552 & WDJ182713.09+040346.79 & 13.92 & DA & \citealt{2011ApJ...743..138G} & $14{,}350\pm220$ & $7.63\pm0.05$ \\
42871199614383616 & WDJ034411.48+150945.81 & 16.81 & DA & \citealt{2015ApJ...815...63A} & $8300\pm130$ & $7.92\pm0.09$ \\
42871268333859584 & WDJ034410.90+151022.33 & 16.45 & DC & \citealt{2015ApJ...815...63A} &  &  \\
4408166832745003904 & WDJ161452.94+001632.49 & 18.86 & DA & \citealt{2019MNRAS.482.5222T} & $8120\pm140$ & $7.72\pm0.21$ \\
4438147422454430464 & WDJ161511.12+062155.66 & 20.24 & DA & \citealt{2019MNRAS.482.4570G} &  &  \\
4438147422454430464 & WDJ161511.12+062155.66 & 20.39 & DA & \citealt{2019MNRAS.482.4570G} &  &  \\
4450425359563998720 & WDJ160935.76+065509.73 & 17.65 & DQ & \citealt{2015MNRAS.446.4078K} &  &  \\
4454302237563072768 & WDJ155715.82+083022.64 & 18.67 & DA & \citealt{2019MNRAS.482.5222T} & $8160\pm100$ & $7.90\pm0.14$ \\
4455997237817243776 & WDJ155903.78+105614.51 & 18.75 & DC:DA: & \citealt{2013ApJS..204....5K} &  &  \\
4571341505129076992 & WDJ170910.38+231933.58 & 17.31 & DC & \citealt{2015MNRAS.446.4078K} &  &  \\
4592910074976281472 & WDJ181331.21+324831.29 & 16.31 & DA & \citealt{2011ApJ...743..138G} & $7600\pm120$ & $7.76\pm0.14$ \\
4592910105037219072 & WDJ181335.45+324843.78 & 16.93 & DA & \citealt{2011ApJ...743..138G} & $6420\pm220$ & $7.61\pm0.56$ \\
4598830738931385984 & WDJ172929.26+291609.73 & 16.96 & DA & \citealt{2020ApJ...898...84K} &  &  \\
4613612951211823104 & WDJ031715.85-853225.56 & 14.75 & DAH & \citealt{2010ANA...524A..36K} &  &  \\
5096781340192835712 & WDJ041026.50-164142.57 & 16.36 & DA & \citealt{2015MNRAS.453.1879K} &  &  \\
5096781344489104000 & WDJ041024.93-164150.43 & 15.48 & DA & \citealt{2013MNRAS.431..240O} &  &  \\
521406968152848640 & WDJ015901.26+685800.45 & 18.48 & DA & \citealt{2020ApJ...898...84K} &  &  \\
5436014972680358272 & WDJ093659.94-372126.91 & 15.14 & DA & \citealt{2011ApJ...743..138G} & $7930\pm110$ & $7.87\pm0.05$ \\
5436014972680358784 & WDJ093659.79-372130.80 & 14.78 & DQ & \citealt{1991AJ....101.1476S} &  &  \\
5649105922480851200 & WDJ085550.71-263745.13 & 16.92 & DA & \citealt{1999ApJS..121....1M} &  &  \\
5649105926779602432 & WDJ085551.03-263750.20 & 16.77 & DA & \citealt{1999ApJS..121....1M} &  &  \\
5762406957886626816 & WDJ090217.30-040655.48 & 13.17 & DA & \citealt{2011ApJ...743..138G} & $24{,}550\pm370$ & $7.89\pm0.05$ \\
582509956042253312 & WDJ084001.81+051628.41 & 17.97 & DA & \citealt{2019MNRAS.482.5222T} & $7770\pm90$ & $7.87\pm0.09$ \\
5874024769842933760 & WDJ145811.64-631733.46 & 17.51 & DA & \citealt{2011ApJ...743..138G} & $10{,}270\pm150$ & $8.19\pm0.05$ \\
5874024769842933760 & WDJ145811.64-631733.46 & 16.83 & DBA & \citealt{2011ApJ...737...28B} & $14{,}050\pm340$ & $7.96\pm0.09$ \\
606938634805314816 & WDJ090402.84+134915.01 & 16.10 & DA & \citealt{2015ApJ...815...63A} & $9100\pm140$ & $7.78\pm0.08$ \\
606938634805314944 & WDJ090402.86+134911.33 & 16.60 & DA & \citealt{2019MNRAS.482.5222T} & $8170\pm100$ & $7.95\pm0.04$ \\
627764274093469696 & WDJ095427.16+195431.15 & 19.07 & DA & \citealt{2019MNRAS.482.5222T} & $7080\pm190$ & $7.74\pm0.34$ \\
627764274093472896 & WDJ095427.29+195448.38 & 19.02 & DA & \citealt{2019MNRAS.482.5222T} & $7310\pm150$ & $7.61\pm0.28$ \\
630770819920096640 & WDJ092513.50+160144.36 & 16.35 & DA & \citealt{2011ApJ...743..138G} & $24{,}670\pm380$ & $8.35\pm0.05$ \\
630770819920096768 & WDJ092513.20+160145.62 & 17.33 & DA & \citealt{1997ApJ...489L..79F} & $25{,}780\pm590$ & $9.04\pm0.07$ \\
641625576666483584 & WDJ094507.90+232723.59 & 17.26 & DA & \citealt{2011ApJ...743..138G} & $7240\pm110$ & $7.81\pm0.09$ \\
641625576666484480 & WDJ094508.66+232731.64 & 17.41 & DA & \citealt{2011ApJ...743..138G} & $6960\pm120$ & $8.15\pm0.14$ \\
646608632082864256 & WDJ094250.63+260059.85 & 14.62 & DA & \citealt{2011ApJ...743..138G} & $70{,}580\pm1450$ & $7.94\pm0.07$ \\
678526595643807872 & WDJ083413.47+242600.78 & 20.19 & DA & \citealt{2019MNRAS.482.4570G} &  &  \\
6917231476602010240 & WDJ210155.82-005745.10 & 17.80 & DA:DC: & \citealt{2013ApJS..204....5K} & $25{,}910\pm350$ & $7.95\pm0.04$ \\
695748628523317376 & WDJ092212.25+292122.09 & 18.45 & DA & \citealt{2019MNRAS.482.5222T} & $7210\pm130$ & $7.87\pm0.23$ \\
713277141676298240 & WDJ085915.51+330637.62 & 18.77 & DA & \citealt{2014MNRAS.440.3184B} & $11{,}120\pm870$ & $7.75\pm0.39$ \\
713277141676298496 & WDJ085915.03+330644.69 & 18.18 & DA & \citealt{2014MNRAS.440.3184B} & $12{,}040\pm820$ & $8.05\pm0.18$ \\
757911884925087104 & WDJ111319.44+323817.72 & 18.79 & DA & \citealt{2019MNRAS.482.5222T} & $6920\pm160$ & $7.73\pm0.30$ \\
757911988004305280 & WDJ111322.54+323858.80 & 18.96 & DA & \citealt{2019MNRAS.482.5222T} & $7670\pm180$ & $8.63\pm0.25$ \\
769201876477782144 & WDJ114424.52+412847.35 & 18.67 & DA & \citealt{2019MNRAS.482.5222T} & $8590\pm100$ & $8.01\pm0.07$ \\
769201880773377920 & WDJ114424.57+412847.24 & 18.80 & DA & \citealt{2019MNRAS.482.5222T} & $8590\pm100$ & $8.01\pm0.07$ \\
779305808516231552 & WDJ104537.32+400535.94 & 18.90 & DA & \citealt{2011ApJ...730..128T} & $7760\pm90$ & $8.02\pm0.10$ \\
779305808516517376 & WDJ104537.43+400535.72 & 18.79 & DA & \citealt{2019MNRAS.482.5222T} & $7740\pm90$ & $7.99\pm0.10$ \\
779767461666035584 & WDJ104630.60+405905.62 & 18.89 & DA & \citealt{2019MNRAS.482.5222T} & $7220\pm120$ & $7.62\pm0.22$ \\
782193985044906752 & WDJ111020.98+451801.84 & 17.03 & DA & \citealt{2015ApJ...815...63A} & $19{,}000\pm300$ & $8.12\pm0.05$ \\
782194019404645632 & WDJ111016.68+451736.51 & 17.79 & DA & \citealt{2015ApJ...815...63A} & $13{,}700\pm370$ & $8.10\pm0.06$ \\
795886439568266368 & WDJ100244.89+360629.77 & 18.90 & DA & \citealt{2019MNRAS.482.5222T} & $11{,}670\pm520$ & $8.29\pm0.15$ \\
795886439568268032 & WDJ100245.87+360653.50 & 18.99 & DA & \citealt{2015ApJ...815...63A} & $11{,}650\pm580$ & $8.26\pm0.21$ \\
798602271945568256 & WDJ092551.68+354000.58 & 17.67 & DA & \citealt{2019MNRAS.482.5222T} & $7060\pm80$ & $7.87\pm0.10$ \\
804040108562044288 & WDJ102141.29+394215.52 & 19.89 & DA & \citealt{2019MNRAS.482.4570G} &  &  \\
804040486519166976 & WDJ102142.06+394225.50 & 20.24 & DA & \citealt{2019MNRAS.482.4570G} &  &  \\
837550886514384128 & WDJ105449.90+530759.28 & 17.97 & DA & \citealt{2019MNRAS.482.5222T} & $10{,}990\pm130$ & $7.94\pm0.05$ \\
837550890811137664 & WDJ105449.12+530715.35 & 17.65 & DA & \citealt{2015ApJ...815...63A} & $14{,}030\pm470$ & $8.07\pm0.06$ \\
842498482680906496 & WDJ110749.56+524651.01 & 18.31 & DA & \citealt{2019MNRAS.482.5222T} & $7090\pm90$ & $7.67\pm0.17$ \\
842504358196167680 & WDJ110751.05+524653.09 & 18.53 & DA & \citealt{2019MNRAS.482.5222T} & $6670\pm220$ & $8.11\pm0.40$ \\
843874349685029504 & WDJ111726.03+544305.01 & 20.42 & DA & \citealt{2016MNRAS.455.3413K} & $7110\pm140$ & $7.79\pm0.26$ \\
857095736490212608 & WDJ105558.54+564543.62 & 19.52 & DA & \citealt{2016MNRAS.455.3413K} & $8360\pm100$ & $8.36\pm0.12$ \\
868275776880649856 & WDJ073935.24+244505.20 & 17.45 & DB & \citealt{2015MNRAS.446.4078K} & $23{,}000\pm1500$ & $7.83\pm0.02$ \\
879036662822100224 & WDJ074853.08+302543.59 & 17.72 & DAH & \citealt{2019MNRAS.482.5222T} &  &  \\
893134841431510016 & WDJ072147.37+322824.46 & 18.09 & DC & \citealt{2013ApJS..204....5K} &  &  \\
895406329374766208 & WDJ073231.34+353543.74 & 20.00 & DA & \citealt{2019MNRAS.482.4570G} &  &  \\
906922545445227648 & WDJ080410.00+355654.53 & 18.32 & DA & \citealt{2019MNRAS.482.5222T} & $7630\pm90$ & $8.17\pm0.14$ \\
906922549740583808 & WDJ080409.95+355659.11 & 18.70 & DA & \citealt{2019MNRAS.482.5222T} & $6930\pm120$ & $7.52\pm0.23$ \\
925644685844771712 & WDJ075412.09+430230.67 & 18.94 & DA & \citealt{2019MNRAS.482.5222T} & $15{,}360\pm410$ & $7.83\pm0.08$ \\
929001804082135168 & WDJ080644.09+444503.19 & 18.22 & DA & \citealt{2014MNRAS.440.3184B} & $12{,}450\pm280$ & $7.99\pm0.07$ \\
929001804082135296 & WDJ080643.64+444501.43 & 18.76 & DA & \citealt{2014MNRAS.440.3184B} & $10{,}100\pm230$ & $7.96\pm0.08$
\enddata
\end{deluxetable*}
\end{document}